\documentclass[lettersize,journal]{IEEEtran}

\usepackage{amsmath,amssymb,graphicx,amsthm,bbm}
\usepackage{latexsym,todonotes}
\usepackage{mathtools}
\usepackage{cite}
\usepackage{color}
\usepackage{arydshln}
\usepackage{subfigure}
\usepackage{bm}
\usepackage{dsfont}
\usepackage{enumitem}
\usepackage{mathrsfs}
\usepackage{epstopdf}
\usepackage[bookmarks=false]{hyperref}

\usepackage{setspace}
\graphicspath{{./figs/}}

\usepackage{textcomp}
\def\BibTeX{{\rm B\kern-.05em{\sc i\kern-.025em b}\kern-.08em
    T\kern-.1667em\lower.7ex\hbox{E}\kern-.125emX}}
\usepackage[]{geometry}
\newcommand{\yuki}[1]{{\color{blue}{#1}}}

\newcommand*{\shen}[1]{\textbf{\textcolor{violet}{#1}}}

\graphicspath{{figures/}}

\mathcode`*=\string"8000
\begingroup
\catcode`*=\active
\xdef*{\noexpand\textup{\string*}}
\endgroup

\newtheorem{theorem}{\bf Theorem}

\newtheorem{proposition}[theorem]{\bf Proposition}
\newtheorem{corollary}{\bf Corollary}

\usepackage[vlined,ruled,linesnumbered]{algorithm2e}
\usepackage{booktabs}
\usepackage{comment}
\usepackage{subcaption}
\usepackage[percent]{overpic}
\begin{document}

\title{Energy Efficient Offloading Policies in Multi-Access Edge Computing Systems with Task Handover}

\author{Ling Hou, Shi Li,~\IEEEmembership{Student Member,~IEEE}, Zhishu Shen,~\IEEEmembership{Member,~IEEE}, Jing Fu,~\IEEEmembership{Member,~IEEE}, Jingjin Wu,~\IEEEmembership{Member,~IEEE,} and Jiong Jin,~\IEEEmembership{Member,~IEEE}
        % <-this % stops a space
%\thanks{This paper was produced by the IEEE Publication Technology Group. They are in Piscataway, NJ.}% <-this % stops a space
%\thanks{Manuscript received April 19, 2021; revised August 16, 2021.}
%\thanks{Ling Hou completed the majority of this paper when he was with the School of Engineering, STEM College, RMIT University, Melbourne, Australia. E-mail: jasonlinghou666@gmail.com}
%\thanks{Jing Fu is with the School of Engineering, STEM College, RMIT University, Melbourne, Australia. E-mail: jing.fu@rmit.edu.au}
\thanks{Ling Hou and Jing Fu are with the School of Engineering, STEM College, RMIT University, Melbourne, Australia. E-mail: jasonlinghou666@gmail.com, jing.fu@rmit.edu.au}
\thanks{Shi Li and Jiong Jin are with the School of Science, Computing and Engineering Technologies, Swinburne University of Technology, Melbourne, Australia. E-mail: 103791828@student.swin.edu.au, jiongjin@swin.edu.au}
\thanks{Zhishu Shen is with the School of Computer Science and Artificial Intelligence, Wuhan University of Technology, Wuhan, China. E-mail: z\_shen@ieee.org}
\thanks {Jingjin Wu is with the Guangdong Key Laboratory of IRADS, Department of Statistics and Data Science, BNU-HKBU United International College, China. E-mail: jj.wu@ieee.org}
}



\maketitle

\begin{abstract}
The rapid growth of mobile devices and the increasing complexity of tasks have made energy efficiency a critical challenge in Multi-Access Edge Computing (MEC) systems. This paper explores energy-efficient offloading strategies in large-scale MEC systems with heterogeneous mobile users, diverse network components, and frequent task handovers to capture user mobility. The problem is inherently complex due to the system's scale, task and resource diversity, and the need to maintain real-time performance. Traditional optimization approaches are often computationally infeasible for such scenarios. To tackle these challenges, we model the offloading problem using the restless multi-armed bandit (RMAB) framework and develop two scalable online policies that prioritize resources based on their marginal costs. The proposed policies dynamically adapt to the system's heterogeneity and mobility while ensuring near-optimal energy efficiency. Through extensive numerical simulations, we demonstrate that the policies significantly outperform baseline methods in power conservation and show robust performance under non-exponentially distributed task lifespans. These results highlight the practical applicability and scalability of our approach in dynamic MEC environments.

\end{abstract}


\section{Introduction} \label{sec:Introduction}
\IEEEPARstart{T}{he} Multi-access Edge Computing (MEC) has emerged as an evolutionary paradigm with the rapid expansion of Information and Communication Technology (ICT) infrastructure~\cite{Taleb2017Multi}. By decentralizing computation and storage closer to the network edge, MEC enhances data processing efficiency while alleviating the load on centralized cloud systems. This edge-cloud synergy not only reduces latency but also offers potential gains in energy efficiency by optimizing task execution at the network's edge rather than relying solely on large-scale data centers. As task complexity grows, optimizing energy consumption across both MEC and cloud networks becomes crucial. Energy efficiency must be achieved while maintaining a balance between minimizing energy consumption and meeting task response time requirements.
%\IEEEPARstart{M}{ulti-access Edge Computing (MEC)} ~\cite{Taleb2017Multi} has emerged as an evolutionary paradigm with the rapid expansion of Information and Communication Technology (ICT). Compared with conventional cloud-based techniques, edge computing decentralizes resources for computation and storage closer to the network edge closer to data sources. It thus enhances the efficiency of data processing and management. Meanwhile, utilizing edge devices for computational tasks can help reduce the enormous amount of energy consumption in large-scale data centers.
%\jing{Emphasize the importance of energy efficiency in MEC+cloud, because if a task cannot be solved at the edge it will consume cloud energy. Don't say battery things, but emphasizing the importance of total energy efficiency. Note that it's not about only minimizing energy consumption but minimizing energy consumption while guarantee response time. I will add response time in the objective function. It's nothing more complex but reviewers care about it a lot.}

One notable challenge in designing efficient task assignment policies in MEC systems is to handle handovers of mobile terminals (MTs). The dynamic nature of task handovers between edge servers and cloud resources, driven by MT mobility, adds complexity to energy-efficient task offloading~\cite{Ho2022joint}. While handovers are an inherent feature of MEC systems, optimizing resource allocation during these events is particularly difficult due to heterogeneous communication channels, user mobility patterns, and the computational demands of diverse applications. 

Existing studies often simplify the heterogeneity of network resources, implicitly assuming either than an MT will experience the same channel condition after a handover~\cite{Deng2023task,Maleki2023handover}, or that the service provider can dynamically allocate a new channel at the moment of handover~\cite{Monir2022seamless,Shu2023joint}. However, the former assumption overlooks the dynamic nature of wireless transmissions, while the latter risks unexpected service interruptions due to unavailable channels during handovers. These limitations are particularly problematic for real-time applications, such as live video analytics or autonomous vehicle navigation, where identifying a new channel during handover can lead to additional latency and energy consumption.

Moreover, as the problem size grows, with a large number of MTs, diverse tasks, and resource types, the increased complexity of control variables due to channel and resource heterogeneity presents a significant challenge. This complexity often renders traditional optimization methods computationally infeasible, particularly in real-time environments where stringent latency constraints must be satisfied. Such challenges hinder the scalability and practical applicability of existing approaches, particularly in large-scale, dynamic MEC systems.

Different from most existing studies that explicitly considered MT handovers~\cite{Ho2022joint,Maleki2023handover,Li020deep}, this work focuses on scenarios where network service providers can accurately predict the future movements and hence handovers of MTs. Such scenarios are applicable in MEC applications like urban traffic flow management~\cite{Davoli2022flow}, emergency response~\cite{Wang2020edge}, and healthcare management~\cite{Hewa2020multi}, where MTs in the same class typically perform similar routine activities, generate homogeneous computation tasks, and follow regular trajectories. By utilizing  the prediction information of MT handovers, this study seeks to optimize the energy efficiency of such a system by appropriately allocating computation and communication resources to incoming tasks. 
%\jing{this paragraph is OK, but move it to relatively later places.}

To address the inherent complexities of this problem, we adopt a stochastic optimization approach that captures the dynamic states of network resources and communication channels to minimize the long-run average power consumption (energy consumption rate) of the MEC system. Unlike conventional methods mentioned earlier, which are often limited by static or simplified mobility assumptions, our approach models the heterogeneous and dynamic nature of MEC environments. Specifically, we formulate the offloading problem using the \emph{restless multi-armed bandit (RMAB)} framework \cite{whittle1988restless}, which computes marginal costs associated with resource allocation decisions. Building on this, we adopt the restless-bandit-based (RBB) resource allocation technique \cite{fu2021restless}, which dynamically assigns and reserves communication and computation resources (if available) upon arrival of a task in the MEC system, according to the type of task. By leveraging the RMAB framework, we dynamically allocate resources while ensuring scalability and robustness, particularly in MEC systems with dense mobile users and frequent task handovers. 
Furthermore, our proposed methodology addresses practical challenges arising from heterogeneous MTs, diverse task types, varying channel conditions, as well as computing/storage components distributed across MTs, network edge, and central cloud. By explicitly considering frequent task handovers and resource heterogeneity, our work provides a scalable and optimal solution to address critical research gaps in energy-efficient resource management for large-scale MEC systems.
%Therefore, we take a different perspective in this paper. As mentioned earlier, the service provider is aware of the trajectories of handover MTs and thus can determine the channel allocation before the handover. We adopt stochastic optimization methods that capture the dynamic states of channels and network resources to minimize the long-run average power consumption (energy consumption rate) of the MEC system.
%In particular, we formulate the offloading problem in the context of the \emph{restless multi-armed bandit} process \cite{whittle1988restless}.
%We adopt the restless-bandit-based (RBB) resource allocation technique \cite{fu2021restless}, which dynamically assigns and reserves communication and computation resources (if available) upon arrival of a task in the MEC system, according to the type of task.

%Our proposed model, analysis, and algorithm facilitate in-depth discussions of highly-complex and dynamic practical scenarios with heterogeneous MTs, task types, communication channels, as well as computing/storage components distributed across MTs, network edge, and central cloud. \yuki{These contributions offer a novel solution to the challenges posed by heterogeneity and frequent task handovers in large-scale MEC systems, addressing energy efficiency and resource management challenges in a more scalable and optimal way.}

The contributions of this paper are summarized as follows:
\begin{itemize}
\item A novel RMAB-based framework for energy-efficient task offloading is proposed in large-scale MEC systems, incorporating user mobility, resource heterogeneity and task handovers into the problem formulation.
\item A class of scalable scheduling policies, termed highest energy efficiency-adjusted capacity coefficient (HEE-ACC) is developed, which prioritize resources with the least marginal costs while ensuring asymptotic optimality for large-scale MEC systems.
\item Among HEE-ACC, two specific policies are introduced: one with proven asymptotic optimality under dominant server or channel capacity, and another that dynamically learns marginal costs from historical data for robust performance in dynamic environments.
%\item\yuki{Two online offloading algorithms are introduced that prioritize the marginal energy costs of computing and communication resources in both edge and cloud networks, ensuring scalability as system size increases.}
%\item\yuki{\shen{[Combine with the previous contribution]}Our approach dynamically adapt to frequent task handovers and varying network conditions, ensuring robust energy optimization as system scale increases.}
\item Extensive simulations demonstrate significant energy conservation compared to baseline methods, and highlight the robustness of HEE-ACC policies to varying shapes of non-exponentially distributed task lifespans. 
\end{itemize}

The remainder of the paper is structured as follows: Section \ref{sec:Related Work} reviews the relevant literature. Section \ref{sec:model} introduces the MEC system model for task handover. Section \ref{sec:problem} outlines the stochastic process and formulates the optimization problem. Section \ref{sec:policy} presents the energy efficient scheduling policies. Section \ref{sec:simulation} showcases the results and analysis. The paper concludes in Section \ref{sec:conclusion}. The main notations used in this paper are listed in Table \ref{table:symbol1} (Appendix~\ref{app:notation}).

\section{Related Work}\label{sec:Related Work}
\subsection{MEC Offloading}
%Recent studies on offloading policies in MEC systems have focused on different aspects, including time-varying nature of transmission channels~\cite{Sun2019online}, partially offloading policies~\cite{Zhao2021energy}, and combined offloading and dispatching processes~\cite{Liu2023joint}. Joint planning of MEC and other techniques, such as edge server deployment~\cite{Song2022Joint}, network slicing~\cite{Xiang2023joint}, and 3D rendering~\cite{Xie2023sharingaware}, has also been considered.
Recent studies on offloading policies in MEC systems have explored various aspects of this problem, including time-varying nature of transmission channels~\cite{Sun2019online}, partially offloading policies~\cite{Zhao2021energy}, and the combination of offloading and dispatching processes~\cite{Liu2023joint}. Additionally, joint planning of MEC and other techniques, such as edge server deployment~\cite{Song2022Joint}, network slicing~\cite{Xiang2023joint}, and 3D rendering~\cite{Xie2023sharingaware}, have been investigated to enhance system performance. While many of these works aim to improve performance metrics like latency or throughput, some also consider energy efficiency as an additional objective.

%As the offloading problem is usually non-convex  due to contradicting objectives and/or constraints, deep reinforcement learning is commonly adopted to obtain the optimal task offloading policy~\cite{Li020deep,Zhao2022deep}. A multi-agent deep reinforcement learning-based Hungarian algorithm was applied in~\cite{Alam2022multiagent} to derive optimal task offloading policy based on a bipartite graph matching problem. Other notable deep-learning mechanisms applied in similar problems include deep Q-network~\cite{Xiong2020Resource,Wang2021Smart}, double deep Q-network~\cite{Ning2019Deep}, and Monte-Carlo tree search~\cite{Chen2019iRAF}. However, common issues with these methods, such as stability, convergence, and computational complexity, could be problematic in dynamic environments like MEC where real-time decision making and reliable system performance are required. 
Task offloading in MEC systems is often a non-convex optimization problem due to conflicting objectives and constraints. To tackle this, many studies have adopted deep reinforcement learning (DRL) approaches to obtain optimal task offloading policies~\cite{Li020deep,Zhao2023deep}. Common DRL methods include deep Q-network (DQN)~\cite{Wang2021Smart}, deep deterministic policy gradient (DDPG)~\cite{hu2021dynamic}, TD3~\cite{wakgra2024multi} etc. Despite the potential of these methods, their high computational complexity and convergence challenges often make them less suitable for real-time and energy-efficient decision-making in dynamic MEC environments, where reliable system performance is critical.

%On the other hand, a number of low-complexity heuristic algorithms have been widely applied to overcome the complexity issue in solving NP-hard offloading problems. In addition to those mentioned earlier in this paper, (e.g.~\cite{Xiang2023joint,Liu2023joint,Xie2023sharingaware,Maleki2023mobility}), a dependency-aware edge-cloud collaborating strategy was proposed in~\cite{Chen2022dependency} to minimize task completion time by always mitigating tasks in the ascending (or descending) order of the expected energy consumption of task execution. In~\cite{Lin2022popularity}, a contextual online vehicular task offloading policy was studied to improve energy efficiency and reduce the delay in vehicular networks through an online bandits approach with quantified popularity of tasks. In~\cite{Chen2021Energy}, the Lyapunov optimization technique was adopted to balance the energy efficiency and the queue length when the channel condition is unknown. In addition, game theory~\cite{Wu2023computation,Nguyen2021Price} and auction theory~\cite{Su2023Truthful} based optimization techniques were applied to maximize the net profit of MTs when they need to bid for renting communication or computation resources. 

In contrast, several low-complexity heuristic algorithms have been proposed to overcome these computational challenges. For example, a dependency-aware edge-cloud collaboration strategy was introduced in~\cite{Chen2022dependency} to minimize task completion time by mitigating tasks in an energy-efficient order. The Lyapunov optimization technique was applied in~\cite{Chen2021Energy} to balance energy efficiency and queue length under unknown channel conditions. Other approaches, including game theory~\cite{Wu2023computation} and auction theory~\cite{Su2023Truthful}, were also proposed to maximize the net profit of MTs by allowing them to bid for communication or computation resources.
\subsection{Task Handover in MEC}
%A number of literature (e.g.~\cite{Ho2022joint, Maleki2023handover, Shu2023joint,Deng2023task,Monir2022seamless,Maleki2023mobility,Wu2023computation}) have considered on the impact of potential handovers of MTs on offloading decisions in MEC systems. However, they tend to focus more on the impact of such handovers on re-allocation of computation-related resources in addition to deciding the offloading destination of computing tasks. For example, constraints of computational capacities at MTs and CPU frequencies were considered in~\cite{Maleki2023handover,Ho2022joint}. For the transmission-related resources, several studies (e.g.,~\cite{Maleki2023mobility,Wu2023computation}) have explicitly considered the constraints related to the availability of transmission channels, such as limited bandwidth. However, the unpredictable nature of wireless transmissions, such as random fading, has been less studied in the context of computational offloading in MEC.
The impact of task handovers in MEC systems, particularly when MTs move between different network zones, has been a growing focus in recent research. While several works (e.g.~\cite{Ho2022joint, Maleki2023handover, Shu2023joint,Deng2023task,Monir2022seamless,Maleki2023mobility,Wu2023computation}) considered the effect of MT handovers on offloading decisions, many of these studies concentrate on re-allocation of computational resources rather than the full offloading process. For example, constraints on MT computational capacities and CPU frequencies were examined in~\cite{Maleki2023handover,Ho2022joint}. 
Some studies focused on the availability of transmission channels, considering bandwidth limitations~\cite{Maleki2023mobility,Wu2023computation}. However, the unpredictable nature of wireless transmissions, such as random fading, has been less studied in the context of task offloading during handovers.

%Handover prediction in mobility aware MEC can help to improve the overall performance of resource allocation policies~\cite{Uniyal2021intelligent}. In~\cite{Yuan2022dynamic}, the authors utilized prediction results by Long Short-Term Memory (LSTM) method to determine the optimal deployment of edge nodes in MEC.  In~\cite{Maleki2023mobility}, a machine-learning prediction mechanism is applied to predict the future specifications of offloaded tasks and reassign the offloading destinations after a fixed number of time slots. Our approach, while sharing certain similarity to~\cite{Maleki2023mobility}, is more flexible as we do not restrict the handover (reassignment) actions to a periodic manner.
More recent research aimed to improve the overall performance of resource allocation by incorporating handover prediction~\cite{Uniyal2021intelligent, Yuan2022dynamic}. For example, a machine learning based prediction mechanism was applied in~\cite{Maleki2023mobility} to forecast future specifications of offloaded tasks and reassign offloading destinations after a set number of time slots. While such methods improve resource allocation performance, they often rely on fixed periodic reassignment intervals, limiting their flexibility in real-time scenarios. 
%However, unlike these approaches, which often rely on fixed periodic reassignment intervals, our approach provides more flexibility by dynamically adjusting task handovers based on real-time conditions.}

%\yuki{\subsection{RMAB Algorithm}}

%Task offloading policies in MEC are heavily influenced by the algorithmic approaches used to balance energy consumption and resource allocations. The aforementioned methods can become prohibitive, especially in large-scale, dynamic environments like MEC.
To address the computational and scalability challenges of MEC offloading in dynamic environments, recent studies have proposed the use of multi-armed bandit (MAB) algorithms~\cite{wang2024adpative,li2024a}. These algorithms are particularly well-suited to balancing exploration and exploitation in resource-constrained environments. For example, a contextual online vehicular task offloading policy was explored in~\cite{Lin2022popularity} using an online bandits approach to improve energy efficiency and reduce delays in vehicular networks. Our approach builds on these concepts by using a RMAB framework, which is more appropriate for dynamic environments where resource availability and task requirements change over time. This RMAB-based method incorporates real-time predictions of task handovers and prioritizes resources based on marginal costs, achieving scalable and energy-efficient resource allocation in large-scale MEC systems.
\subsection{Discussion}
To summarize, compared with existing studies aiming at improving energy efficiency in MEC, our study offers a more comprehensive, holistic and nuanced approach. We address key limitations in current research by explicitly modeling the dynamic and heterogeneous nature of MEC systems, particularly in scenarios with frequent task handovers and real-time adaptability. Our proposed RMAB-based framework optimizes energy efficiency by leveraging predictive information about MT handovers and dynamically allocating resources. This newly proposed low-complexity algorithm effectively bridges the gap in previous studies and enhances the scalability and practicality of energy-efficient resource management in large-scale MEC environments.
 %We optimize energy efficiency in MEC systems by considering the energy profiles of a wider range of system components and mechanisms. Additionally, our work specifically addresses scenarios where the service provider can predict the future handover of MTs. Our newly proposed low-complexity algorithm caters for our model formulations and takes advantage of the predicted handover information and is able to further improve the overall energy efficiency, effectively bridging a gap in previous studies.

\section{System Model}\label{sec:model}

Let $\mathbb{R}_0$, $\mathbb{R}_+$ and $\mathbb{N}_+$ represent the sets of non-negative reals, positive reals and positive integers, respectively. For any  $N\in\mathbb{N}_+$, we use $[N]$ to represent a set $\{1,2,\ldots,N\}$.

%============================================================================================================
\subsection{Network Model}
Consider a wireless network with orthogonal sub-channels, which establishes  connections between MTs and communication nodes (CNs), such as base stations and access points.
Computing tasks generated by MTs can be appropriately offloaded to CNs through the sub-channels. As CNs are equipped and/or connected (through wired connections) with more storage, computing and networking resources (CPU, GPU, RAM, disk I/O, etc.), they are generally considered preferable for executing computationally intensive tasks than personal MTs. This offloading operation can liberate MTs from being overloaded and achieve higher energy efficiency in the network.
We refer to all these storage, computing and networking equipped and/or connected with the CNs as the \emph{service components} (SCs), which are classified in \emph{groups} (referred to as \emph{SC groups} thereafter) based on their functionalities and geographical locations. 
We denote $K$ as the number of SC groups at the network edge, and let $C_k\in\mathbb{N}_+$ represent the \emph{capacity} of SC groups $k\in[K]$, which is the total number of SC units in the SC group. % as the \emph{capacity} of the SC group. 
CNs are connected to the central cloud via wired connections, enabling additional task offloading when required. Tasks offloaded by MTs can eigher be processed at the network edge by SC groups or forwarded to the cloud for completion. We assume that the cloud has infinite capacity for all SC groups.
%\yuki{[? CNs or SCs]}Through wired cables, CNs are connected to the central cloud, where tasks can be further offloaded to. We assume that the cloud has infinite capacity for all SC groups.
%As mentioned earlier, the MTs keep offloading tasks which can be either completed at the network edge or further offloaded to the cloud. 

Consider a case where MTs, with similar hardware/software profiles, in the same geographical area have homogeneous channel conditions for transmitting to and from CNs in a certain \emph{destination area}, labeled by $\ell$, which locates a set of SC groups $\mathscr{K}_{\ell}\subset [K]$. 
There are in total $L\in\mathbb{N}_+$ such destination areas (that is, $\ell\in[L]$) at the network edge, and we assume without loss of generality that all $\mathscr{K}_{\ell}$ are non-empty and mutually exclusive for different $\ell\in[L]$.
We label all such channels used to connect the $L$ areas, each of which is equipped with SC groups $\mathscr{K}_{\ell}$ ($\ell\in[L]$), as $i=1,2,\ldots,I$.
Each channel $i\in[I]$ can support at most $N_i\in\mathbb{N}_+$ orthogonal sub-channels due to technical constraints such as total bandwidth limits.
When an MT wishes to utilize computational/storage resources in area $\ell$, a subset of the $I$ channels are \emph{eligible} to support the connection between the MT and the CNs in area $\ell$. 
If this MT is too far away and cannot be connected to any CN in area $\ell$, then none of the $I$ channels can be used between the MT and any SC group in $\mathscr{K}_{\ell}$.  
We assume that $I \geq L$ - for each area $\ell$, the SC groups may be connected via more than one wireless channels.
\begin{comment}
\begin{figure*}[t]
\centering
\subfigure[]{\includegraphics[width=0.4\linewidth]{figs/offloading1.png}\label{fig:example:1}}
\subfigure[]{\includegraphics[width=0.4\linewidth]{figs/offloading2.png}\label{fig:example:2}}
\caption{A simple example for the network system (a) at the start of task handover, (b) at the end of task handover. CNs 1 and 2, some SC units, and a sub-channel in
each of the channels are highlighted in red, as they
are occupied/reserved by a task generated by the moving MT. \shen{[FIG TO BE UPDATED]}\label{fig:example}}
\end{figure*}
\end{comment}

\begin{figure}[t]
\centering
\includegraphics[width=0.85\linewidth]{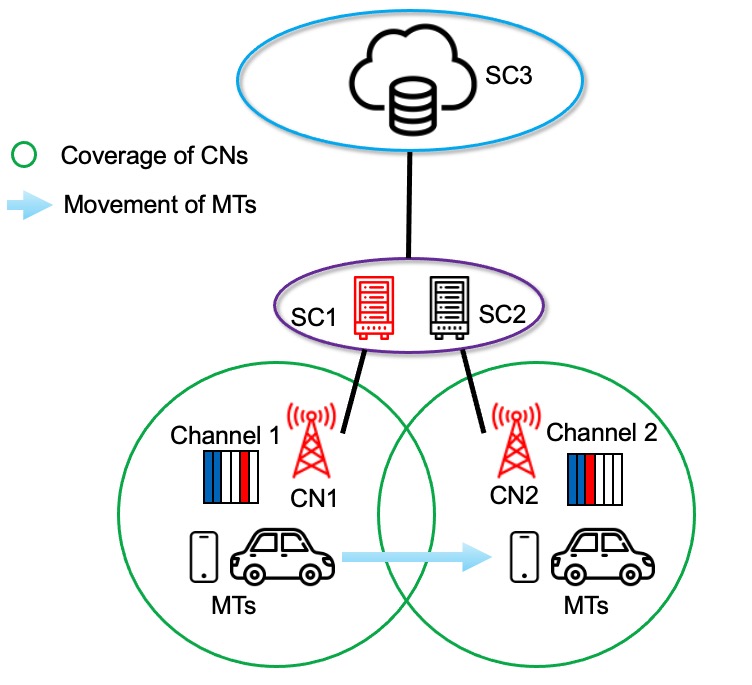}
\caption{A simple handover example for the network system. CN1 and CN2, SC1, and a sub-channel in each of the channels are highlighted in red, as they are occupied/reserved by a task generated by the moving MTs. \label{fig:example}}
\end{figure}

Consider a simple example of the network model in \figurename~\ref{fig:example}, where there are $K=2$ SC groups (SC1 and SC2) locate in the same area $\ell=1$ (with $L=1$) with $I=2$ channels, used by two base stations (CN1 and CN2).
For MTs with similar hardware/software profiles, if they are in the coverage of CN 1, or equivalently the coverage of Channel 1, they can connect to both SC groups in the area $\ell=1$ through Channel 1. Similarly, the connections can be established between the SC groups and the MTs through Channel 2. 
In this case, both channels are eligible to establish connections to the SC groups in the area $\ell=1$.

Consider $J\in\mathbb{N}_+$ classes of moving MTs that can be classified by the differences in their locations, moving speeds and directions, application styles and relevant requirements on computing resources.
In particular, we consider a scenario where MTs in the same class can upload/download data through the same set of eligible channels and potentially connect to the same set of SC groups. In this context, a moving MT may change its class from time to time.
For instance, in \figurename~\ref{fig:example}, all MTs, belonging to the same class when they start moving from the left circle, may be assigned to different classes after they have arrived in the right circle. The MTs can be classified manually by the service provider and/or automatically through conventional techniques, such as signature matching~\cite{jing2011efficient}, 
%\yuki{and Support Vector Machine (SVM) methods~\cite{li2007accurate}} 
based on historical features of the network traffic and mobility prediction techniques (for handover), such as~\cite{magnano2015novel}. 

We refer to the offloaded tasks generated by MTs in class $j\in[J]$ as $j$-tasks. We consider tasks of the same type to be those generated by MTs in the same class, with similar moving directions and speeds, with the same application styles, and with the same requirements on SCs. 
For a certain type of task, not all SC groups at the network edge are necessarily able to serve it due to geographical or functional dis-match.
If a $j$-task is accommodated by an SC in group $k\in[K]$ that is able to serve it at the network edge, $w_{j,k}\in[C_k]$ units of the SC are occupied by this task until it is completed.
Occupied SC units will be simultaneously released upon the completion of the task, and can be reused for future tasks~\cite{Gillam2018}.
If a $j$-task cannot be served by units in SC group $k$, because of un-matched geographical locations or required functions, define $w_{j,k} \rightarrow +\infty$, or any other integer greater than $C_k$, to prohibit these tasks from being assigned there.

\vspace{-0.3cm}
\subsection{Edge Offloading with Task Handover}
With $I$ channels in total, an MT offloading a $j$-task transmits through a sub-channel of channel $i\in[I]$ at an achievable rate
\begin{equation}\label{eqn:transmission_rate}
\mu_{i,j}\coloneqq \left\{
\begin{array}{l l}
B_i \log_2\left(1+\frac{p_{i,j}h_{i,j}}{N_0}\right), &\text{if }\frac{p_{i,j}h_{i,j}}{N_0} \ge 20\mathrm{dB}, \\
0 & \text{otherwise.}
\end{array}
\right.
\end{equation}
where $p_{i,j}\in\mathbb{R}_0$, $h_{i,j}\in\mathbb{R}_+$, $B_i\in\mathbb{R}_+$ and $N_0\in\mathbb{R}_+$ are the transmission power, channel gain, bandwidth of each channel and the noise power, respectively~\cite{Mukherjee2018, Lee2017}.
If the signal-to-noise-ratio (SNR) $(p_{i,j}h_{i,j})/N_0$ is smaller than $20$dB, then reliable connections cannot be established, and the channel is regarded as not available.

We consider the situation where offloaded tasks with relatively small sizes are dominant in the network, so that the computing power of SCs located at the edge of the network or in the central cloud are sufficient to complete the services of the requests in a relatively short period of time. In this sense,
the duration of computational operations of offloaded tasks is generally much shorter than the transmission time between the MTs and the CNs or the cloud;
that is, the processing time of an offloaded task is dominated by the transmission time between the MT and the CN for tasks completed at the network edge, and the transmission time between the MT and the central cloud via the network edge for tasks offloaded to the central cloud.
In this context, a wireless sub-channel between the MT and the CN needs to be reserved for data transmission for the entire period of the offloaded task, no matter whether the task is addressed at the network edge or in the central cloud~\cite{You2017}. If no such an available sub-channel (all sub-channels are fully occupied), the MT cannot offload this task to any CN, and the task has to be computed locally. In this case, the task never reaches the network edge, and any scheduling policies employed at the network edge will not affect the power consumption. 

%Therefore, we will not consider the situation where the task is computed locally in this paper.

In a MEC network, MTs move across different areas, requiring connections to different CNs through different sub-channels.
%MTs are moving around across different areas that request connections to different CNs through different sub-channels.
When an MT is sending data to or receiving results from an SC, it may change the connected CNs when moving into different places covered by different CNs, leading to task handover between CNs, as illustrated in \figurename~\ref{fig:example}.
For the tasks generated by such MTs, other CNs with available sub-channels in their moving directions should be prepared and reserved to ensure the quality of the connections.
For instance, in \figurename~\ref{fig:example}, CNs 1 and 2, some SC units, and a sub-channel in each of the channels are highlighted in red, meaning that they are occupied/reserved by a task generated by the moving MTs.

\subsection{Cloud Offloading}
For the tasks that are offloaded to the cloud, two transmission segments, namely from the MT to the network edge and from the network edge to the cloud, are requested. For notational consistency, define a special SC group $K+1$, which has infinite capacity $C_{K+1} \to +\infty$, for the SC units in the cloud. As the CNs and the cloud are usually connected by a cable backbone network, we can assume that the transmission time between the network edge and the cloud, denoted by $D_0$, is the same for all tasks of the same class. Therefore, we denote the expected duration of $j$-tasks computed in the cloud and transmitted via an CN in destination area $\ell$ through a sub-channel in channel $i$ as $ 1/\mu_{i,j}+D_0$~\cite{Deng2015}. Note that $K+1$ does not belong to any destination area $\ell\in[L]$ located at the network edge. 

%We use the notation $\ell(K+1)$ only to indicate that the connection between a CN in the area $\ell(K+1)$ and the cloud is set up.

\vspace{-0.2cm}
\subsection{Energy Efficiency}
Power consumption of the network edge consists of \emph{static} and \emph{operational} power.
Static power is the essential consumption incurred when an SC group is activated (which means the hardware components associated with the SC group must be powered on and connected to the network); while the operational power is consumed when SC units are engaged in processing offloading tasks. 
In a dynamic system such as a wireless network, the amount of operational power consumption is affected by the real-time utilization rate of the SCs.
Let $\varepsilon_k^0\in\mathbb{R}_+$ and $\varepsilon_k\in\mathbb{R}_+$ represent the amounts of the static power and the operational power consumption per SC unit in group $k\in[K]$, and define the expected power consumption of an SC unit for computing a $j$-task in the cloud as $\bar{\varepsilon}_j\in\mathbb{R}_+$. Similar power consumption models have been empirically justified and widely applied in existing research~\cite{ Wu2020}.
%~\cite{Jalali2016, Wu2020}.
Because of its dependencies on the dynamic states of the network, a detailed discussion about the power consumption of the network edge will be provided in Section~\ref{sec:problem}.

%Intuitively, a larger amount of power is consumed for transmission when a task is offloaded further, and . Therefore, for all $j\in[J]$ and $k\in[K]$, we have $\varepsilon_k <\bar{\varepsilon}_j$.

\section{Stochastic Process and Optimization}\label{sec:problem}

Define $\mathscr{K} \coloneqq [K]\cup\{K+1\}$ as the set of SC groups in the edge and the cloud.
Consider random variables $X_{i,i',j,k}(t)\in\mathbb{N}_0$, $i,i'\in[I]$, $j\in[J]$, $k\in\mathscr{K}$, that represent the number of $j$-tasks being served by SC units in group $k$ and occupying two sub-channels in channels $i$ and $i'$ at time $t\geq 0$. 
When the $j$-task starts to be transmitted to the network, it uses channel $i$; while the corresponding MT will move during the data transmission and computing process and end up using channel $i'$ to complete the data transmission and to receive computing results from the network.
For instance, in \figurename~\ref{fig:example}, Channels 1 and 2 correspond to such $i$ and $i'$, respectively.
We refer to such channels $i$ and $i'$ as the \emph{starting} and \emph{ending channel}, respectively.
If MTs of a specific class $j$ move relatively slowly and do not need to change the communication channel when processing their tasks, then the tasks may be transmitted via the same starting and ending channels (that is, $i=i'$).
Because of the limited capacities of channels and SCs, these variables should satisfy
\begin{equation}\label{eqn:capacity_constraint:1}
\sum\limits_{j\in[J]} w_{j,k}\sum\limits_{\begin{subarray}~i,i'\in[I]:\\\mu_{i,j}\mu_{i',j}>0\end{subarray}}X_{i,i',j,k}(t) \leq C_k,~\forall k\in[K],~t \geq 0,
\end{equation}
and
\begin{multline}\label{eqn:capacity_constraint:2}
\sum\limits_{k\in\mathscr{K}} \sum\limits_{i'\in[I]}\sum\limits_{j\in[J]}\Bigl(X_{i,i',j,k}(t) +X_{i',i,j,k}(t) \Bigr)\leq N_i,\\~\forall i\in[I],~t\geq 0.
\end{multline}
Constraints~\eqref{eqn:capacity_constraint:1} and \eqref{eqn:capacity_constraint:2} are led by the limited capacities of SC groups and wireless channels, respectively.

Let $\bm{X}(t)=(X_{i,i',j,k}(t):i,i'\in[I],j\in[J],k\in\mathscr{K})$, and the state space of the process $\{\bm{X}(t),\ t\geq 0\}$ (the set involves all possible values of $\bm{X}(t)$ for $t\geq 0$) be
\begin{equation}
\mathscr{X}\coloneqq\prod\limits_{\begin{subarray}~i,i'\in[I],\\ j\in[J],\\k\in\mathscr{K}\end{subarray}}\left\{0,1,\ldots,\min\Bigl\{\lfloor \frac{C_k}{w_{j,k}}\rfloor,N_i,N_{i'}\Bigr\}\right\},
\end{equation}
where $\prod$ represents the Cartesian product.
Note that although $\mathscr{X}$ is larger than the set of possible values of $\bm{X}(t)$, the process $\bm{X}(t)$ will be constrained by Constraints~\eqref{eqn:capacity_constraint:1} and \eqref{eqn:capacity_constraint:2} in our optimization problem defined later in this section.

Upon each arrival of a user task, a scheduling policy selects a channel-SC tuple $(i,i',k)$ to serve it. 
Define action variables $a_{i,i',j,k}(\bm{x})\in\{0,1\}$, $\bm{x}\in\mathscr{X}$, as a function of the state for each $i,i'\in[I]$, $j\in[J]$ and $k\in\mathscr{K}$.
If $a_{i,i',j,k}(\bm{x})=1$, then two sub-channels of channels $i$ and $i'$ and $w_{j,k}$ units of SC in group $k$ are selected to serve an incoming $j$-task when $\bm{X}(t)=\bm{x}$; otherwise, the tuple of channels $i$ and $i'$ and SC group $k$ is not selected.
An incoming $j$-task, when $\bm{X}(t)=\bm{x}$, means the first $j$-task coming after and excluding time $t$, and the process $\bm{X}(t)$ is defined as left continuous in $t\geq 0$.
%We define the action space, which is the set involving all possible values of action variables, as
%\begin{equation}
%\mathscr{A} = \{0,1\}^{I\times J\times (K+1)}.
%\end{equation}
Let $\bm{a}(\bm{x})=(a_{i,i',j,k}(\bm{x}):i,i'\in[I],j\in[J],k\in\mathscr{K})$, $\bm{x}\in\mathscr{X}$.
For all the action variables, it should further satisfy
\begin{equation}\label{eqn:action_constraint}
\sum\limits_{\begin{subarray}~i,i'\in[I]:\\\mu_{i,j}\mu_{i',j} > 0\end{subarray}}\sum\limits_{k\in\mathscr{K}}a_{i,i',j,k}(\bm{X}(t)) \leq 1,~\forall j\in[J],t\geq 0.
\end{equation}
In Constraint~\eqref{eqn:action_constraint}, when a $j$-task coming at time $t$, at most one tuple $(i,i',k)$ will be selected, representing sub-channels for data transmission and SC units for computation to serve it. If no tuple $(i,i',k)$ with $\mu_{i,j}>0$ and $\mu_{i',j} >0$ is selected, then this task is blocked by the network and has to be processed or dropped by the associated MT.
The blocking event happens due to the limited capacities of the channels in $[I]$, although we assume an infinite computing capacity for the SC units in the cloud.
We provide a diagram of the decision making process in Appendix~\ref{app:diagram}.

If, at time $t$, a newly arrived $j$-task is decided to be served by the channel-SC tuple $(i,i',k)$, then $X_{i,i',j,k}(t)$ increments by one; if a $j$-task is completed and leaves the system at time $t$, then $X_{i,i',j,k}(t)$ decrements by one.

For a newly arrived $j$-task served by the channel-SC tuple $(i,i',k)$, its lifespan is considered as an independently and identically distributed random variable with mean $1/u_j(i,i',k)$, where $u_j(i,i',k)$ is determined by the profiles of the associated MT and the intrinsic parameters of the communications scenarios such as the moving speed of the MT, the transmission rates of the connected sub-channels $\mu_{i,j}$ and $\mu_{i',j}$ and the actual processing time of the task. 
Assume without loss of generality, for $i,i'\in[I]$ and $k\in\mathscr{K}$, if $\mu_{i,j} = 0$ or $\mu_{i',j} = 0$, then $u_j(i,i',k) \equiv 0$; otherwise, $u_j(i,i',k) > 0$.

We consider a realistic case with a large number of MTs that generate tasks independently and are classified into different classes. The arrivals of tasks in class $j\in[J]$ are considered to follow a Poisson process with the mean rate $\lambda_j\in\mathbb{R}_+$, which is appropriate for a large number of independent MTs sharing similar stochastic properties~\cite{Lee2017}.
We consider Poisson arrivals for the clarity of analytical descriptions, while our scheduling policy proposed in Section~\ref{sec:policy} is not limited to the Poisson case and applies to a wide range of practical scenarios. 
In Section~\ref{sec:simulation}, the effectiveness of the proposed policy is demonstrated through extensive simulation results with time-varying arrival rates that are able to capture scenarios with busy and idle periods of the network system.  

In conjunction with the mean arrival rates and the expected lifespans of different tasks being served by different SCs through different channels, the action variables determine the transition rates of the system state, $\bm{X}(t)$, at each time $t$. They further affect the long-run probabilities of different states that incur different energy consumption rates.

A scheduling policy, denoted by $\phi$, is determined by the action variables for all states $\bm{x}\in\mathscr{X}$ and selects a channel-SC tuple $(i,i',k)$ upon each arrival of a user task.
We add a superscript and rewrite the action variables as $a^{\phi}_{i,i',j,k}(\bm{x})$, which represents the action variables under policy $\phi$.
To ensure the fairness for all arrived tasks, in this paper, we consider those policies $\phi$ that rejects/blocks a new task if and only if there is no vacant channel-SC tuple to locate it.
Let $\Phi$ represent the set of all such policies $\phi$ determined by action variables $a^{\phi}_{i,i',j,k}(\bm{x})$, $i,i'\in[I]$, $j\in[J]$ and $k\in\mathscr{K}$.
Similarly, since the stochastic process $\{\bm{X}(t),t\geq 0\}$ is conditioned on the underlying policy, we rewrite it as $\bm{X}^{\phi}(t)$. 

We aim to minimize the long-run average power consumption (energy consumption rate) of the network. That is,
\begin{multline}\label{eqn:obj}
\min\limits_{\phi\in\Phi}\lim\limits_{T\rightarrow \infty} \frac{1}{T}\mathbb{E}\int_0^T \sum\limits_{i,i'\in[I]}\sum\limits_{j\in[J]}\sum\limits_{k\in[K]}\varepsilon_k w_{j,k}X^{\phi}_{i,i',j,k}(t)\ dt \\+\sum\nolimits_{k\in[K]} \varepsilon_k^0
+ \mathcal{E}^{\phi}_c,
\end{multline}
where the long-run average power consumption for computing tasks offloaded to the cloud is given by %\vspace{-0.3cm}
\begin{multline}\label{eqn:obj:cloud_consumption}
\mathcal{E}^{\phi}_c\coloneqq \\
\sum\limits_{j\in[J]}\bar{\varepsilon}_j\lim\limits_{T\rightarrow \infty} \frac{1}{T}\mathbb{E}\int_0^T\sum\limits_{i,i'\in[I]}u_j(i,i',K+1) X^{\phi}_{i,i',j,K+1}(t)\ dt, %\vspace{-1cm}
\end{multline}
subject to Constraints~\eqref{eqn:action_constraint}, \eqref{eqn:capacity_constraint:1} and \eqref{eqn:capacity_constraint:2}.

The first and second items in Objective~\eqref{eqn:obj} are the long-run average operational and static power consumption of the SC groups at the edge network, respectively, and the last term, described in Eq.~\eqref{eqn:obj:cloud_consumption}, stands for the long-run average power consumption for computing tasks offloaded to the cloud. Recall that $\varepsilon_k$ ($k\in[K])$ is the energy consumption rate per SC unit of group $k$, while $\bar{\varepsilon}_j$ is the energy consumption per $j$-task that is processed and completed by computing components in the cloud network.

We refer to the minimization problem described in Eq.~\eqref{eqn:obj}, Constraints~\eqref{eqn:action_constraint}, \eqref{eqn:capacity_constraint:1} and \eqref{eqn:capacity_constraint:2} as the \emph{task offloading scheduling problem} (TOSP). 
The TOSP consists of $I^2J(K+1)$ parallel \emph{bandit processes}, $\{X^{\phi}_{i,i',j,k}(t), t\geq 0\}$, each of which is a Markov decision process (MDP) with binary actions~\cite{gittins2011multiarmed}.
The parallel bandit processes are coupled by Constraints~\eqref{eqn:action_constraint}, \eqref{eqn:capacity_constraint:1} and \eqref{eqn:capacity_constraint:2}.
The TOSP is complicated by its large state space $\mathscr{X}$, which increases exponentially in $I,J$ and $K$ and prevents conventional optimizers for MDP, such as value iteration and reinforcement learning, from being applied directly. 
The TOSP also extends the unrealistic assumption in \cite{wang2018energy} that considers only communication channels without any MEC servers or components. It follows with a substantially complicated problem with respect to both analytical and numerical analysis.

%forms an instance of the \emph{resource allocation problem} discussed in \cite{fu2021restless}, which consists of parallel \emph{restless multi-armed bandit problems} (RMABPs) coupled by capacity constraints.

%We introduce a \emph{scaling parameter} $h\in \mathbb{N}_+$ to emphasize the scale of the network. Define $\lambda_j \coloneqq h\lambda_j^0$
%for all $j\in[J]$, $\varepsilon^0_k \coloneqq h\tilde{\varepsilon}^0_k$ and $C_k \coloneqq hC^0_k$ for all $k\in[K]$, where $\lambda_j^0, \tilde{\varepsilon}^0_k, C^0_k \in\mathbb{R}_+$.
%A large $h$ is more appropriate for modeling the network environment in a densely populated area. 
\vspace{-0.5cm}
\section{Scheduling Policy}\label{sec:policy}
%\yuki{This section will adapt the restless bandit technique proposed in~\cite{fu2021restless}. Then we will propose a scalable scheduling policy in a greedy manner and demonstrate its near-optimality.}
This section first presents the restless bandit technique, then describes our proposed scalable scheduling policy in a greedy manner and demonstrates its near-optimality.

The restless bandit technique in \cite{fu2021restless} provides a sensible way to decompose all the bandit processes coupled by the constraints involving both state and action variables into $I^2J(K+1)$ independent processes.
The marginal cost of selecting a certain tuple of the SC and the communication channels $(i,i',k)$ ($i,i'\in[I],k\in\mathscr{K}$) is then quantified by an offline-computed real number with significantly reduced computational complexity.
Based on the marginal costs of all the possible channel-SC tuples, a heuristic scheduling policy can be proposed by prioritizing those with the least marginal costs. 
Following the tradition of the restless multi-armed bandit (RMAB) problem initially proposed in \cite{whittle1988restless}, we refer to the marginal cost as the \emph{index} of the associated tuple of SC and communication channels.
For any task in class $j$, each of the index is offline-computed by optimizing the bandit process associated with the channel-SC tuple $(i,i',k)$, for which the computational complexity is only linear to $\min\{\lfloor C_k/w_{j,k} \rfloor,N_i, N_{i'}\}$. 
The resulting scheduling policy is hence applicable to realistically large network systems without requesting excessively large computational or storage power.
More importantly, under provided conditions, we prove that such a scalable scheduling policy approaches optimality as the size of the system tends to infinity. We refer to a detailed exposition in Proposition~\ref{prop:asym_opt}.
Unlike the canonical resource allocation problem studied in \cite{fu2021restless}, the TOSP focuses on the edge computing systems with mobile MTs and task handover between different CNs. 
In particular, for the TOSP, the marginal costs for the tuples involve unknown parameters and not directly computable.

%In this section, we will adapt the restless bandit technique proposed in \cite{fu2021restless} to the TOSP. We propose a scalable scheduling policy in a greedy manner and demonstrate its near-optimality.
\vspace{-0.5cm}
\subsection{Randomization and Relaxation}
%\vspace{-0.7cm}
Along with the Whittle relaxation technique~\cite{whittle1988restless}, we randomize the action variables and relax Constraints~\eqref{eqn:action_constraint}, \eqref{eqn:capacity_constraint:1} and \eqref{eqn:capacity_constraint:2} to %\vspace{-0.3cm}
\begin{equation}\label{eqn:action_constraint:relax}
    \sum\limits_{\begin{subarray}~(i,i',k)\in[I]^2\times\mathscr{K}:\\\mu_{i,j}\mu_{i',j}>0\end{subarray}}
    \lim\limits_{t\rightarrow \infty}\mathbb{E}\Bigl[a_{i,i',j,k}^{\phi}\bigl(\bm{X}(t)\bigr)\Bigr] \leq 1, \forall j\in[J],
\end{equation}
\begin{equation}\label{eqn:capacity_constraint1:relax}
    \sum\limits_{\begin{subarray}~(i,i',j)\in[I]^2\times[J]:\\\mu_{i,j}\mu_{i',j}>0\end{subarray}}w_{j,k}\lim\limits_{t\rightarrow \infty}\mathbb{E}\Bigl[X^{\phi}_{i,i',j,k}(t) \Bigr]\leq C_k, \forall k\in[K],%\vspace{-0.3cm}
\end{equation}
and%\vspace{-0.3cm}
\begin{multline}\label{eqn:capacity_constraint2:relax}
    \sum\limits_{i'\in[I]}\sum\limits_{j\in[J]}\sum\limits_{k\in\mathscr{K}}\Bigl(\lim\limits_{t\rightarrow \infty}\mathbb{E}\bigl[X^{\phi}_{i,i',j,k}(t)\bigr]+\lim\limits_{t\rightarrow \infty}\mathbb{E}\bigl[X_{i',i,j,k}(t)\bigr]\Bigr)\\ \leq N_i,\forall i\in[I],%\vspace{-0.3cm}
\end{multline}
respectively. 
For $i,i'\in[I]$, $j\in[J]$, $k\in\mathscr{K}$, 
define
\begin{equation}
    \mathscr{X}_{i,i',j,k}\coloneqq \Bigl\{0,1,\ldots,\min\bigl\{\lfloor C_k/w_{j,k}\rfloor, N_i, N_{i'}\bigr\}\Bigr\},
\end{equation}
and for $x\in\mathscr{X}_{i,i',j,k}$, define 
\begin{equation}
    \alpha^{\phi}_{i,i',j,k}(x)\coloneqq \lim_{t\rightarrow \infty} \mathbb{E}[a^{\phi}_{i,i',j,k}(\bm{X}^{\phi}(t)) | X^{\phi}_{i,i',j,k}(t) = x],
\end{equation} 
which takes values in $[0,1]$.
Let $\bm{\alpha}^{\phi}_{i,i',j,k}\coloneqq (\alpha^{\phi}_{i,i',j,k}(x): x\in\mathscr{X}_{i,i',j,k})$, and let
$\tilde{\Phi}$ represent the set of all the policies determined by the randomized action variables $\bm{\alpha}^{\phi}_{i,i',j,k}$ for all $i,i'\in[I]$, $j\in[J]$, and $k\in\mathscr{K}$. We refer to the problem %\vspace{-0.3cm}
\begin{multline}\label{eqn:obj:relax}
    \min\limits_{\phi\in\tilde{\Phi}} \lim\limits_{T\rightarrow \infty}\frac{1}{T}\mathbb{E}\int_0^T \sum\limits_{i,i'\in[I]}\sum\limits_{j\in[J]}\sum\limits_{k\in[K]}\varepsilon_kw_{j,k}X^{\phi}_{i,i',j,k}(t)~dt\\ + \sum\nolimits_{k\in[K]}\varepsilon_k^0 + \mathcal{E}^{\phi}_c,
\end{multline}
subject to Constraints~\eqref{eqn:action_constraint:relax}, \eqref{eqn:capacity_constraint1:relax} and \eqref{eqn:capacity_constraint2:relax} as the \emph{relaxed problem}.
Objective \eqref{eqn:obj:relax} is derived from Objective \eqref{eqn:obj} by replacing $\Phi$ with $\tilde{\Phi}$.
Since $\Phi \subset \tilde{\Phi}$ and Constraints~\eqref{eqn:action_constraint}, \eqref{eqn:capacity_constraint:1} and \eqref{eqn:capacity_constraint:2} are more stringent than Constraints \eqref{eqn:action_constraint:relax}, \eqref{eqn:capacity_constraint1:relax} and \eqref{eqn:capacity_constraint2:relax}, any policy applicable to the TOSP will also be applicable to the relaxed problem; while, a policy for the relaxed problem is not necessarily applicable to the TOSP.
It follows that the relaxed problem achieves a lower bound for the minimum of TOSP.

\vspace{-0.2cm}
Consider a stable system with existing stationary distribution $\pmb{\pi}^{\phi}\in[0,1]^{|\mathscr{X}|}$ in the long-run case under a given policy $\phi\in\tilde{\Phi}$, we write the dual function of the relaxed problem as 
\begin{multline}\label{eqn:dual_func}
L(\pmb{\nu},\pmb{\gamma},\bm{\eta}) \\
= \min\limits_{\phi\in\tilde{\Phi}}\biggl[
\sum\limits_{i,i'\in[I]}\sum\limits_{j\in[J]}\sum\limits_{k\in[K]} \varepsilon_k w_{j,k}\sum_{x\in\mathscr{X}_{i,i',j,k}}\pi^{\phi}_{i,i',j,k}(x)x\\
+ \sum\limits_{k\in[K]}\varepsilon_k^0 \\
+ \sum\limits_{i,i'\in[I]}\sum\limits_{j\in[J]}\bar{\varepsilon}_j u_j(i,i',K+1) \sum\limits_{x\in\mathscr{X}_{i,i',j,K+1}}\pi^{\phi}_{i,i',j,K+1}(x)x\\
+\sum\limits_{j\in[J]}\nu_j\Bigl(\sum\limits_{\begin{subarray}~(i,i',k)\in[I]^2\times\mathscr{K}:\\u_j(i,i',k)>0\end{subarray}}\sum\limits_{x\in\mathscr{X}_{i,i',j,k}}\pi^{\phi}_{i,i',j,k}(x)\alpha^{\phi}_{i,i',j,k}(x)-1\Bigr)\\
+\sum\limits_{k\in[K]}\gamma_k\Bigl(\sum\limits_{\begin{subarray}~(i,i',j)\in[I]^2\times[J]:\\u_j(i,i',k)>0\end{subarray}}w_{j,k}\sum\limits_{x\in\mathscr{X}_{i,i',j,k}}\pi^{\phi}_{i,i',j,k}(x)x-C_k\Bigr)\\
+\sum\limits_{i\in[I]}\eta_i\Bigl(\sum\limits_{j\in[J]}\sum\limits_{i'\in[I]}\sum\limits_{k\in\mathscr{K}}\bigl(\sum\limits_{x\in\mathscr{X}_{i,i',j,k}}\pi^{\phi}_{i,i',j,k}(x)x\\+\sum\nolimits_{x\in\mathscr{X}_{i',i,j,k}}\pi^{\phi}_{i',i,j,k}(x)x\bigr)-N_i\Bigr)\biggr],
\end{multline}
where $\pmb{\nu}\in\mathbb{R}_0^J$, $\pmb{\gamma}\in\mathbb{R}_0^K$, and $\bm{\eta}\in \mathbb{R}_0^I$ are the Lagrange multipliers for Constraints~\eqref{eqn:action_constraint:relax}, \eqref{eqn:capacity_constraint1:relax}, and \eqref{eqn:capacity_constraint2:relax}, respectively.
Following the Whittle relaxation~\cite{whittle1988restless} and the restless bandit technique for resource allocation~\cite{fu2021restless}, the minimization in Eq.~\eqref{eqn:dual_func} can be decomposed into $I^2 J(K+1)$ independent \emph{sub-problems} that are, for $(i,i',j,k)\in [I]^2\times[J]\times[K]$,
\begin{multline}\label{eqn:sub-problem}
    L_{i,i',j,k}(\nu_j,\gamma_k,\eta_i,\eta_{i'})\coloneqq
    \min\nolimits_{\phi\in\tilde{\Phi}}L^{\phi}_{i,i',j,k}(\nu_j,\gamma_k,\eta_i,\eta_{i'})\\
    = \min\nolimits_{\phi\in\tilde{\Phi}}\varepsilon_kw_{j,k}\sum\nolimits_{x\in\mathscr{X}_{i,i',j,k}}\pi^{\phi}_{i,i',j,k} x\\
    +\nu_j\sum\nolimits_{x\in\mathscr{X}_{i,i',j,K+1}}\pi^{\phi}_{i,i',j,k}(x)\alpha^{\phi}_{i,i',j,k}(x)\\
    +\gamma_{k} w_{j,k}\sum\nolimits_{x\in\mathscr{X}_{i,i',j,k}}\pi^{\phi}_{i,i',j,k}(x)x\\
    +(\eta_i+\eta_{i'})\sum\nolimits_{x\in\mathscr{X}_{i,i',j,k}}\pi^{\phi}_{i,i',j,k}(x)x ,
\end{multline}
and, for $(i',i,j) \in I^2\times [J]$,
\begin{multline}\label{eqn:sub-problem:K+1}
    L_{i,i',j,K+1}(\nu_j,\gamma_{K+1},\eta_i,\eta_{i'})\\\coloneqq
    \min_{\phi\in\tilde{\Phi}}L^{\phi}_{i,i',j,K+1}(\nu_j,\gamma_{K+1},\eta_i,\eta_{i'})\\
    = \min\nolimits_{\phi\in\tilde{\Phi}}\bar{\varepsilon}_ju_j(i,i',j,K+1)\sum\limits_{x\in\mathscr{X}_{i,i',j,K+1}}\pi^{\phi}_{i,i',j,K+1}(x) x\\
    +\nu_j\sum\nolimits_{x\in\mathscr{X}_{i,i',j,K+1}}\pi^{\phi}_{i,i',j,K+1}(x)\alpha^{\phi}_{i,i',j,K+1}(x)\\
    +(\eta_i+\eta_{i'})\sum\nolimits_{x\in\mathscr{X}_{i,i',j,K+1}}\pi^{\phi}_{i,i',j,K+1}(x)x ,
\end{multline}
where recall that each policy $\phi\in\tilde{\Phi}$ is determined by the action variables $\alpha^{\phi}_{i,i',j,k}(x)$ for $x\in\mathscr{X}_{i,i',j,k}$.
In particular,
\begin{multline}
    L(\pmb{\nu},\pmb{\gamma},\bm{\eta}) = \sum\limits_{i,i'\in[I]}\sum\limits_{j\in[K]}\sum\limits_{k\in\mathscr{K}}L_{i,i',j,k}(\nu_j,\gamma_k,\eta_i,\eta_{i'}) \\+ \sum\limits_{k\in[K]}\varepsilon_k^0 - \sum\limits_{j\in[J]}\nu_j - \sum\limits_{k\in[K]}C_k\gamma_k - \sum\limits_{i\in[I]}\eta_i.
\end{multline}

In this context, the computational complexity of each sub-problem is linear to the size of $\mathscr{X}_{i,i',j,k}$ - alternatively, $\min\bigl\{\lfloor C_k/w_{j,k}\rfloor, N_i,N_{i'}\bigr\}$. 
Together with the independence between those sub-problems, the computational complexity to achieve the minimum in Eq.~\eqref{eqn:dual_func} is linear in $I^2$, $J$ and $K$.
We have the following corollaries of~\cite[Proposition 1]{fu2021restless}.
\begin{corollary}\label{coro:indexability}
For $i,i'\in[I]$, $j\in[J]$, and $k\in\mathscr{K}$ with $\mu_{i,j}\mu_{i',j} > 0$ and given $\nu_j,\gamma_k,\eta_i,\eta_{i'}\in\mathbb{R}_0$, a policy $\phi$ determined by the action variables $\bm{\alpha}^{\phi}_{i,i',j,k}\in[0,1]^{|\mathscr{X}_{i,i',j,k}|}$ is optimal to the sub-problem associated with $(i,i',j,k)$, if, for any $x\in\mathscr{X}_{i,i',j,k}$ and $k\in[K]$,
\begin{equation}\label{eqn:indexbility:1}
    \alpha^{\phi}_{i,i',j,k}(x)\begin{cases}
        1, & \text{if } \nu_j > \psi_j(i,i',k),\\
        \in[0,1], &
        \text{if } \nu_j = \psi_j(i,i',k),\\
        0,& \text{otherwise,} 
    \end{cases}
\end{equation}
where
\begin{multline}\label{eqn:indexability:2}
\psi_j(i,i',k)\coloneqq \frac{\lambda_j}{u_j(i,i',k)}\varepsilon_k w_{j,k}\mathds{1}\{k<K+1\}\\ + \lambda_j\bar{\varepsilon}_j\mathds{1}\{k=K+1\}\\+\bigl(1+\frac{\lambda_j}{u_j(i,i',k)}\bigr)\bigl(w_{j,k}\gamma_k\mathds{1}\{k<K+1\} + \eta_i + \eta_{i'}\bigr)
\end{multline}
\end{corollary}
Corollary~\ref{coro:indexability} indicates the existence of a \emph{threshold-style} policy, satisfying Eq.~\eqref{eqn:indexbility:1}, that is optimal to the sub-problem associated with $(i,i',j,k)$.
Although there also exists a threshold-style policy that achieves the minimum of the sub-problems, the minimum of the sub-problems is only a lower bound of the minimum of the original TOSP described in Objective~\eqref{eqn:obj}, Constraints~\eqref{eqn:action_constraint}, \eqref{eqn:capacity_constraint:1} and \eqref{eqn:capacity_constraint:2}, and the threshold-style policy described in Eq.~\eqref{eqn:indexbility:1} is in general not applicable to the TOSP.
The key is to establish a connection between the sub-problems and the TOSP, and then we can translate the threshold-style policy to those applicable, scalable and near-optimal to the original TOSP.

Following the tradition of the restless bandit studies in the past decades, we refer to the real number $\psi_j(i,i',k)$ as the \emph{index} associated with the bandit process $\{X^{\phi}_{i,i',j,k}(t),t\geq 0\}$, which intuitively represents the marginal cost of selecting the tuple $(i,i',k)$ to serve a $j$-task. 
For the threshold-style, optimal policy, tuples $(i,i',k)$  with smaller indices are prioritized to have $a^{\phi}_{i,i',j,k}(X^{\phi}_{i,i',j,k}(t)) = 1$ than those with larger indices.

In \cite{whittle1988restless}, Whittle proposed the well-known Whittle index policy that always prioritizes bandit processes with the highest/lowest indices and conjectured asymptotic optimality of the Whittle index policy to the original problem.
In subsequent studies such as \cite{fu2021restless,fu2020energy}, the Whittle index policy has been proved to be asymptotically optimal in special cases and/or numerically demonstrated to be near-optimal.

However, unlike the past work, exempli gratia, \cite{wang2018energy,fu2020energy}, the index $\psi_j(i,i',k)$ described in Eq.~\eqref{eqn:indexability:2} is not directly computable. 
It is dependent on the unknown multipliers $\gamma_k$, $\eta_i$ and $\eta_{i'}$ due to the inevitable capacity constraints described in Constraints \eqref{eqn:capacity_constraint:1} and \eqref{eqn:capacity_constraint:2}. 
These capacity constraints substantially complicate the analysis of the TOSP, and, more importantly, prevent existing theorems related to bounded performance degradation from being directly applied to the TOSP.

In Section~\ref{subsec:greedy}, we will discuss  the methods to approximate the indices. Based on the approximated indices (representing the marginal costs), we propose heuristic policies applicable and scalable to the original TOSP, and demonstrate their effectiveness with respect to energy efficiency in Section~\ref{sec:simulation}.
%\vspace{-0.3cm}

\subsection{Highest Energy Efficiency with Adjusted Capacity Coefficients (HEE-ACC)}
\label{subsec:greedy}
%\IncMargin{1em}
\begin{algorithm}[t]\small
\linespread{1}\selectfont
%\SetKwFunction{IndexPolicy}{$(\bm{\eta}_e)\gets$ IndexPolicy}
\SetKwProg{Fn}{Function}{}{End}
\SetKwInOut{Input}{Input}
\SetKwInOut{Output}{Output}
%\SetAlgoLined
%\DontPrintSemicolon

%\Input{$\bm{S}^{\rm IND}(t)=\bm{s}=(s_{i,j}:i\in[I],j\in[J])$, $\mathscr{I}^{\rm IND}_j(t)=\mathscr{I}_j$ for all $j\in[J]$, and $\eta_{i,i',j}\bigl(\bm{S}^{\rm IND}(t),t\bigr) = \eta_{i,i',j}(s_{i,j})$ for all $j\in[J]$ and $i,i'\in[I]$.} 
\Input{Indices $\psi_j(i,i',k)$ for all tuples $[I]^2\times[J]\times\mathscr{K}$, adjusted capacity coefficients $\pmb{\gamma}$ and $\bm{\eta}$, and system state $\bm{X}^{\text{HEE-ACC}}(t)$.}
\Output{Scheduling actions $\pmb{a}^{\text{HEE-ACC}}\bigl(\bm{X}^{\text{HEE-ACC}}(t)\bigr)\coloneqq \Bigl(a^{\text{HEE-ACC}}_{i,i',j,k}\bigl(\bm{X}^{\text{HEE-ACC}}(t)\bigr): (i,i',j,k)\in[I]^2\times [J]\times \mathscr{K}\Bigr)$}
%\tcc*{$Q_1,\hat{\lambda}_1, \bm{a}_1$ are the updated Q function, the resulting estimated Whittle}
\Fn{HEE-ACC}{
	Initialize $\pmb{a}^{\text{HEE-ACC}}\bigl(\bm{X}^{\text{HEE-ACC}}(t)\bigr)\gets \bm{0}$\;
    For $j\in[J]$, build the minimum heap $\mathcal{H}_j$ of all the tuples $(i,i',k)\in[I]^2\times\mathscr{K}$ with $u_j(i,i',k)>0$ according to their indices $\psi_j(i,i',k)$.\;
    \tcc*{Tie cases are broken by selecting the tuples with the smallest expected lifespans.}\;
	\For{$\forall j\in [J]$}{
        $(\bar{i},\bar{i}',\bar{k})\gets$ the root node of the minimum heap $\mathscr{H}_j$\label{line:selected}\;
        \While{$(\bar{i},\bar{i}',\bar{k})\notin \mathscr{T}_j\bigl(\bm{X}^{\text{HEE-ACC}}(t)\bigr)$}{
            $\mathscr{H}_j$ pop heap\;
            $(\bar{i},\bar{i}',\bar{k})\gets$ the root node of the updated $\mathscr{H}_j$\;
        }
        $a^{\text{HEE-ACC}}_{\bar{i},\bar{i}',j,\bar{k}}\bigl(\bm{X}^{\text{HEE-ACC}}(t)\bigr) \gets 1$\;
    }
    \Return $\pmb{a}^{\text{HEE-ACC}}\bigl(\bm{X}^{\text{HEE-ACC}}(t)\bigr)$\;
}
\caption{Implementing HEE-ACC}\label{algo:HEE-ACC}
\end{algorithm}

Let $\bm{\nu}=\bm{\nu}^*$, $\pmb{\gamma}=\pmb{\gamma}^*$ and $\bm{\eta}=\bm{\eta}^*$ represent the optimal dual variables of the relaxed problem described in Objective~\eqref{eqn:obj:relax}, Constraints~\eqref{eqn:action_constraint:relax}, \eqref{eqn:capacity_constraint1:relax} and \eqref{eqn:capacity_constraint2:relax}.
Based on \cite[Theorem EC.1]{fu2021restless} and Corollary~\ref{coro:indexability}, we have the following proposition.
\begin{proposition}\label{prop:asym_opt}
If, for $\bm{\nu}=\bm{\nu}^*$, $\pmb{\gamma}=\pmb{\gamma}^*$ and $\bm{\eta}=\bm{\eta}^*$, a policy $\phi\in\tilde{\Phi}$ satisfying Eq.~\eqref{eqn:indexbility:1}  is optimal to the relaxed problem described in Objective~\eqref{eqn:obj:relax}, Constraints~\eqref{eqn:action_constraint:relax}, \eqref{eqn:capacity_constraint1:relax} and \eqref{eqn:capacity_constraint2:relax}, then there exists a policy $\varphi$ proposed based on the indices $\psi_j(i,i',k)$ with plugged in $\pmb{\gamma}=\pmb{\gamma}^*$ and $\bm{\eta}=\bm{\eta}^*$ such that $\varphi$ approaches optimality of TOSP (described in Objective~\eqref{eqn:obj}, Constraints~\eqref{eqn:action_constraint}, \eqref{eqn:capacity_constraint:1} and \eqref{eqn:capacity_constraint:2}) as $\lambda_j$ ($j\in[J]$), $C_k$ ($k\in[K]$), and $N_i$ ($i\in[I]$) tend to infinity proportionately. 
\end{proposition}
The proof of Proposition~\ref{prop:asym_opt} is provided in Appendix~\ref{app:prop:asym_opt}.
We say such a policy $\varphi$ is \emph{asymptotically optimal} to TOSP.
Asymptotic optimality is appropriate for large-scale systems with highly dense and heterogeneous mobile users, MEC servers/service components, base stations, access points, et cetera.

Recall that, from Corollary~\ref{coro:indexability}, the threshold-style policy $\phi$ satisfying Eq.~\eqref{eqn:indexbility:1} minimizes the right-hand side of the dual function in Eq.~\eqref{eqn:dual_func} (that is, minimize all the sub-problems), but such a policy is in general not applicable to the TOSP, and this minimum is only a lower bound of the minimum of the TOSP.
The threshold-style policy $\phi$ reveals intrinsic features of the bandit processes and quantifies them through the indices, $\psi_j(i,i',k)$, representing the marginal costs of selecting certain bandit processes.
These indices can then be utilized to construct effective policices, such as the $\varphi$ in Proposition~\ref{prop:asym_opt}, applicable to the original TOSP.
However, the exact values of the indices that are able to lead to asymptotic optimality remain an open question because of the unknown $\pmb{\gamma}$ and $\bm{\eta}$.

%To asymptotic optimality, we wish to obtain the optimal values of the dual variables $\pmb{\gamma}=\pmb{\gamma}^*$ and $\bm{\eta}=\bm{\eta}^*$. 

Intuitively, the Lagrange multipliers $\pmb{\gamma}$ and $\bm{\eta}$ represent the marginal budgets for running out the network capacities described in Constraints~\eqref{eqn:capacity_constraint1:relax} and \eqref{eqn:capacity_constraint2:relax}, respectively.
When an SC or a communication channel exhibits heavy traffic, it should become less popular to incoming tasks, which can be adjusted by the attached multipliers $\pmb{\gamma}$ or $\bm{\eta}$ to the indices.
In other words, the multiplier associated with a capacity constraint of an SC or a channel with heavy traffic is expected to be large; while for those SCs or channels with light traffic or sufficiently large capacities, the corresponding multipliers should remain zero.
We refer to $\pmb{\gamma}\in\mathbb{R}_0^K$ and $\bm{\eta}\in\mathbb{R}_0^I$ as the \emph{capacity coefficients} attached to the indices.

On the other hand, observing $\psi_j(i,i',k)$ in Eq.~\eqref{eqn:indexability:2}, apart from the items related to $\pmb{\gamma}$ and $\bm{\eta}$, the remaining item is, for $i,i'\in[I]$, $j\in[J]$, and $k\in\mathscr{K}$
\begin{equation}\label{eqn:ratio}
\mathcal{R}_j(i,i',k)=\begin{cases}
    \frac{\lambda_j}{u_j(i,i',k)} \varepsilon_k w_{j,k}, & \text{if } k\in[K],\\
    \lambda_j \bar{\varepsilon}_j, &\text{otherwise},
\end{cases}    
\end{equation}
which is in fact the expected power consumption per unit service rate on SC $k$ contributed by the $j$-tasks, or equivalently the reciprocal of the achieved service rate per unit power consumption.
We refer to the achieved service rate per unit power consumption of a certain SC as its \emph{energy efficiency}.
Less marginal costs of selecting certain tuples $(i,i',k)$ imply SCs with higher energy efficiencies.
The phenomenon coincides with a special case of the RMAB process studied in \cite{fu2020energy} but further considers the heterogeneous task requirements and complex capacity constraints over heterogeneous network resources.

In this context, for the TOSP described in Objective~\eqref{eqn:obj}, Constraints~\eqref{eqn:action_constraint}, \eqref{eqn:capacity_constraint:1} and \eqref{eqn:capacity_constraint:2}, we propose a policy that always prioritizes tuples $(i,i',k)$ with higher $\psi_j(i,i',k)$ for the $j$-tasks, for which the capacity coefficients $\pmb{\gamma}$ and $\bm{\eta}$ are given a prior through sensible algorithms. 
We refer to this policy as the \emph{Highest Energy Efficiency with Adjusted Capacity Coefficients} (HEE-ACC).
More precisely, for $j\in[J]$ and $\bm{x}\in\mathscr{X}$, define a subset of tuples $\mathscr{T}_j(\bm{x})\subset [I]^2\times\mathscr{K}$ such that, for any $(i,i',k)\in\mathscr{T}_j(\bm{x})$, $u_j(i,i',k)>0$,
\begin{equation}\label{eqn:capacity_test:1}
\sum\limits_{i_1,i'_1\in[I]}\sum\limits_{j_1\in[J]}x_{i_1,i'_1,j_1,k} + w_{j,k} \leq C_k,
\end{equation}
\begin{equation}\label{eqn:capacity_test:2}
\sum\limits_{i_1\in[I]}\sum\limits_{j_1\in[J]}\sum\limits_{k_1\in\mathscr{K}}\bigl(x_{i,i_1,j_1,k_1}+x_{i_1,i,j_1,k_1}\bigr)+ 1 \leq N_i,
\end{equation}
and 
\begin{equation}\label{eqn:capacity_test:3}
\sum\limits_{i_1\in[I]}\sum\limits_{j_1\in[J]}\sum\limits_{k_1\in\mathscr{K}}\bigl(x_{i',i_1,j_1,k_1}+x_{i_1,i',j_1,k_1}\bigr)+ 1 \leq N_{i'}.
\end{equation}
The set $\mathscr{T}_j(\bm{x})$ is the set of available tuples $(i,i',j)$ that can serve a new $j$-task with complied capacity constraints~\eqref{eqn:capacity_constraint:1} and \eqref{eqn:capacity_constraint:2} when $\bm{X}^{\phi}(t) = \bm{x}$.
The action variables for HEE-ACC are given by
\begin{multline*}
    a^{\text{HEE-ACC}}_{i,i',j,k}\bigl(\bm{X}^{\text{HEE-ACC}}(t)\bigr) \\= \begin{cases}
        1,& \text{if } (i,i',k) = \arg\min\nolimits_{(i,i',k)\in\mathscr{T}_j(\bm{X}^{\text{HEE-ACC}}(t))}\psi_j(i,i',k),\\
        0, & \text{otherwise},
    \end{cases}
\end{multline*}
where if $\arg\min$ returns more than one tuples $(i,i',k)$, then we select the one with the highest $u_j(i,i',k)$.
In Algorithm~\ref{algo:HEE-ACC}, as an example, we provide the pseudo-code of implementing HEE-ACC. Note that Algorithm~\ref{algo:HEE-ACC} is a possible but not unique way of implementing HEE-ACC.
The computational complexity of implementing HEE-ACC is at most linear logarithmic to the number of possible tuples $|\{(i,i',k)\in I^2\times\mathscr{K}|u_j(i,i',k) > 0\}|$, mainly dependent on the complexity of finding the tuple with smallest index and checking tuples' availability with respect to the capacity constraints.

It remains to adjust the values of the capacity coefficients $\pmb{\gamma}$ and $\bm{\eta}$, which should be adversely affected by the SCs and channels' remaining capacities.
\subsubsection{HEE-ACC-zero}\label{subsubsec:HEE-ACC-zero} 
In a simple case with sufficiently large capacities, we can directly set $\pmb{\gamma}=\bm{0}$ and $\bm{\eta}=\bm{0}$, for which HEE-ACC is equivalent to a policy that always selects the most energy-efficient channel-SC tuples. 
We refer to such a policy as the \emph{HEE-ACC-zero} policy, which implies the zero capacity coefficients.
Given the simplicity of the HEE-ACC-zero policy, from \cite[Corollary EC.1]{fu2021restless}, when the capacity constraint over an SC (in Constraint~\eqref{eqn:capacity_constraint:1}) or a channel (in Constraint~\eqref{eqn:capacity_constraint:2}) is dominant, HEE-ACC-zero is near-optimal - it approaches optimality as the problem size becomes sufficiently large. 

Mathematically, we say SC $k\in[K]$ is dominant for the resource tuple $(i,i',k)$ if $N_i,N_{i'}\rightarrow +\infty$. 
Similarly, a channel $i$ (or $i'\in[I]$) is dominant when $C_k \rightarrow +\infty$ and $N_{i'}\rightarrow +\infty$ (or $N_i\rightarrow +\infty)$.
We have the following proposition.
\begin{proposition}\label{prop:asym_opt_zero}
If, for each $j\in[J]$ and resource tuple $(i,i',k)\in[I]^2\times[K]$ with $\mu_j(i,i',k)>0$, the SC $k$, channel $i$, or channel $i'$ is dominant, 
the blocking probabilities of all task classes $j\in[J]$ are positive in the asymptotic regime, and  $w_{j,k}(1+\frac{\lambda_j}{\mu_j(i,i',k)})$ is a constant for all $j\in[J]$ and tuples $(i,i',k)\in[I]^2\times[K]$ with $\mu_j(i,i',k)>0$, then HEE-ACC-zero approaches optimality of TOSP (described in Objective~\eqref{eqn:obj}, Constraints~\eqref{eqn:action_constraint}, \eqref{eqn:capacity_constraint:1} and \eqref{eqn:capacity_constraint:2}) as $\lambda_j$ ($j\in[J]$) and the capacities of the dominant SCs or channels of all the resource tuples tend to infinity proportionately.
\end{proposition}
The proof of Proposition~\ref{prop:asym_opt_zero} is provided in Appendix~\ref{app:prop:asym_opt_zero}. 
Proposition~\ref{prop:asym_opt_zero} imposes a hypothesis, requesting the existence of a dominant SC $k\in[K]$ or channel $i\in[I]$ for each resource tuple such that $C_k$ or $N_i$ is significantly less than the capacities of the non-dominant SC and/or channel(s) of the same tuple. 
It indicates a situation where the capacity constraint associated with the dominant SC group $k$ (in Constraint~\eqref{eqn:capacity_constraint:1}) or channel $i$ (in Constraint~\eqref{eqn:capacity_constraint:2}) frequently achieves equality while there is negligible chance to invoke the capacity constraints for the non-dominant SC groups and channels.
Note that, for a resource tuple $(i,i',k)$, the definition about dominant SC $k$ (or channel $i$) in Proposition~\ref{prop:asym_opt_zero}, i.e., $N_i, N_{i'} \rightarrow +\infty$ (or $C_k,N_{i'}\rightarrow +\infty$), are stated for rigorous descriptions in mathematics, which, in practice, can be interpreted as a case with $C_k \ll N_i,N_{i'}$ (or $N_i\ll C_k,N_{i'}$).

Although, for the more general case with no dominant capacity constraint, HEE-ACC-zero neglects the effects of traffic densities imposed on different SC groups and channels,
it is neat with proven near-optimality in special cases and is easy to implement. 

\begin{algorithm}[t]\small
\linespread{1}\selectfont
%\SetKwFunction{IndexPolicy}{$(\bm{\eta}_e)\gets$ IndexPolicy}
\SetKwProg{Fn}{Function}{}{End}
\SetKwInOut{Input}{Input}
\SetKwInOut{Output}{Output}
%\SetAlgoLined
%\DontPrintSemicolon

\Input{The learned capacity coefficients $\pmb{\gamma}$ and $\bm{\eta}$ at time $t$, variables $\bm{M}(t)$ and system state $\bm{X}^{\text{HEE-ALRN}}(t)$.}
\Output{Scheduling actions $\pmb{a}^{\text{HEE-ALRN}}\bigl(\bm{X}^{\text{HEE-ALRN}}(t)\bigr)\coloneqq \Bigl(a^{\text{HEE-ALRN}}_{i,i',j,k}\bigl(\bm{X}^{\text{HEE-ALRN}}(t)\bigr): (i,i',j,k)\in[I]^2\times [J]\times \mathscr{K}\Bigr)$}
%\tcc*{$Q_1,\hat{\lambda}_1, \bm{a}_1$ are the updated Q function, the resulting estimated Whittle}

\Fn{HEE-ALRN}{
	Initialize $\pmb{a}^{\text{HEE-ALRN}}\bigl(\bm{X}^{\text{HEE-ALRN}}(t)\bigr)\gets \bm{0}$\;
    For $j\in[J]$, build the minimum heap $\mathcal{H}_j$ of all the tuples $(i,i',k)\in[I]^2\times\mathscr{K}$ with $u_j(i,i',k)>0$ according to their indices $\psi_j(i,i',k)$ with the input $\pmb{\gamma}$ and $\bm{\eta}$.\;
%    \tcc*{Tie cases are broken by selecting the tuples with the smallest expected lifespans.}\;
	\For{$\forall j\in [J]$}{
        $(\bar{i},\bar{i}',\bar{k})\gets$ the root node of the minimum heap $\mathscr{H}_j$\;
        \If{$(\bar{i},\bar{i}',\bar{k})\in \mathscr{T}_j\bigl(\bm{X}^{\text{HEE-ALRN}}(t)\bigr)$}{\label{line:selected6}
            \If {$k\in[K]$}{
                $\gamma_{\bar{k}}\gets \max\{0,\gamma_{\bar{k}}-\Delta_{\gamma}\}$\;
            }
            $\eta_{\bar{i}}\gets \max\{0,\eta_{\bar{i}}-\Delta_{\eta}\}$\;
            $\eta_{\bar{i}'}\gets \max\{0,\eta_{\bar{i}'}-\Delta_{\eta}\}$\;
        }
        \While{$(\bar{i},\bar{i}',\bar{k})\notin \mathscr{T}_j\bigl(\bm{X}^{\text{HEE-ALRN}}(t)\bigr)$}{
            $\mathscr{H}_j$ pop heap\;
            $(\bar{i},\bar{i}',\bar{k})\gets$ the root node of the updated $\mathscr{H}_j$\;
            \uIf{$(\bar{i},\bar{i}',\bar{k})\in \mathscr{T}_j\bigl(\bm{X}^{\text{HEE-ALRN}}(t)\bigr)$}{
                \If {$k\in[K]$}{
                    $\gamma_{\bar{k}}\gets \max\{0,\gamma_{\bar{k}}-\Delta_{\gamma}\}$\;
                }
                $\eta_{\bar{i}}\gets \max\{0,\eta_{\bar{i}}-\Delta_{\eta}\}$\;
                $\eta_{\bar{i}'}\gets \max\{0,\eta_{\bar{i}'}-\Delta_{\eta}\}$\;
            }\Else{
                If \eqref{eqn:capacity_test:1}, \eqref{eqn:capacity_test:2} and/or \eqref{eqn:capacity_test:3} is violated by setting $k=\bar{k}$, $i=\bar{i}$ and $i'=\bar{i}'$, then increment 
                $M_{\bar{k}}(t)$, $M_{K+\bar{i}}(t)$ and/or
                    $M_{K+\bar{i}'}(t)$ by one.\label{line:selected20}\;
                }
            }%end while
            If the increment adjustment is triggered by the updated $M_{\bar{k}}(t)$, $M_{K+\bar{i}}(t)$ or $M_{K+\bar{i}'}(t)$, then, based on the sub-gradients described in \eqref{eqn:sub-gradient:1} and \eqref{eqn:sub-gradient:2}, update the values of $\bm{M}(t)$, $\pmb{\gamma}$ and $\bm{\eta}$ accordingly.\;
 %           \If {$M_{\bar{k}}(t) > \bar{M}$}{
%                \If {$\Lambda_{\bar{k}}^{\gamma}\bigl(\pmb{\nu}^*(\pmb{\gamma},\bm{\eta}),\pmb{\gamma},\bm{\eta}\bigr) < 0$}{$\gamma_{\bar{k}}\gets \gamma_{\bar{k}} + \Delta^+_{\gamma}$\;}
 %               $M_{\bar{k}}(t)\gets 0$\;
 %           }
%            \If{$M_{K+\bar{i}JL+jL+\ell(\bar{k})}(t)>\bar{M}$}{
                %\If{$\Lambda_{\bar{i},j,\ell(\bar{k})}^{\eta}\bigl(\pmb{\nu}^*(\pmb{\gamma},\bm{\eta}),\pmb{\gamma},\bm{\eta}\bigr) < 0$}{
 %                   $\eta_{\bar{i},j,\ell(\bar{k})}\gets \eta_{\bar{i},j,\ell(\bar{k})} + \Delta^+_{\eta}$\;
 %               }
 %               $M_{K+\bar{i}JL+jL+\ell(\bar{k})}(t)\gets0$\;
 %           }
 %           \If{$M_{K+\bar{i}'JL+jL+\ell(\bar{k})}(t)>\bar{M}$}{
                %\If{$\Lambda_{\bar{i}',j,\ell(\bar{k})}^{\eta}\bigl(\pmb{\nu}^*(\pmb{\gamma},\bm{\eta}),\pmb{\gamma},\bm{\eta}\bigr) < 0$}{
 %                   $\eta_{\bar{i}',j,\ell(\bar{k})}\gets \eta_{\bar{i}',j,\ell(\bar{k})} + \Delta^+_{\eta}$\;
 %               }
 %               $M_{K+\bar{i}'JL+jL+\ell(\bar{k})}(t)\gets0$\;
%            }
        $a^{\text{HEE-ALRN}}_{\bar{i},\bar{i}',j,\bar{k}}\bigl(\bm{X}^{\text{HEE-ALRN}}(t)\bigr) \gets 1$\;
    }
    \Return $\pmb{a}^{\text{HEE-ALRN}}\bigl(\bm{X}^{\text{HEE-ALRN}}(t)\bigr)$\;
}
\caption{Implementing HEE-ALRN}\label{algo:HEE-ACC-learning}
\end{algorithm}

\subsubsection{HEE-ALRN}

Recall that the minimum of the sub-problems is a lower bound of the minimum of the relaxed TOSP problem.
As stated in Proposition~\ref{prop:asym_opt}, when the relaxed problem achieves strong duality in the asymptotic regime, there exists a policy prioritizes tuples $(i,i',k)$ with the lowest $\psi_j(i,i',k)$ that is asymptotically optimal to TOSP. 
In particular, in this case with achieved asymptotic optimality, the capacity coefficients $\pmb{\gamma}$ and $\bm{\eta}$ should be equal to the optimal dual variables $\pmb{\gamma}^*$ and $\bm{\eta}^*$ of the relaxed problem in the asymptotic regime.
Nonetheless, due to the complexity of the relaxed problem in both the asymptotic and non-asymptotic regime, the strong duality and the exact values of the dual variables remain open questions.
Here, we aim at a diminishing gap between the performance of the sub-problems and the relaxed problem and approximate the values of $\pmb{\gamma}^*$ and $\bm{\eta}^*$ through a learning method with closed-loop feedback.

To emphasize the effects of different $\pmb{\gamma}$ and $\bm{\eta}$, we rewrite HEE-ACC as HEE-ACC$(\pmb{\gamma},\bm{\eta})$.
We explore the values of $\pmb{\gamma}\in\mathbb{R}_0^K$ and $\bm{\eta}\in\mathbb{R}_0^I$ while exploiting HEE-ACC$(\pmb{\gamma},\bm{\eta})$ with the most updated capacity coefficients in TOSP.
We iteratively learn the values of $\pmb{\gamma}\in\mathbb{R}_0^K$ and $\bm{\eta}\in\mathbb{R}_0^I$ in the vein of the well-known gradient descent algorithm \cite{boyd2004convex} based on the observed traffic tensity upon the SCs and channels.

In particular, 
let $\Phi(\pmb{\nu},\pmb{\gamma},\bm{\eta})$ represent the set of all policies satisfying \eqref{eqn:indexbility:1} with given $\pmb{\nu}\in\mathbb{R}_0^J$, $\pmb{\gamma}\in\mathbb{R}_0^K$ and $\bm{\eta}\in\mathbb{R}_0^I$.
Define
\begin{multline}\label{eqn:nu_star}
    \pmb{\nu}^*(\pmb{\gamma},\bm{\eta}) \coloneqq \sup \Bigl\{\pmb{\nu}\in\mathbb{R}_0^J \Bigl|\\\exists \phi\in\Phi(\pmb{\nu},\pmb{\gamma},\bm{\eta}),~\text{Constraint~\eqref{eqn:action_constraint:relax} is satisfied. }\Bigr\},
\end{multline}
and let $\varphi^*(\pmb{\gamma},\bm{\eta})$ represent the only policy in $\Phi\bigl(\pmb{\nu}^*(\pmb{\gamma},\bm{\eta}),\pmb{\gamma},\bm{\eta}\bigr)$. 
We provide the pseudo-code of computing $\pmb{\nu}^*(\pmb{\gamma},\bm{\eta})$ and the action variables for $\varphi^*(\pmb{\gamma},\bm{\eta})$ with given $\pmb{\gamma}\in\mathbb{R}_0^K$ and $\bm{\eta}\in\mathbb{R}_0^I$ in Appendix~\ref{app:algo:nu_star}.%Algorithm~\ref{algo:nu_star}.

Since dual function $L(\pmb{\nu},\pmb{\gamma},\bm{\eta})$ is continuous, piece-wise linear and concave in $\pmb{\nu}$, $\pmb{\gamma}$ and $\pmb{\eta}$, $-L(\pmb{\nu},\pmb{\gamma},\bm{\eta})$ is sub-differentiable with existing sub-gradients at $\gamma_k=\gamma$ for $k\in[K]$,
\begin{multline}\label{eqn:sub-gradient:1}
    \Lambda_k^{\gamma}(\pmb{\nu},\pmb{\gamma},\bm{\eta})\coloneqq \lim\limits_{\gamma_k\downarrow \gamma}\nabla_{\gamma_k}L(\pmb{\nu},\pmb{\gamma},\bm{\eta}) \\=
    \sum\nolimits_{\begin{subarray}~(i,i',j)\in[I]^2\times[J]:\\ u_j(i,i',k)\end{subarray}}w_{j,k}\sum\limits_{x\in\mathscr{X}_{i,i',j,k}}\pi^{\varphi(\pmb{\nu},\pmb{\gamma},\bm{\eta})}_{i,i',j,k}(x)x - C_k,
\end{multline}
where $\varphi(\pmb{\nu},\pmb{\gamma},\bm{\eta})$ is a policy in $\Phi(\pmb{\nu},\pmb{\gamma},\bm{\eta})$, and at $\eta_i=\eta$ for $i\in[I]$,
\begin{multline}\label{eqn:sub-gradient:2}
   \Lambda_i^{\eta}(\pmb{\nu},\pmb{\gamma},\bm{\eta})\coloneqq \lim\limits_{\eta_{i,j,\ell}\downarrow \eta}\nabla_{\eta_i}L(\pmb{\nu},\pmb{\gamma},\bm{\eta}) \\=\sum\limits_{(i',j,k)\in[I]\times[J]\times\mathscr{K}}\Bigl(\sum\limits_{x\in\mathscr{X}_{i,i',j,k}}\pi^{\varphi(\pmb{\nu},\pmb{\gamma},\bm{\eta})}_{i,i',j,k}(x)x\\+\sum\limits_{x\in\mathscr{X}_{i',i,j,k}}\pi^{\varphi(\pmb{\nu},\pmb{\gamma},\bm{\eta})}_{i',i,j,k}(x)x\Bigr)-  N_i.    
\end{multline}
%Let $\pmb{\Lambda}(\pmb{\nu},\pmb{\gamma},\bm{\eta})\coloneqq \bigl(\Lambda_{\gamma_k}(\pmb{\nu},\pmb{\gamma},\bm{\eta}): k\in[K]; \Lambda_{\eta_{i,j,\ell}}(\pmb{\nu},\pmb{\gamma},\bm{\eta}):(i,j,\ell)\in[I]\times[J]\times[L]\bigr)$.
%To approximate optimal dual variables, we aim to minimize  $\lVert\pmb{\Lambda}(\pmb{\nu}^*(\pmb{\gamma},\bm{\eta}),\pmb{\gamma},\bm{\eta})\rVert$ by appropriately adjusting the values of $\pmb{\gamma}\in \mathbb{R}_0^K$ and $\bm{\eta}\in\mathbb{R}_0^{IJL}$.

We implement the HEE-ACC policy with some prior values of the capacity coefficients $\pmb{\gamma}\in\mathbb{R}_0^K$ and $\bm{\eta}\in\mathbb{R}_0^I$ and initialize $\pmb{\gamma} = \pmb{\gamma}_0$ and $\bm{\eta} = \bm{\eta}_0$.
HEE-ACC is implemented in the way as described in Algorithm~\ref{algo:HEE-ACC}.
Recall that we aim to approximate the optimal dual variables $\pmb{\gamma}^*$ and $\bm{\eta}^*$ in the asymptotic regime, which should satisfy the complementary slackness conditions of the relaxed problem described in Objective~\eqref{eqn:obj:relax}, Constraints~\eqref{eqn:action_constraint:relax}, \eqref{eqn:capacity_constraint1:relax} and \eqref{eqn:capacity_constraint2:relax}. 
That is, for the SCs and channels exhibiting light traffic, the corresponding capacity coefficients should remain zero, while for SCs and channels with heavy traffic, the capacity coefficients are likely to be incremented. 

Upon the arrival of a $j$-task, if tuple $(i,i',k)$ is selected by the currently employed HEE-ACC$(\pmb{\gamma},\bm{\eta})$ and successfully accepts the newly arrived $j$-task, then update the value of $\gamma_k$, $\eta_i$ and $\eta_{i'}$ to $\max\{0,\gamma_k - \Delta_{\gamma}\}$, $\max\{0,\eta_i-\Delta_{\eta}\}$ and $\max\{0,\eta_{i'}-\Delta_{\eta}\}$, respectively, where $\Delta_{\gamma}$ and $\Delta_{\eta}$ are hyper-parameters used to adjust the step size of the learning process. We refer to the decrement process of the capacity coefficients as the \emph{decrement adjustment}. 
On the other hand, we keep a vector $\bm{M}(t) \coloneqq \bigl(M_n(t):n\in[K+I]\bigr)\in\mathbb{N}_0^{K+I}$ to count the occurrences of rejecting a task due to violated capacity constraints.
The vector $\bm{M}(t)$ is initialized to be $\bm{0}$.
In Line~\ref{line:selected6} of Algorithm~\ref{algo:HEE-ACC-learning}, a tuple is selected based on HEE-ACC$(\pmb{\gamma},\bm{\eta})$. However, in Line~\ref{line:selected20} of Algorithm~\ref{algo:HEE-ACC-learning}, if the selected tuple is unable to serve the j-task due to violated capacity constraint(s), the values $M_{\bar{k}}(t)$, $M_{K+\bar{i}}(t)$ and/or $M_{K+\bar{i}'}(t)$ are incremented by one to indicate which constraint has been violated (Constraints~\eqref{eqn:capacity_test:1}, \eqref{eqn:capacity_test:2} and/or \eqref{eqn:capacity_test:3}, respectively).
%If, based on HEE-ACC$(\pmb{\gamma},\bm{\eta})$, a tuple is selected in Line~\ref{line:selected} of Algorithm~\ref{algo:HEE-ACC}; \shen{but, due to violated capacity constraint(s) over SC $k$ and/or channel(s) $i$, failed to serve the corresponding task, then $M_k(t)$ and/or $M_{K+i}(t)$ increment by one. [??]}
When $M_n(t)$ for some $n\in[K+I]$ reaches a pre-determined threshold, $\bar{M}\in\mathbb{N}_+$, we re-set $M_n(t)$ to zero and trigger an \emph{increment adjustment} of the capacity coefficients. 
In particular, with hyper-parameter $\Delta^+_{\gamma}\in\mathbb{R}_+$ and $\Delta^+_{\eta}\in\mathbb{R}_+$, if $n\in[K]$ and $ \Lambda_k^{\gamma}\bigl(\pmb{\nu}^*(\pmb{\gamma},\bm{\eta}),\pmb{\gamma},\bm{\eta}\bigr) > 0$, then increment $\gamma_n$ by $\Delta^+_{\gamma}$;
if $n\in [K+I]\backslash[K]$ and $\Lambda_i^{\eta}\bigl(\pmb{\nu}^*(\pmb{\gamma},\bm{\eta}),\pmb{\gamma},\bm{\eta}\bigr)>0$ with $i = n-K$, then increment $\eta_i$ by $\Delta^+_{\eta}$;
otherwise, do not change the capacity coefficients.
With the adjusted capacity coefficients $\pmb{\gamma}$ and $\bm{\eta}$, we continue implementing the HEE-ACC$(\pmb{\gamma},\bm{\eta})$ policy and keep refining the coefficients subsequently. We refer such a HEE-ACC$(\pmb{\gamma},\bm{\eta})$ policy, for which the coefficients $\pmb{\gamma}$ and $\bm{\eta}$ are learnt and updated through historical observations, as the \emph{HEE-ACC-learning} (HEE-ALRN) policy.
We provide the pseudo-code of implementing HEE-ALRN in Algorithm~\ref{algo:HEE-ACC-learning}.

%\subsubsection{HEE-ACC-dynamic}

%\section{Asymptotic Optimality}\label{sec:asym_opt}
%\section{Offloading with On-demand Connections}\label{sec:on-demand-offloading}

\section{Results and Analysis}\label{sec:simulation}
\begin{figure*}[t]
\centering
\subfigure[Power conservation]{
    \includegraphics[width=0.32\linewidth]{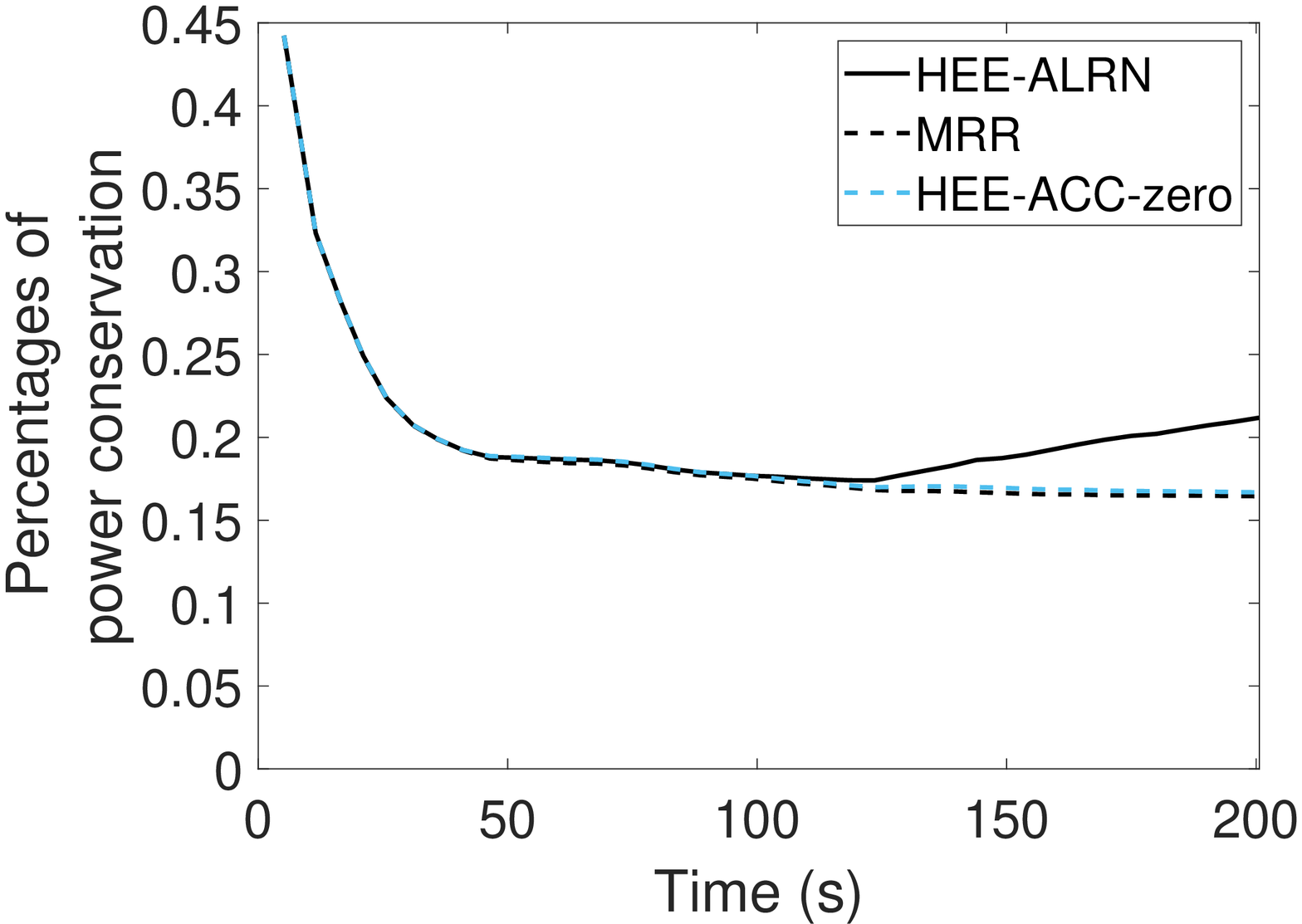}
    \label{fig:rho7.5:power-conservation-seed50:timeline}
}\hspace{-0.008\linewidth}
\subfigure[Throughput per unit power]{
    \begin{overpic}[width=0.32\linewidth]{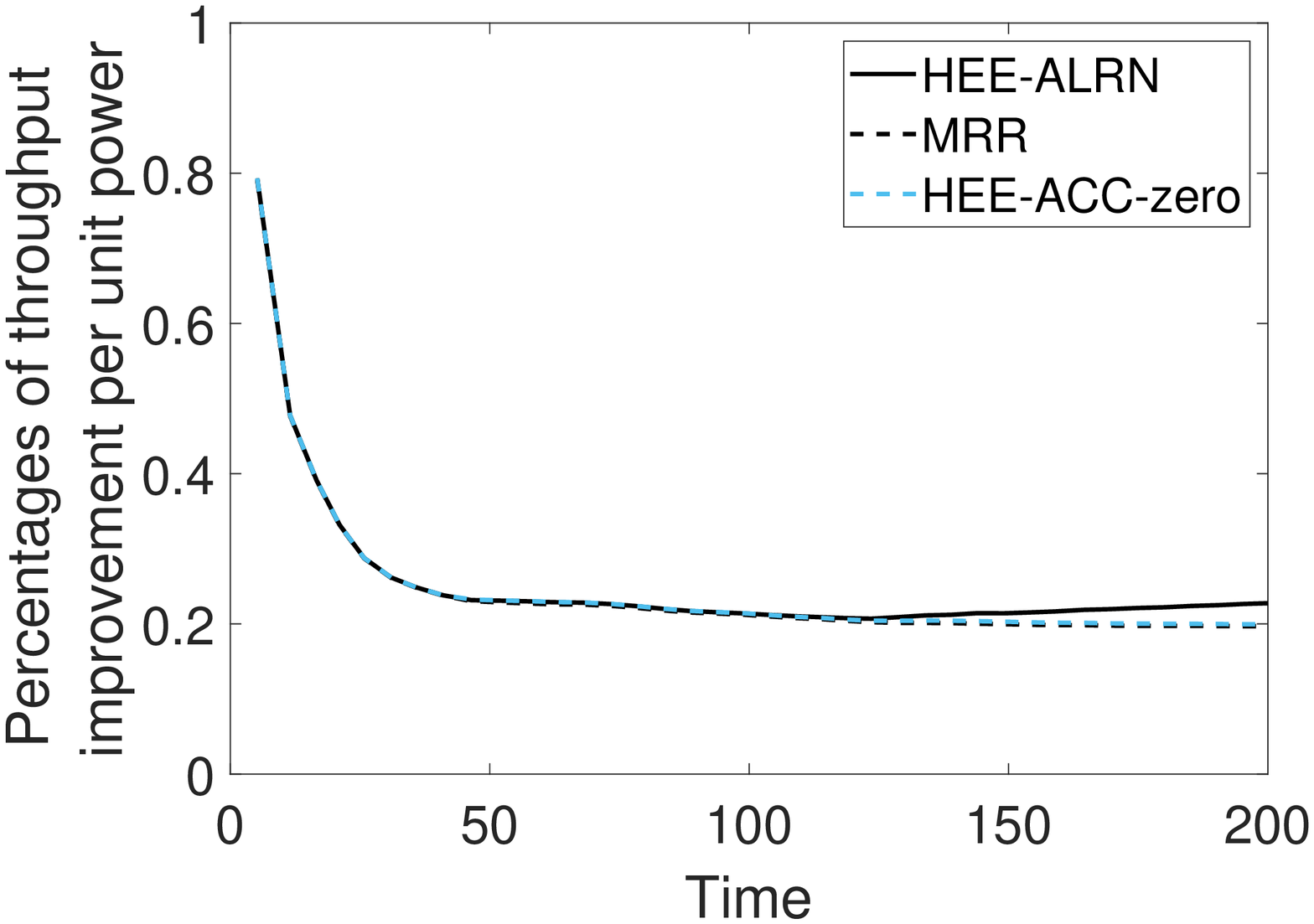}
        \put(59,1.2){\textcolor{black}{\footnotesize(s)}} 
    \end{overpic}
    \label{fig:rho7.5:throughput-power-seed50:timeline}
}\hspace{-0.008\linewidth}
\subfigure[Average delay]{
    \includegraphics[width=0.32\linewidth]{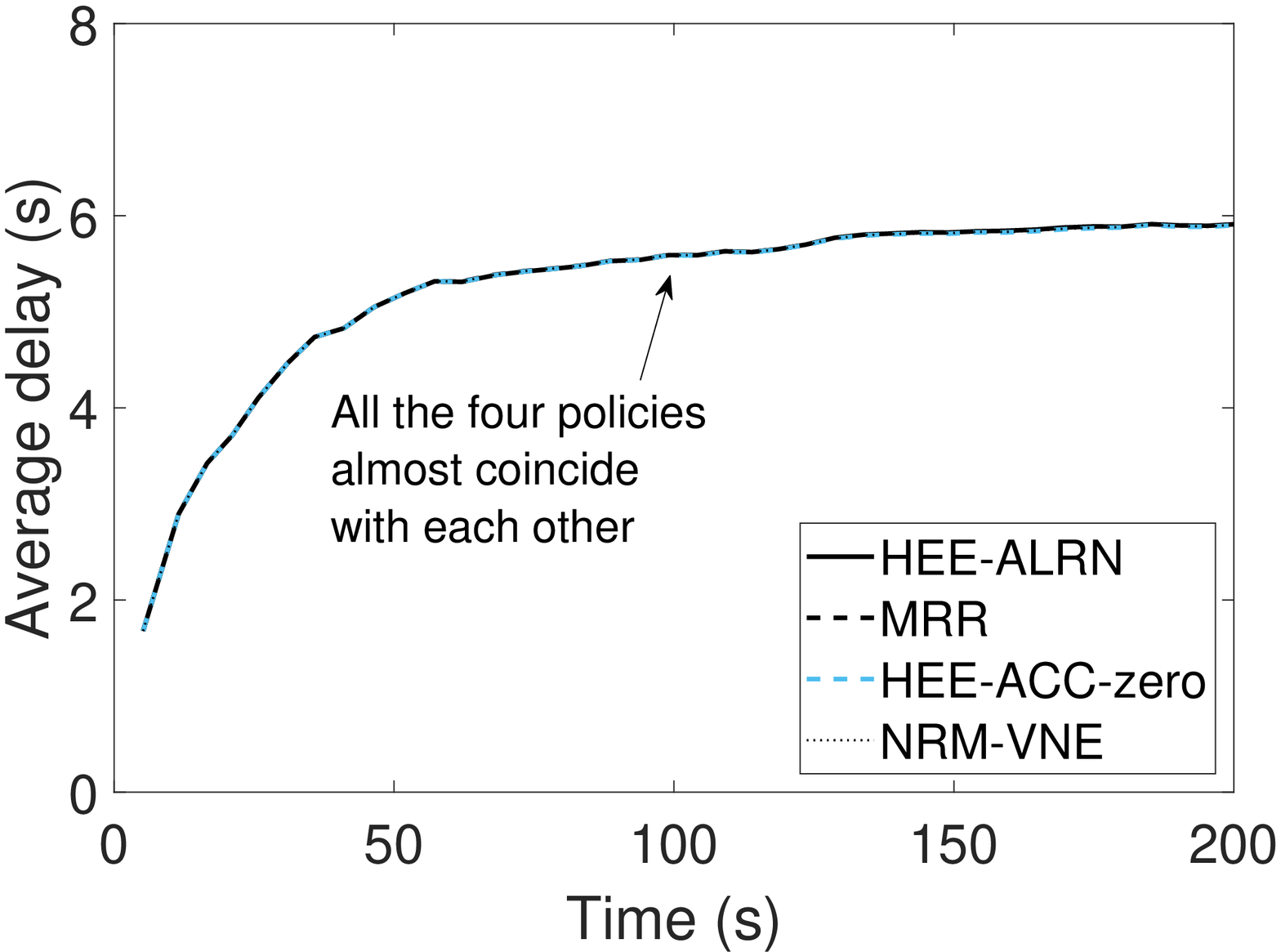}
    \label{fig:rho7.5:throughput-seed50:timeline}
}
\caption{Performance evaluation of HEE-ACC-zero and HEE-ALRN against the timeline, where $\rho = 7.5$.}
\label{fig:seed50-rho7.5:timeline}
\end{figure*}

\begin{figure*}[t]
\centering
%\begin{minipage}[]{\textwidth}
%\centering
%\subfigure[]{\includegraphics[width=0.32\linewidth]{figs/timeline-seed50-rho5-power-percentage.eps}
%\label{fig:rho5:power-conservation-seed50:timeline}}
%\end{minipage}
%\begin{minipage}[]{0.66\textwidth}
%\subfigure[]{\includegraphics[width=0.32\linewidth]{figs/timeline-seed50-rho5-power-throughput.eps}\label{fig:rho5:throughput-power-seed50:timeline}}
%\subfigure[]{\includegraphics[width=0.32\linewidth]{figs/timeline-seed50-rho5-throughput.eps}\label{fig:rho5:throughput-seed50:timeline}}
%\caption{Performance evaluation of HEE-ACC-zero and HEE-ALRN against the timeline, where $\rho = 5$.}
%\label{fig:seed50-rho5:timeline}
%\end{minipage}
\begin{comment}
\begin{minipage}[]{\textwidth}
\centering
\subfigure[]{\includegraphics[width=0.32\linewidth]{figs/timeline-seed50-rho7.5-power-percentage.eps}
\label{fig:rho7.5:power-conservation-seed50:timeline}
}
%\end{minipage}
%\begin{minipage}[]{0.66\textwidth}
\subfigure[]{\includegraphics[width=0.32\linewidth]{figs/timeline-seed50-rho7.5-power-throughput.eps}\label{fig:rho7.5:throughput-power-seed50:timeline}}
\subfigure[]{\includegraphics[width=0.32\linewidth]{figs/timeline-seed50-rho7.5-delay.eps}\label{fig:rho7.5:throughput-seed50:timeline}}
\caption{Performance evaluation of HEE-ACC-zero and HEE-ALRN against the timeline, where $\rho = 7.5$. (a)Power conservation, (b)throughput per unit power, (c)average delay.}
\label{fig:seed50-rho7.5:timeline}
\end{minipage}
\end{comment}

\begin{minipage}[]{\textwidth}
\centering
\subfigure[Power conservation]{\includegraphics[width=0.32\linewidth]{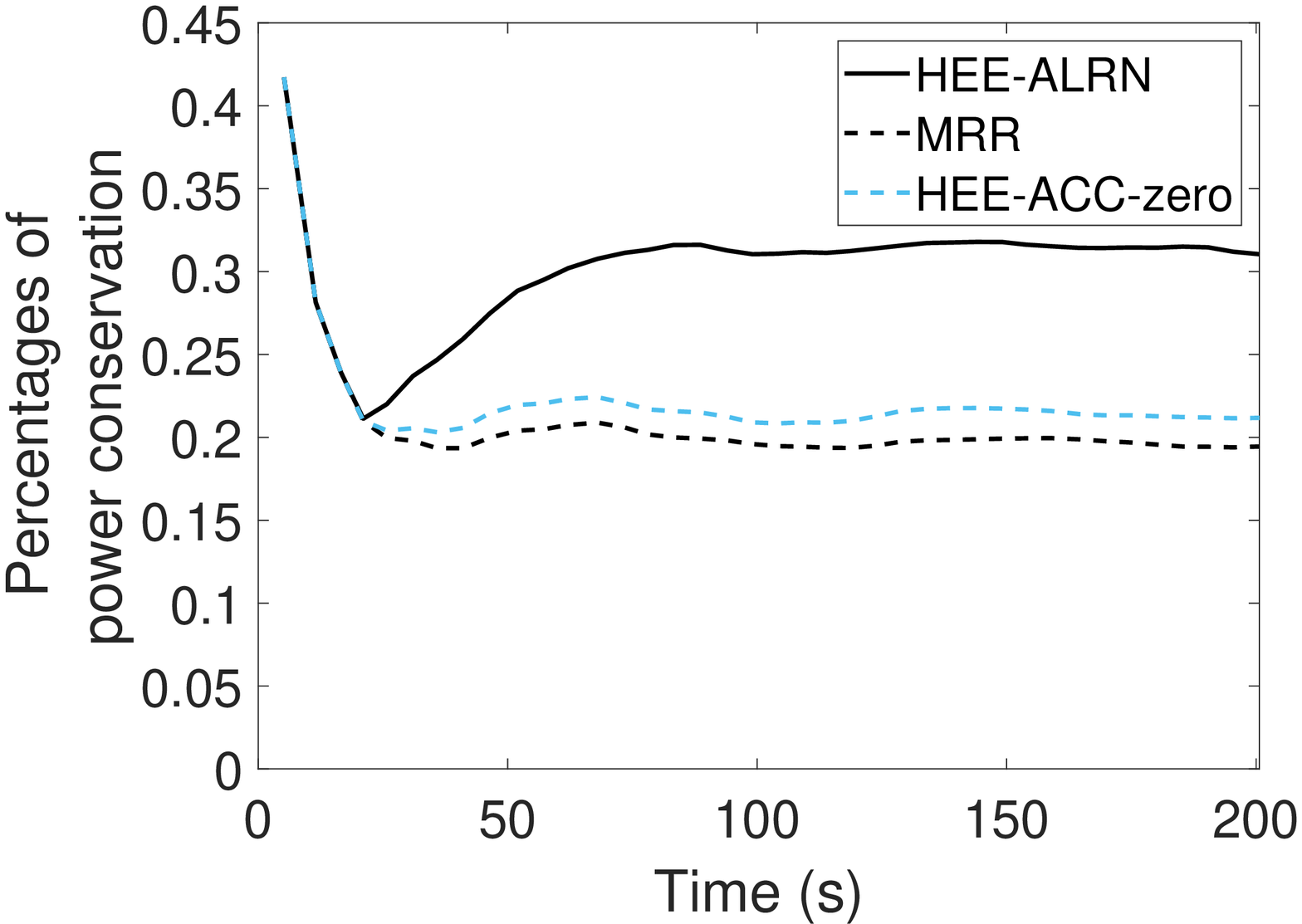}
\label{fig:rho10:power-conservation-seed50:timeline}}
%\end{minipage}
%\begin{minipage}[]{0.66\textwidth}
\subfigure[Throughput per unit power]{\includegraphics[width=0.32\linewidth]{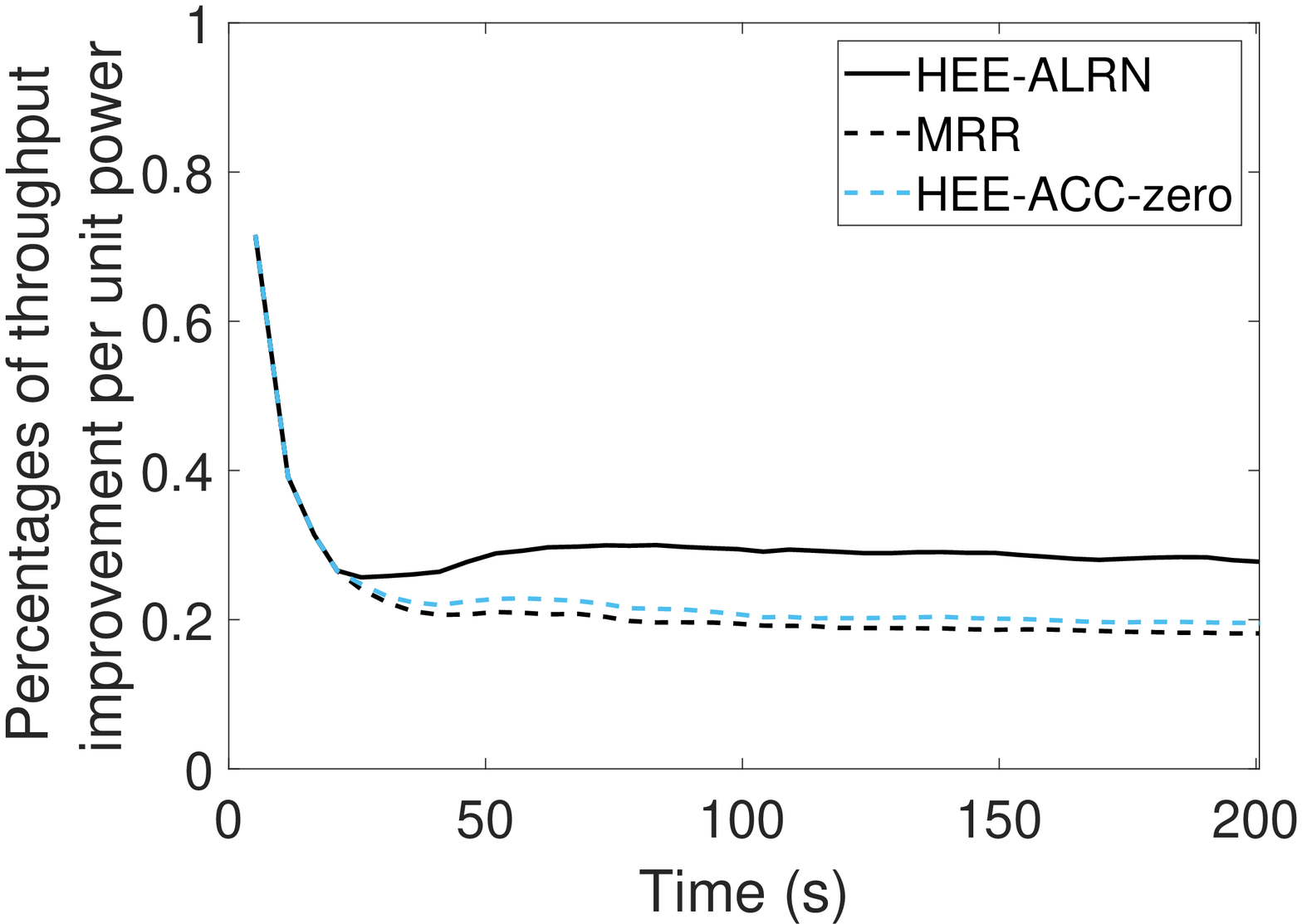}\label{fig:rho10:throughput-power-seed50:timeline}}
\subfigure[Average delay]{\includegraphics[width=0.32\linewidth]{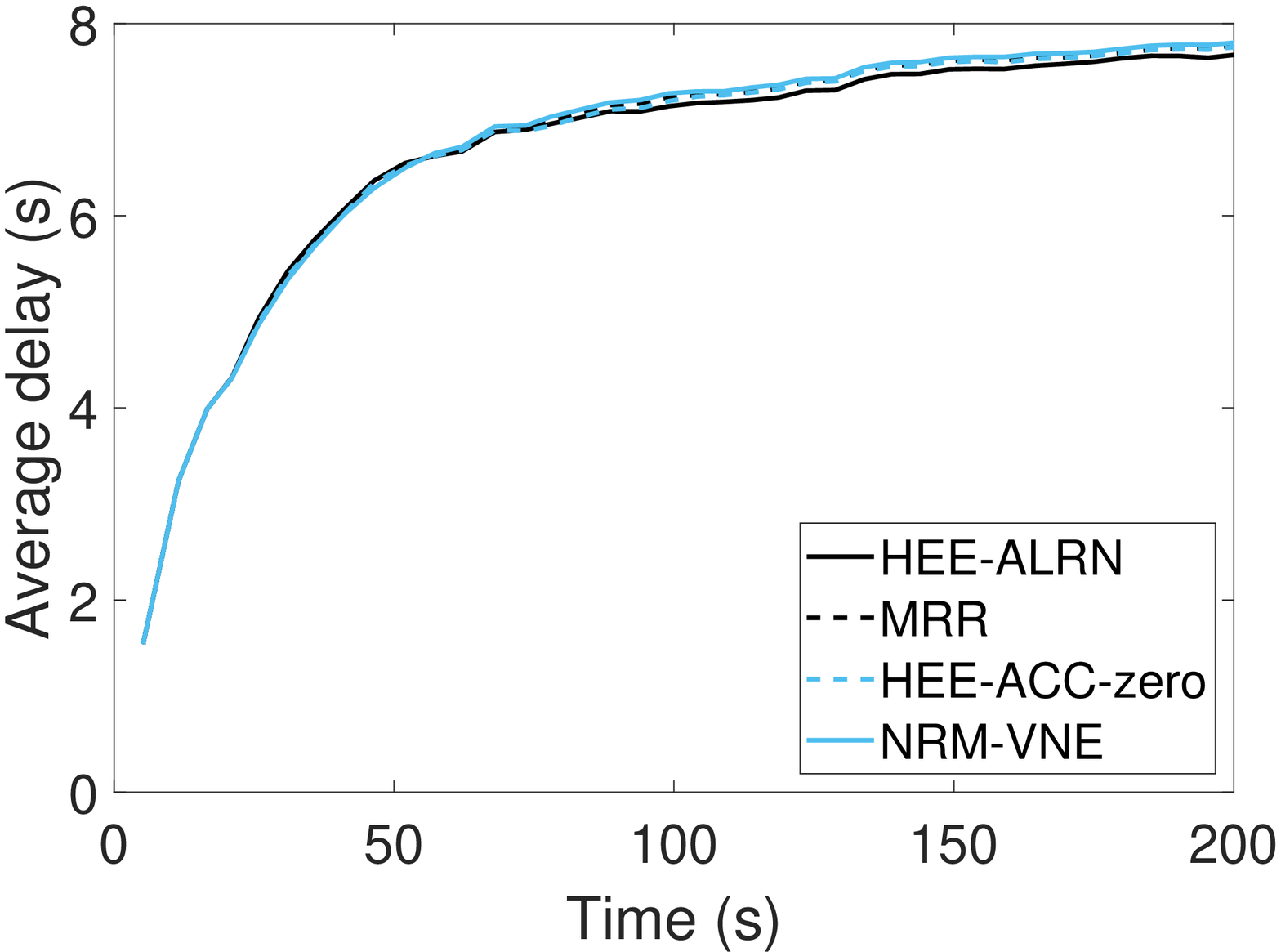}\label{fig:rho10:throughput-seed50:timeline}}
\caption{Performance evaluation of HEE-ACC-zero and HEE-ALRN against the timeline, where $\rho = 10$.}
\label{fig:seed50-rho10:timeline}
\end{minipage}
\end{figure*}

\begin{figure*}[t]
\centering
%\begin{minipage}[]{\textwidth}
%\centering
%\subfigure[]{\includegraphics[width=0.32\linewidth]{figs/timeline-varying-rho1-seed50-power-percentage.eps}
%\label{fig:varying-time:rho1:power-conservation-seed50:timeline}}
%\subfigure[]{\includegraphics[width=0.32\linewidth]{figs/timeline-varying-rho1-seed50-power-throughput.eps}\label{fig:varying-time:rho1:throughput-power-seed50:timeline}}
%\subfigure[]{\includegraphics[width=0.32\linewidth]{figs/timeline-varying-rho1-seed50-delay.eps}\label{fig:varying-time:rho1:throughput-seed50:timeline}}
%\caption{Performance evaluation of HEE-ACC-zero and HEE-ALRN against the timeline, where different task groups have different offered traffic intensities with parameter $p=1$ (light traffic).}
%\label{fig:varying-time:seed50-rho1:timeline}
%\end{minipage}
\begin{minipage}[]{\textwidth}
\centering
\subfigure[Power conservation]{\includegraphics[width=0.32\linewidth]{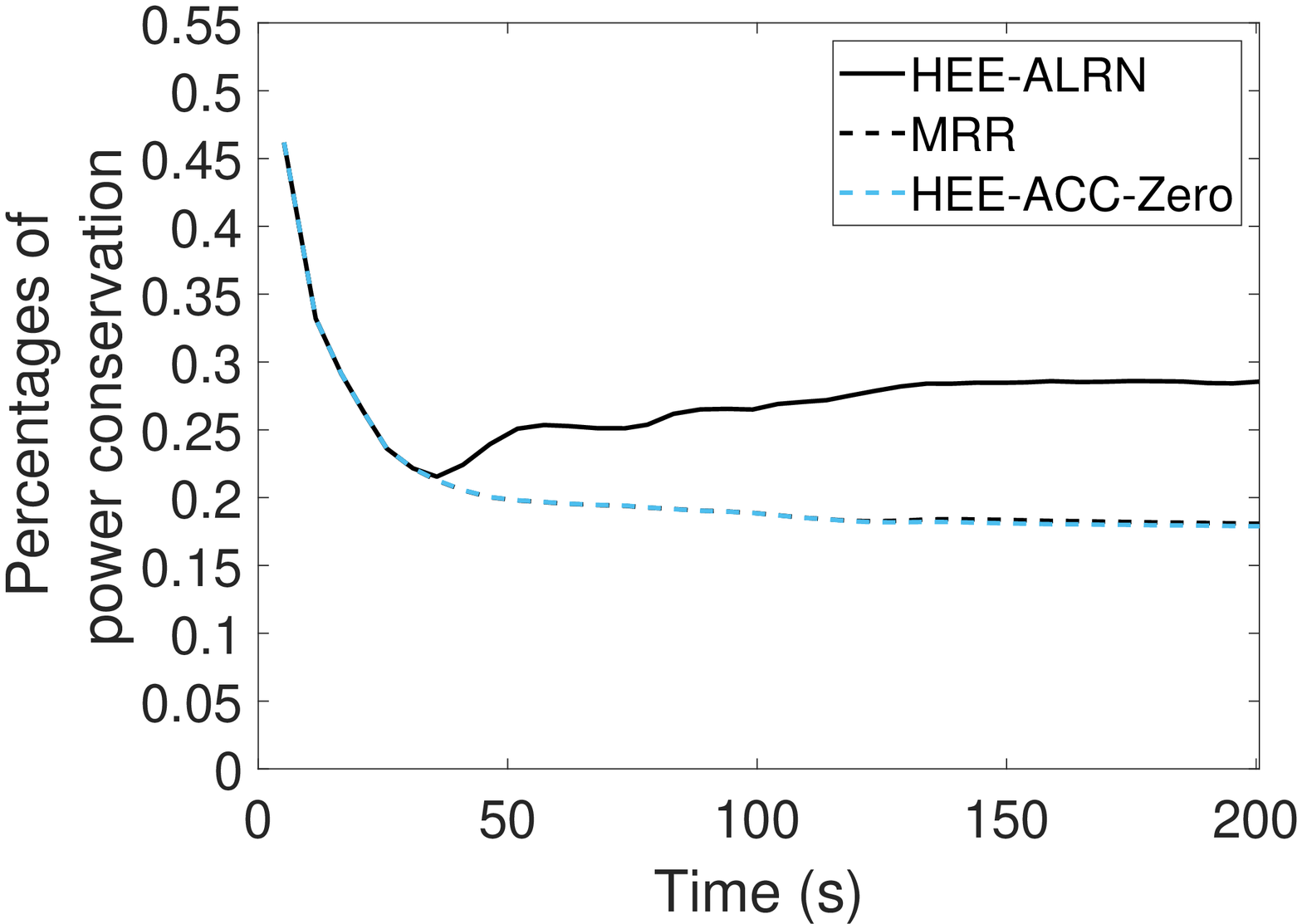}
\label{fig:varying-time:rho1.5:power-conservation-seed50:timeline}}
%\end{minipage}
%\begin{minipage}[]{0.66\textwidth}
\subfigure[Throughput per unit power]{\includegraphics[width=0.32\linewidth]{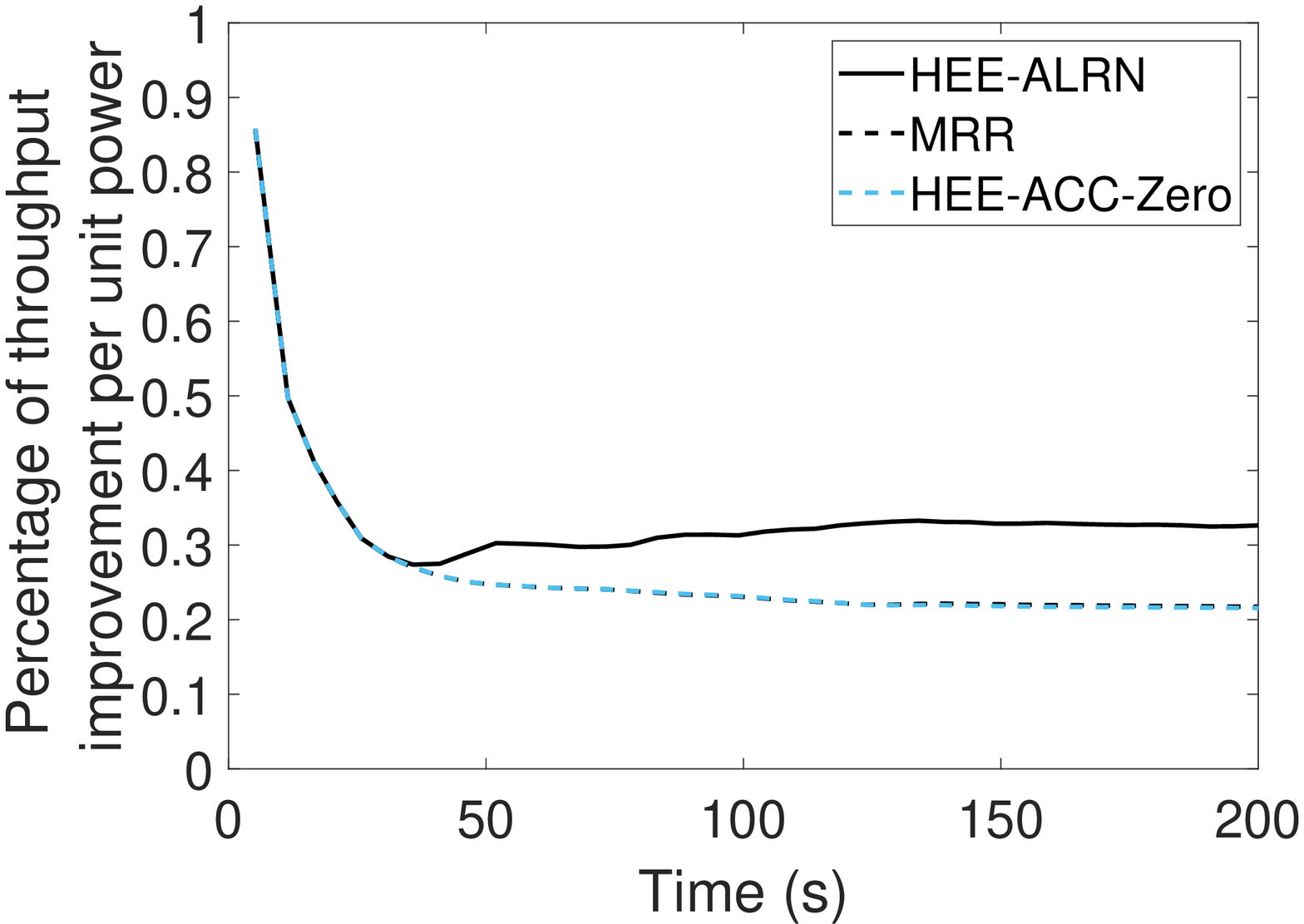}\label{fig:varying-time:rho1.5:throughput-power-seed50:timeline}}
\subfigure[Average delay]{\includegraphics[width=0.32\linewidth]{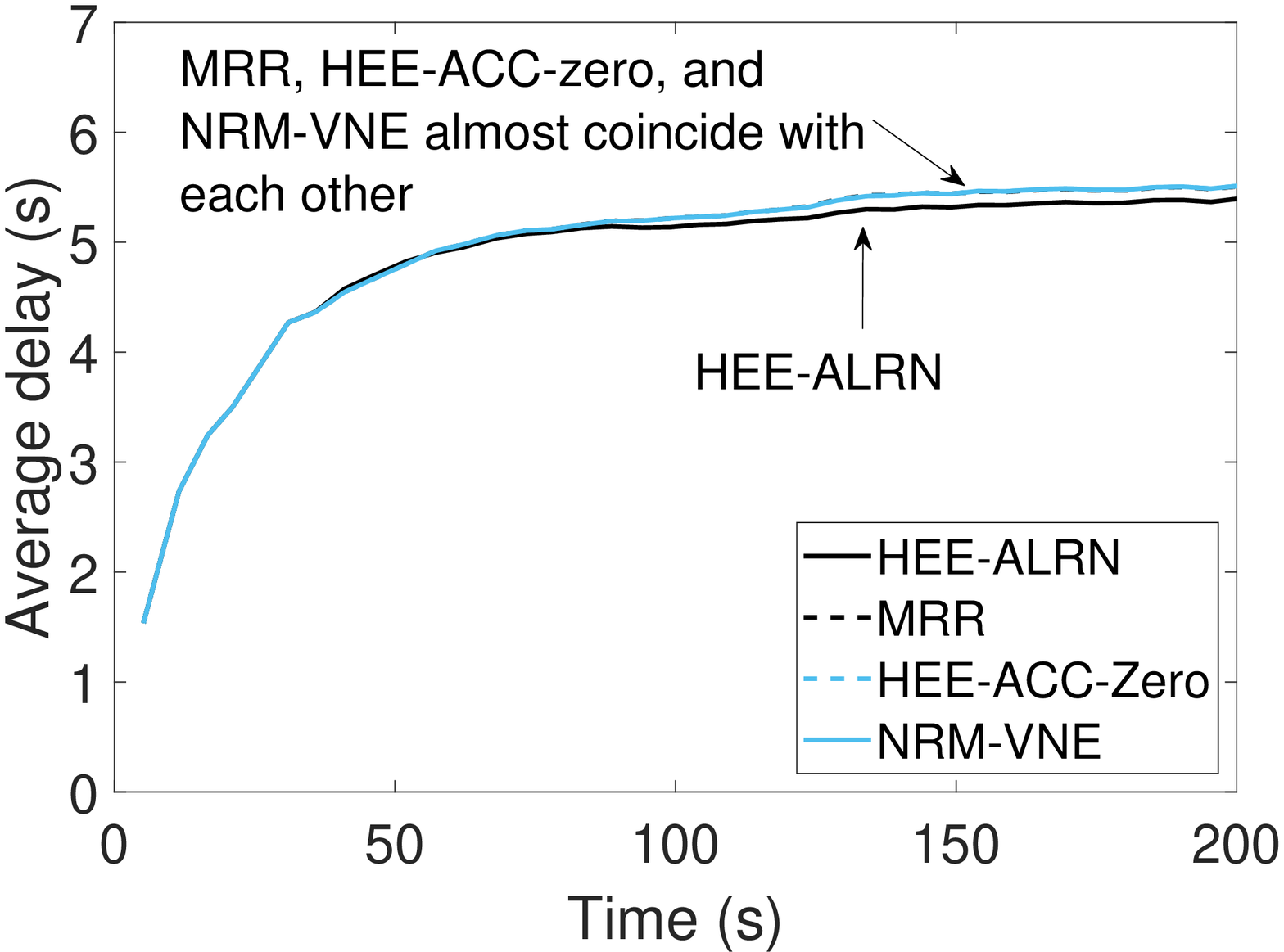}\label{fig:varying-time:rho1.5:throughput-seed50:timeline}}
\caption{Performance evaluation of HEE-ACC-zero and HEE-ALRN against the timeline, where different task groups have different offered traffic intensities.}
\label{fig:varying-time:seed50-rho1.5:timeline}
\end{minipage}
\end{figure*}

\begin{figure*}[t]
\centering
%\begin{minipage}[]{\textwidth}
%\centering
%\subfigure[]{\includegraphics[width=0.32\linewidth]{figs/scaler-rho5-seed50-power.eps}
%\label{fig:power-conservation-seed50:scaler}}
%\subfigure[]{\includegraphics[width=0.32\linewidth]{figs/scaler-rho5-seed50-throughput-power.eps}\label{fig:throughput-power-seed50:scaler}}
%\subfigure[]{\includegraphics[width=0.32\linewidth]{figs/scaler-rho5-seed50-throughput.eps}\label{fig:throughput-seed50:scaler}}
%\caption{Performance evaluation of HEE-ACC-zero and HEE-ALRN against the scaling parameter, where $\rho = 5$.}
%\label{fig:seed50-rho5:scaler}
%\end{minipage}
\begin{minipage}[]{\textwidth}
\centering
\subfigure[Power conservation]{\includegraphics[width=0.32\linewidth]{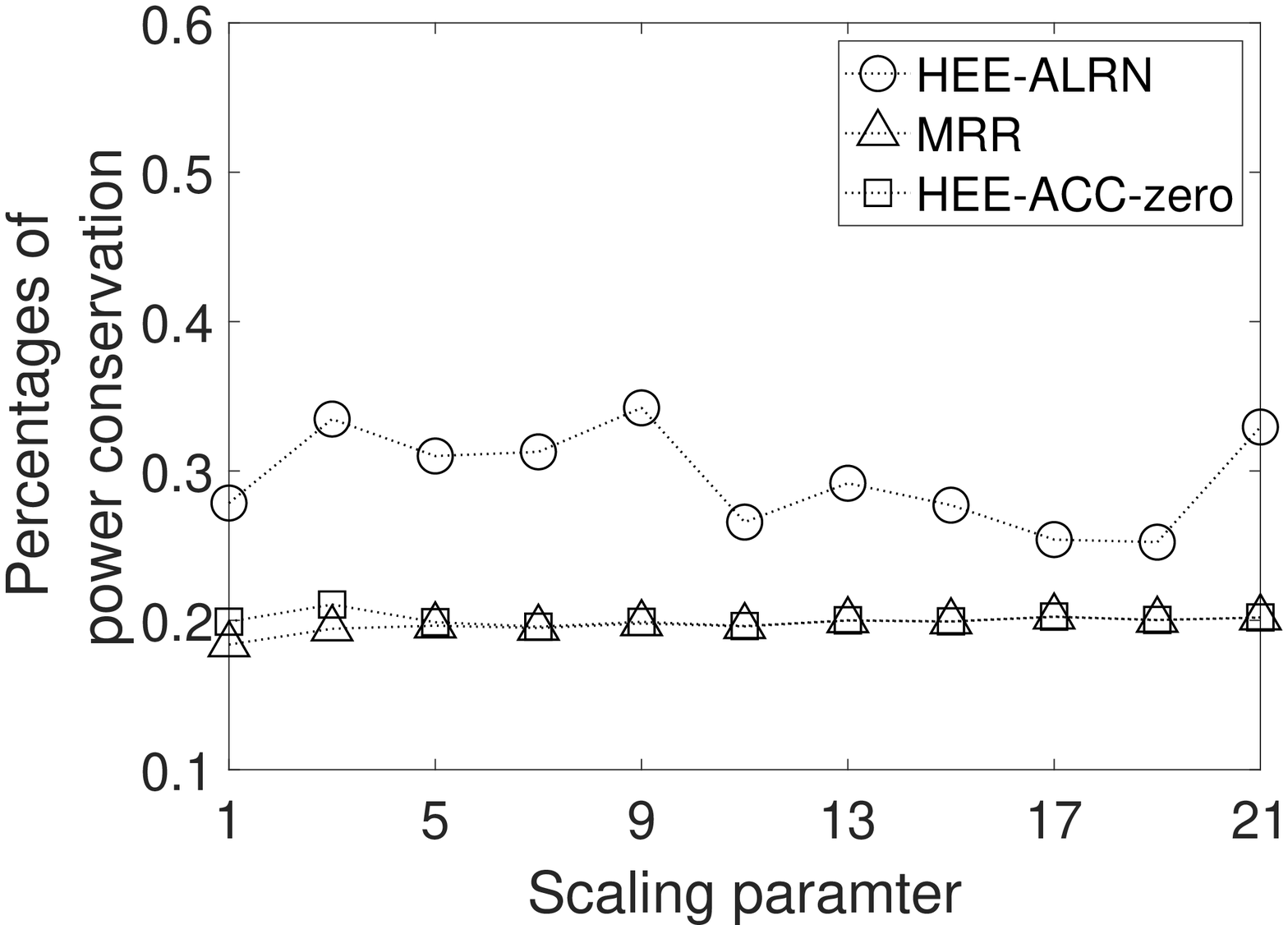}
\label{fig:power-conservation-seed50-rho7.5:scaler}}
%\end{minipage}
%\begin{minipage}[]{0.66\textwidth}
\subfigure[Throughput per unit power]{\includegraphics[width=0.32\linewidth]{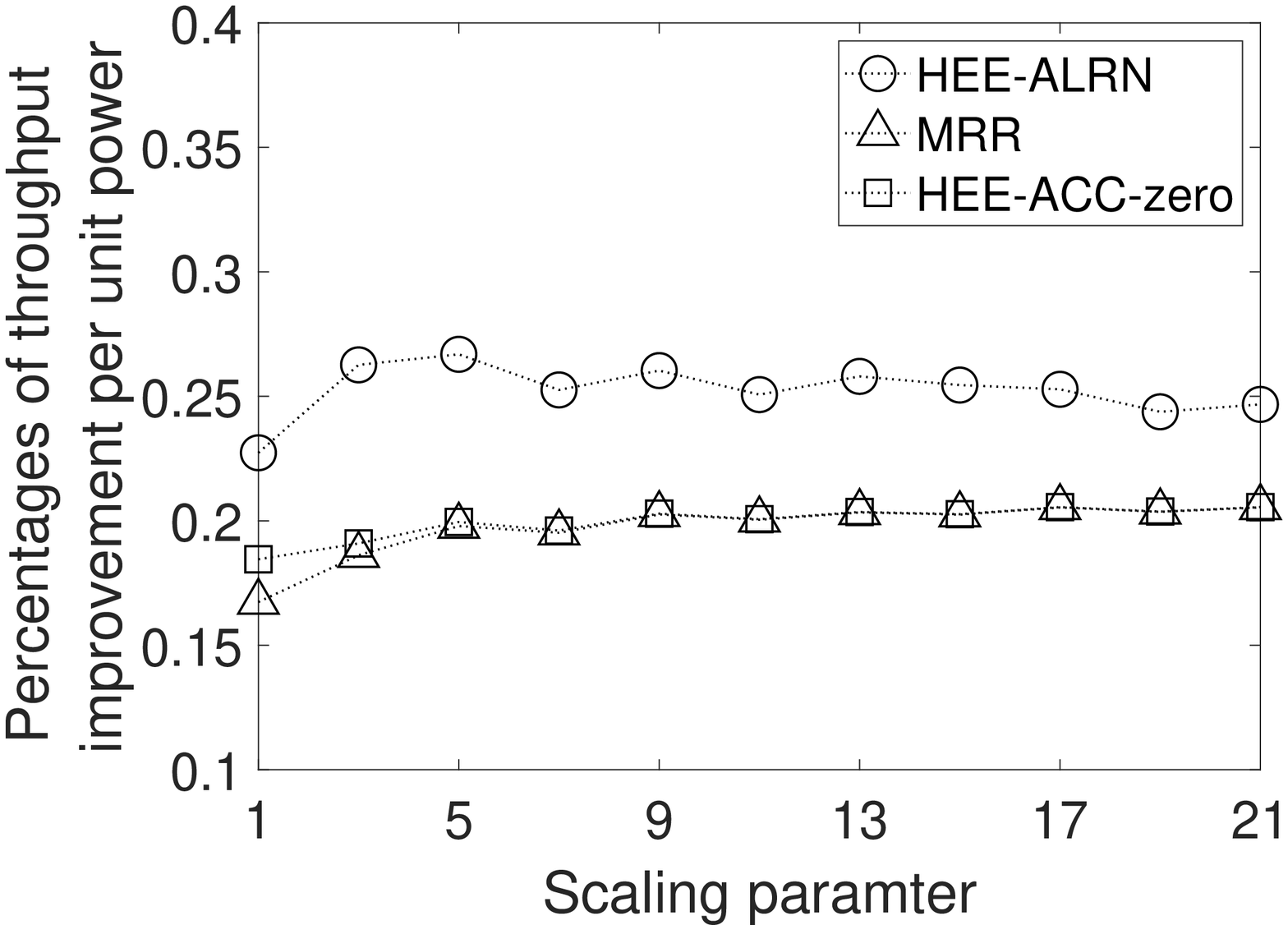}
\label{fig:throughput-power-seed50-rho7.5:scaler}}
\subfigure[Average delay]{\includegraphics[width=0.32\linewidth]{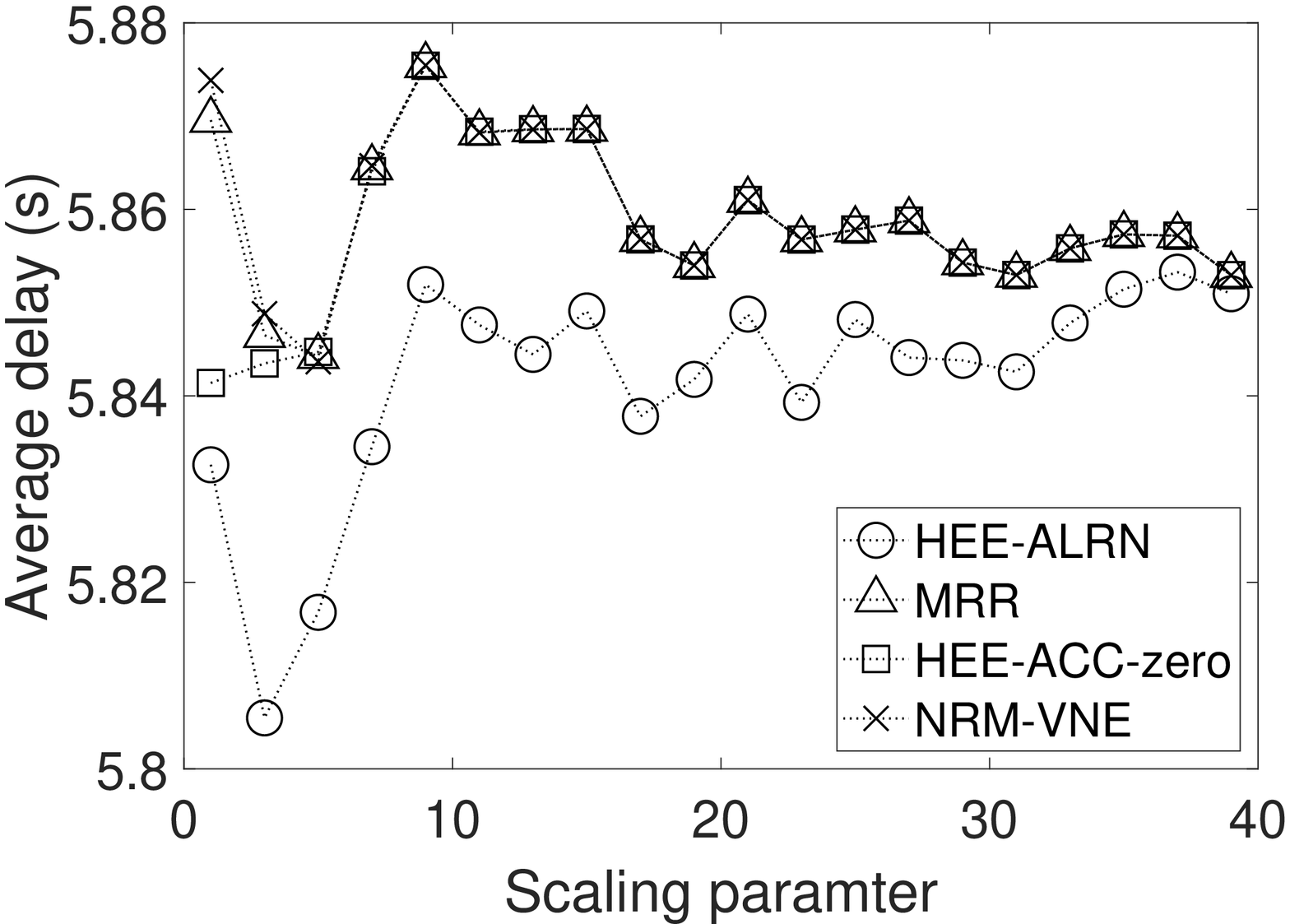}
\label{fig:delay-seed50-rho7.5:scaler}}
\caption{Performance evaluation of HEE-ACC-zero and HEE-ALRN against the scaling parameter, where $\rho = 7.5$.}
\label{fig:seed50-rho7.5:scaler}
\end{minipage}
\end{figure*}

\subsection{Simulation Setup}
We consider an MEC network with 20 different destination areas ($L=20$), three different groups of SCs ($K=3$), installed on the edge of the Internet, and infinitely many SC units located in the cloud. 
Each destination area includes a CN that serves $h\in\mathbb{N}_+$ wireless channels for various MTs, and is connected with the SCs in the edge network through wired connections. 
In this context, there are $I=20h$ wireless channels, each of which is potentially shared by multiple offloading requests with capacity $N_i$ ($i\in[I]$) listed in Appendix~\ref{app:simulation}.
We set the computing capacity of each SC group $C_k=C^0_k h$ ($k\in[K]$) where $C^0_k$ are positive integers randomly generated from $\{5,6,\ldots,10\}$.

There are four different classes of MTs ($J=4$) that keep sending tasks to the MEC system.
We set the arrival rates of the requests $\lambda_j$ ($j\in[J]$) to be $\lambda^0_j h$, where $\lambda^0_j$ is randomly generated from $[1,1.5]$.
Consider the expected lifespan of a $j$-task by setting $\mu_j(i,i',k) = \lambda_j/\rho$ with given parameter $\rho$. We will specify different $\rho$ in several different simulation scenarios. 
The specified $\lambda_j$ ($j\in[J]$) for the tested simulation runs are listed in Appendix~\ref{app:simulation}.
We refer to the parameter $h\in\mathbb{N}_+$ as the \emph{scaling parameter}, which implies the size of the MEC system and the scale of the offloading problem.
When $h$ becomes large, the studied MEC system is appropriate for an urban area with a highly dense population and compatibly many MEC servers or other computing components located near mobile users.

The wireless channels between different CNs and MTs may be interfered and become unstable due to distances, obstacles, geographical conditions, signal power, et cetera.
The starting and ending channels for each MT in class $j$ are selected from eligible candidates with $\mu_{i,j}>0$, where the movement of the MT determines the eligibility. 
We provide in Appendix~\ref{app:simulation} the sets of eligible candidates for the starting and ending channels of the MTs in each of the four classes, as well as the remaining parameter $w_{j,k}$ for all $j\in[J]$ and $k\in[K]$. In particular, if requests in class $j$ cannot be served by SCs in group $k$, we set $w_{j,k}\rightarrow +\infty$.
\subsection{Baseline Policies}
We numerically demonstrate the effectiveness of  HEE-ACC-zero and HEE-ALRN by comparing them with two baseline policies:
%The HEE-ACC-zero and HEE-ARLN policies proposed in this paper are compared to two baseline policies:
Maximum expected Revenue Rate (MRR) \cite{fu2020resource} and Node Ranking Measurement virtual network embedding (NRM-VNE) \cite{zhang2017virtual}.
MRR and NRM-VNE are both priority-style policies that always select the channel-SC tuple with the highest instantaneous \emph{revenue rate per unit requirement} and the highest product of the SC and the channels' remaining capacities, respectively.
We adapt MRR and NRM-VNE to our MEC system by considering the wireless channels and SC units as substrate network resources/physcial nodes used to support the incoming requests.
In particular, from \cite{fu2020resource}, MRR was demonstrated to be a promising policy that aims to maximize the long-run average revenue, which is equivalent to the minimization of the long-run average power consumption discussed in this paper. 
More precisely, for the MEC system in this paper, the instantaneous revenue per unit requirement of MRR for the tuple $(i,i',k)$ and $j$-tasks reduces to 
\begin{multline}
    -\frac{\lambda_jw_{j,k}\varepsilon_k}{(\lambda_j+\mu_j(i,i',k))(w_{j,k}+2)}\mathds{1}\{k<K+1\}\\-\frac{\lambda_j\mu_j(i,i',k)\bar{\varepsilon}_j}{(\lambda_j+\mu_j(i,i',k)(w_{j,k}+2)}\mathds{1}\{k=K+1\}.
\end{multline}
The NRM-VNE was proposed in \cite{zhang2017virtual} to avoid traffic congestion and aimed to achieve a maximized throughput.

In Sections~\ref{subsec:exp} and \ref{subsec:heavy-tail}, we explore scenarios with exponentially and non-exponentially, respectively, distributed task lifespans.
In Section~\ref{subsec:exp}, we further consider time-varying real-world workloads by incorporating Google cluster traces~\cite{clusterdata:Wilkes2011,clusterdata:Reiss2011} in our simulations.

The $95\%$ confidence intervals of all the simulated results in this section, based on the Student t-distribution, are within $\pm3\%$ of the observed mean.

\subsection{Exponentially Distributed Lifespans}\label{subsec:exp}
\begin{figure*}[t]
\centering
\begin{minipage}[]{\textwidth}
\centering
\subfigure[Power conservation]{\includegraphics[width=0.32\linewidth]{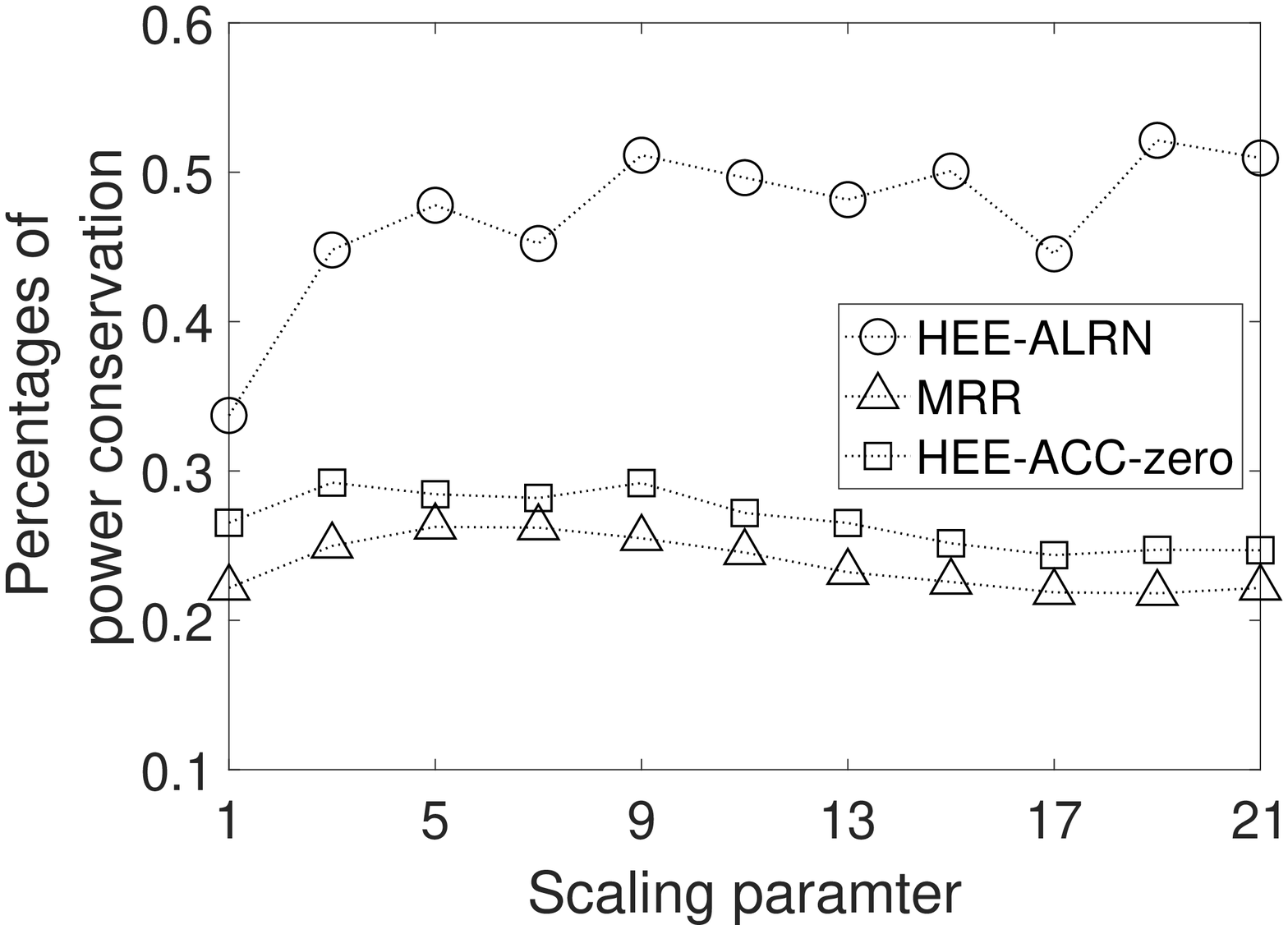}
\label{fig:power-conservation-seed50-rho10:scaler}}
%\end{minipage}
%\begin{minipage}[]{0.66\textwidth}
\subfigure[Throughput per unit power]{\includegraphics[width=0.32\linewidth]{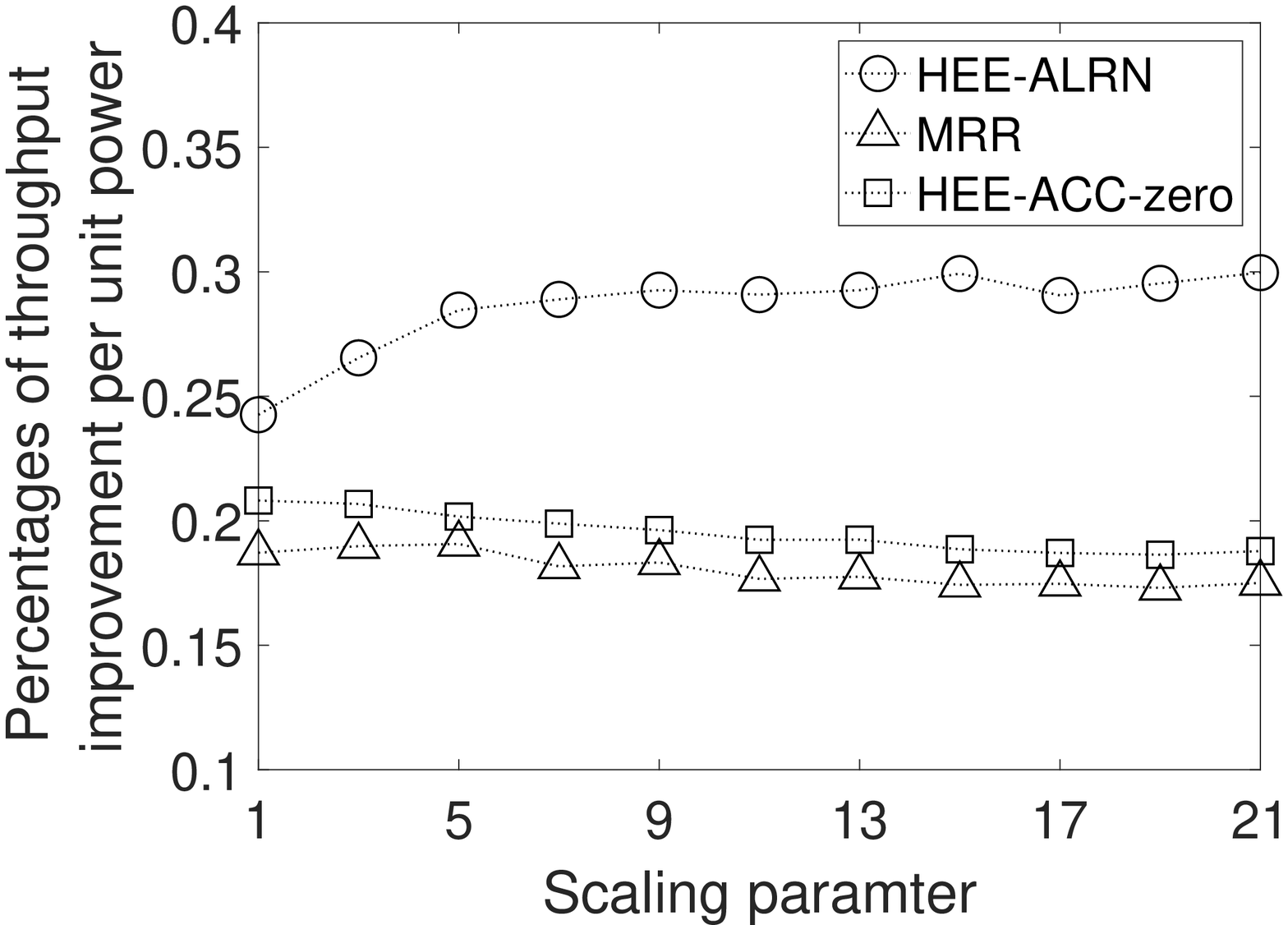}\label{fig:throughput-power-seed50-rho10:scaler}\label{fig:throughput-seed50-rho10:scaler}}
\subfigure[Average delay]{\includegraphics[width=0.32\linewidth]{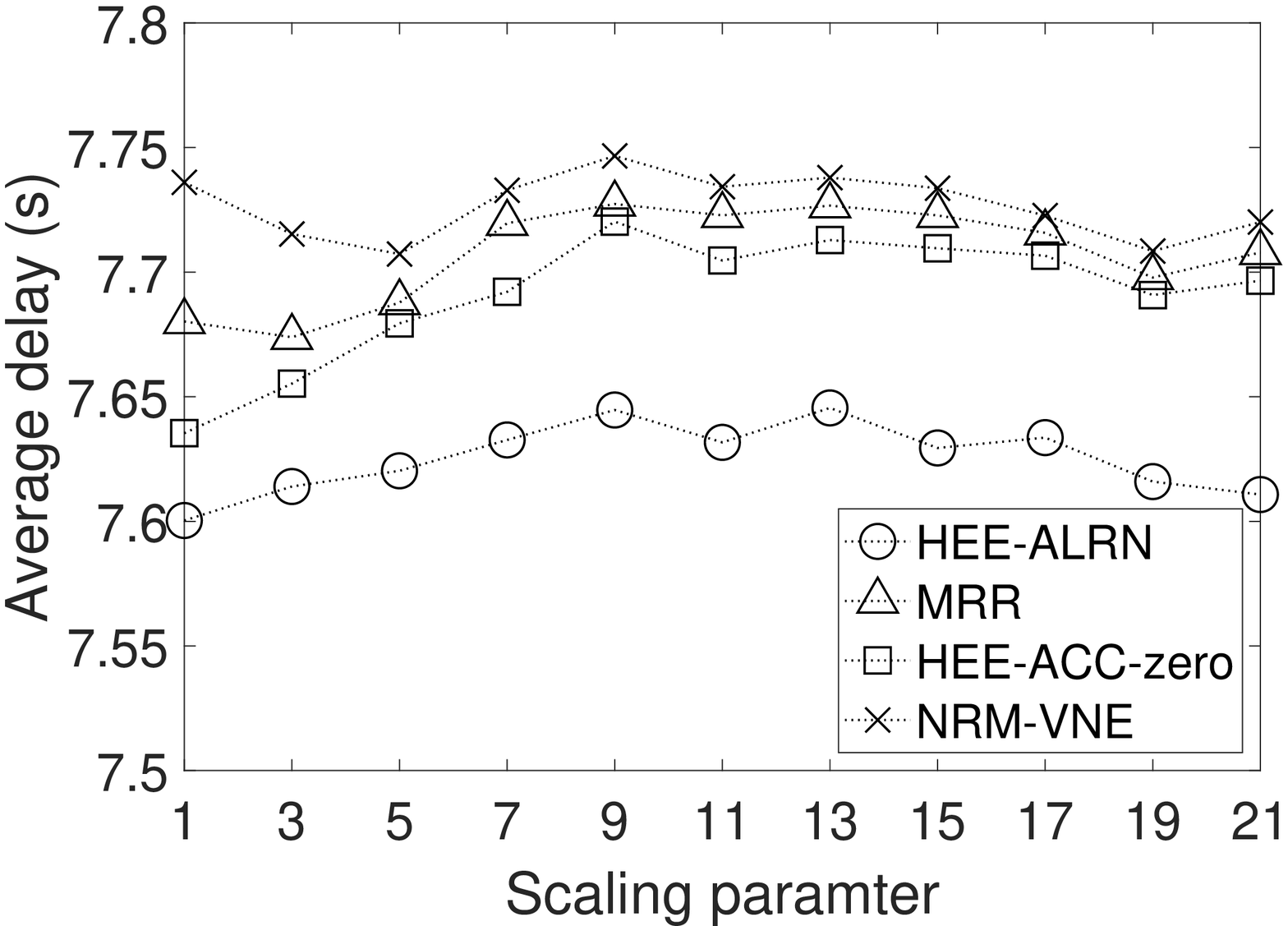}\label{fig:delay-seed50-rho10:scaler}}
\caption{Performance evaluation of HEE-ACC-zero and HEE-ALRN against the scaling parameter, where $\rho = 10$.}
\label{fig:seed50-rho10:scaler}
\end{minipage}
\end{figure*}

\begin{figure*}[t]
\centering
\subfigure[Power conservation]{\includegraphics[width=0.32\linewidth]{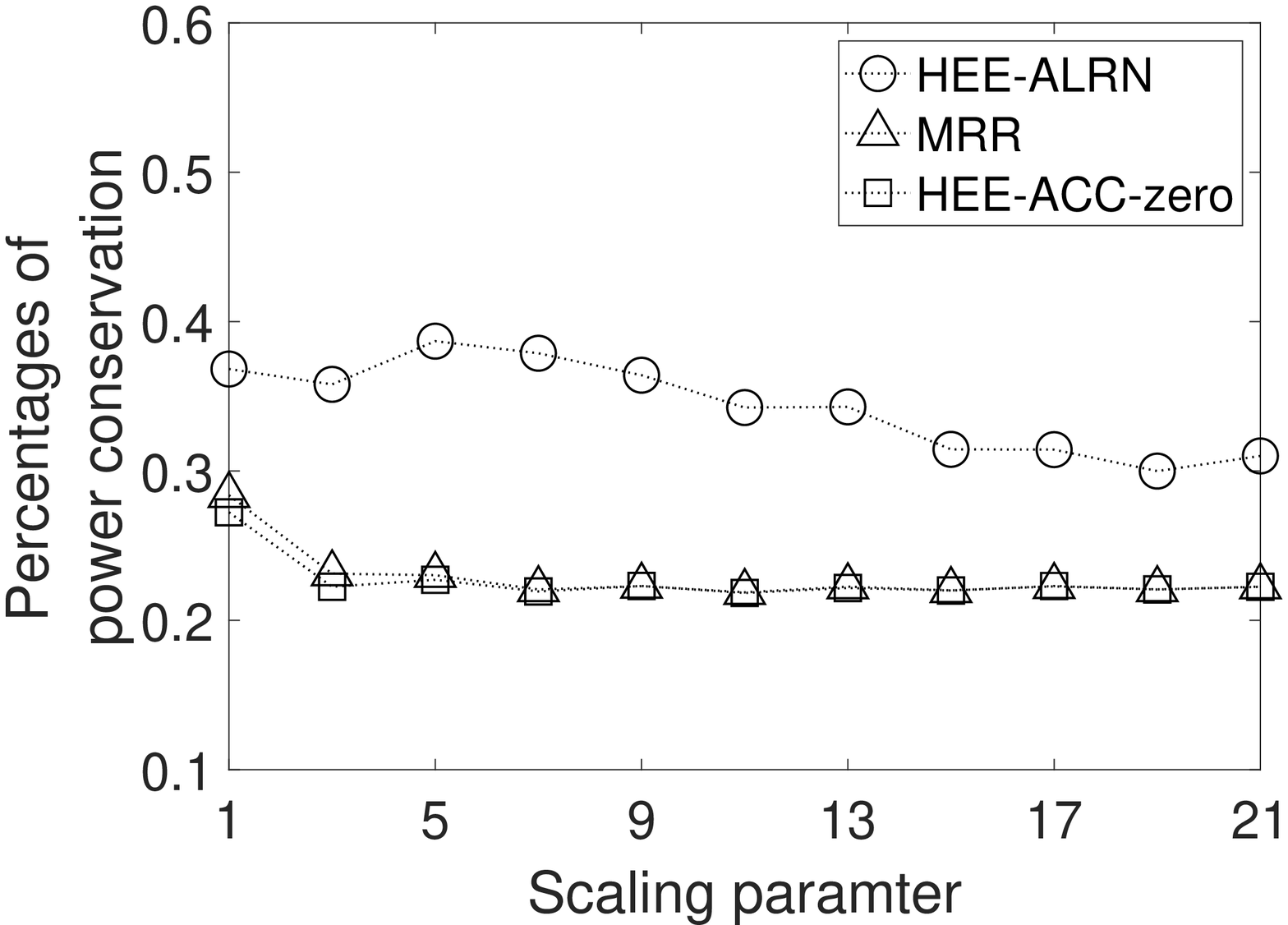}
\label{fig:power-conservation-seed50:scaler}}
%\end{minipage}
%\begin{minipage}[]{0.66\textwidth}
\subfigure[Throughput per unit power]{\includegraphics[width=0.32\linewidth]{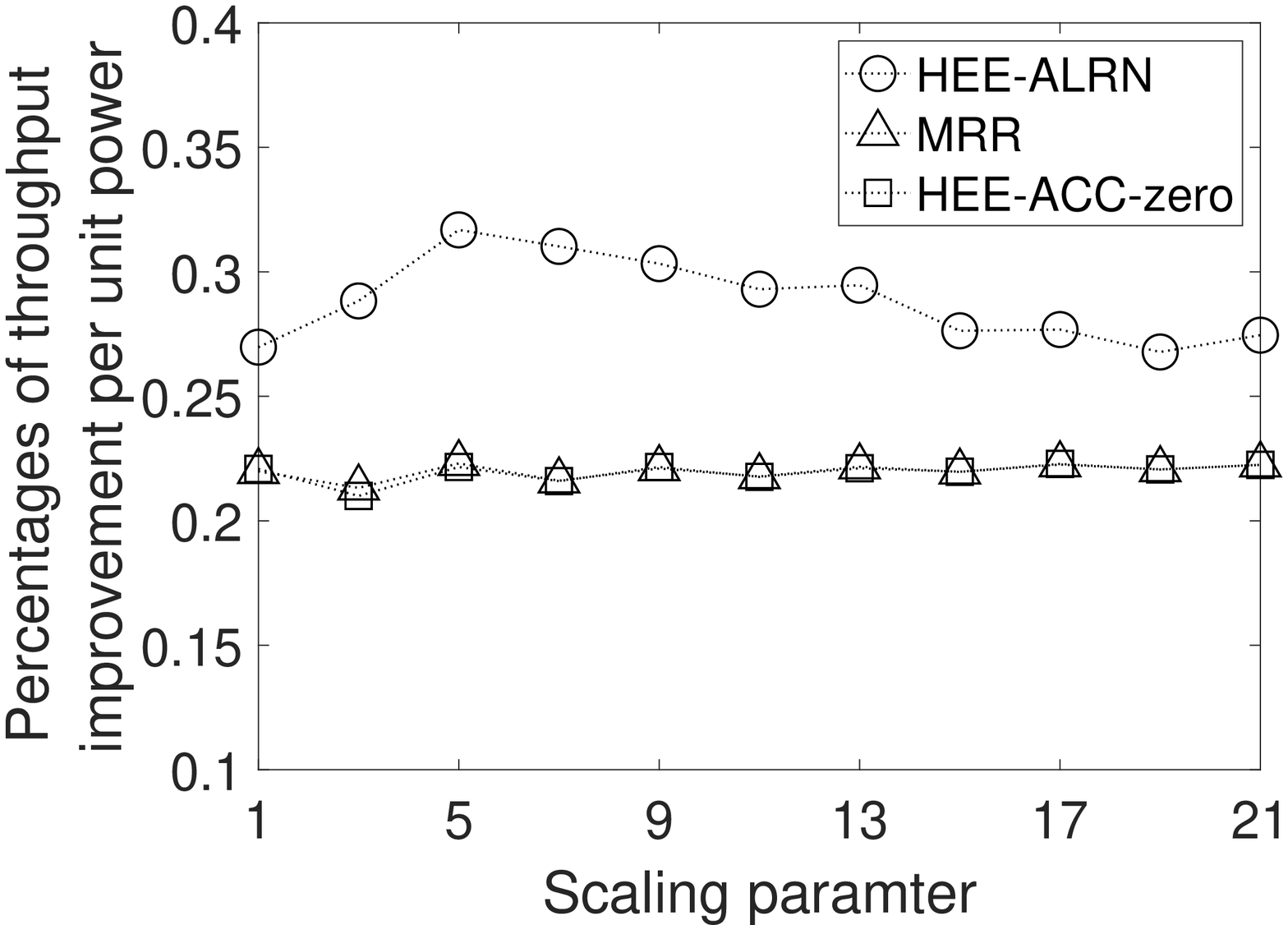}
\label{fig:throughput-power-seed50-rho1.5:scaler}}
\subfigure[Average delay]{\includegraphics[width=0.32\linewidth]{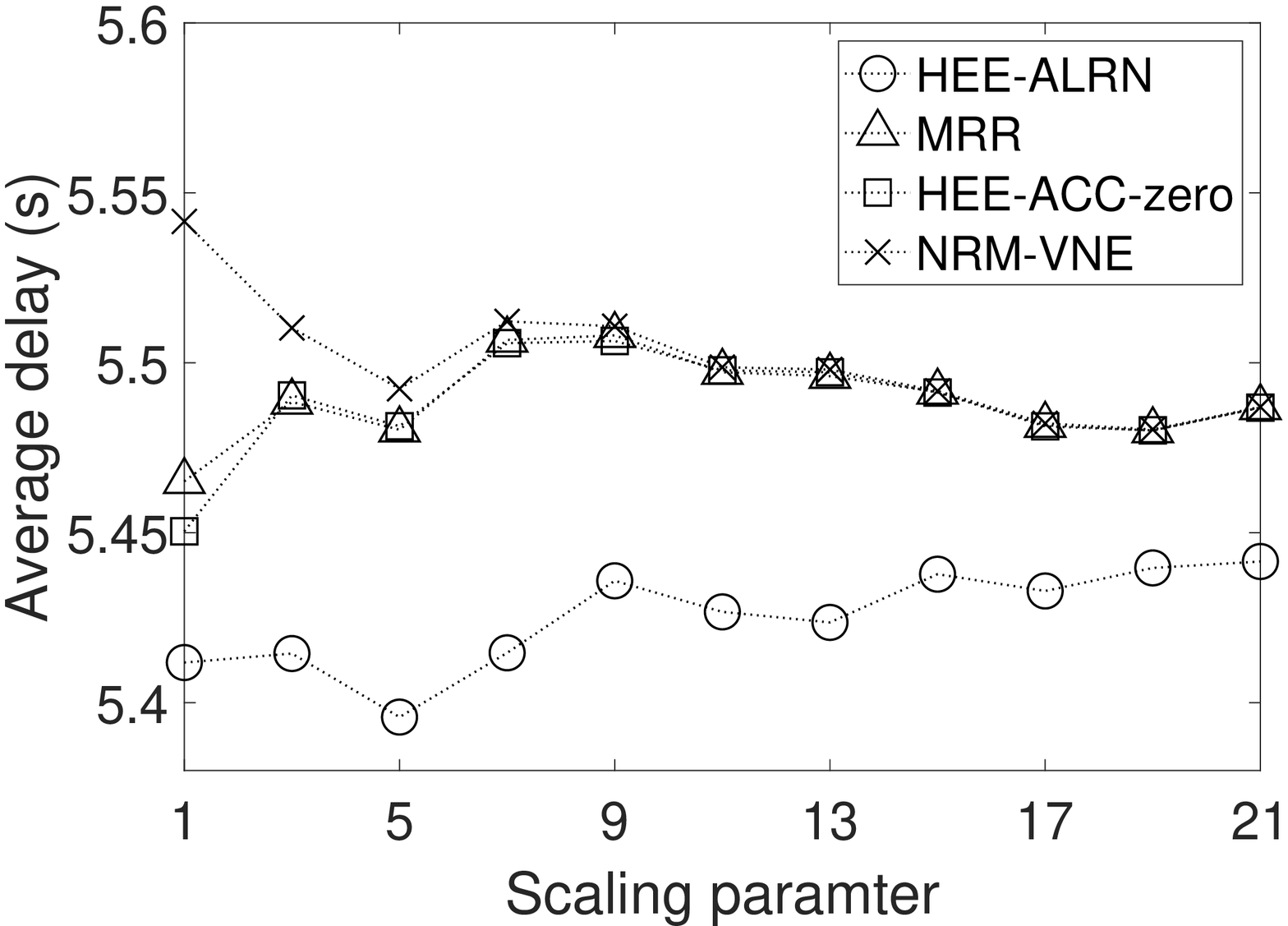}\label{fig:delay-seed50-rho1.5:scaler}}
\caption{Performance evaluation of HEE-ACC-zero and HEE-ALRN against the scaling parameter, where different task groups have different offered traffic intensities.}
\label{fig:seed50-varying-rho1.5:scaler}
%\end{figure*}

%\begin{figure*}[t]
%\centering
\subfigure[Absolute power conservation]{\includegraphics[width=0.32\linewidth]{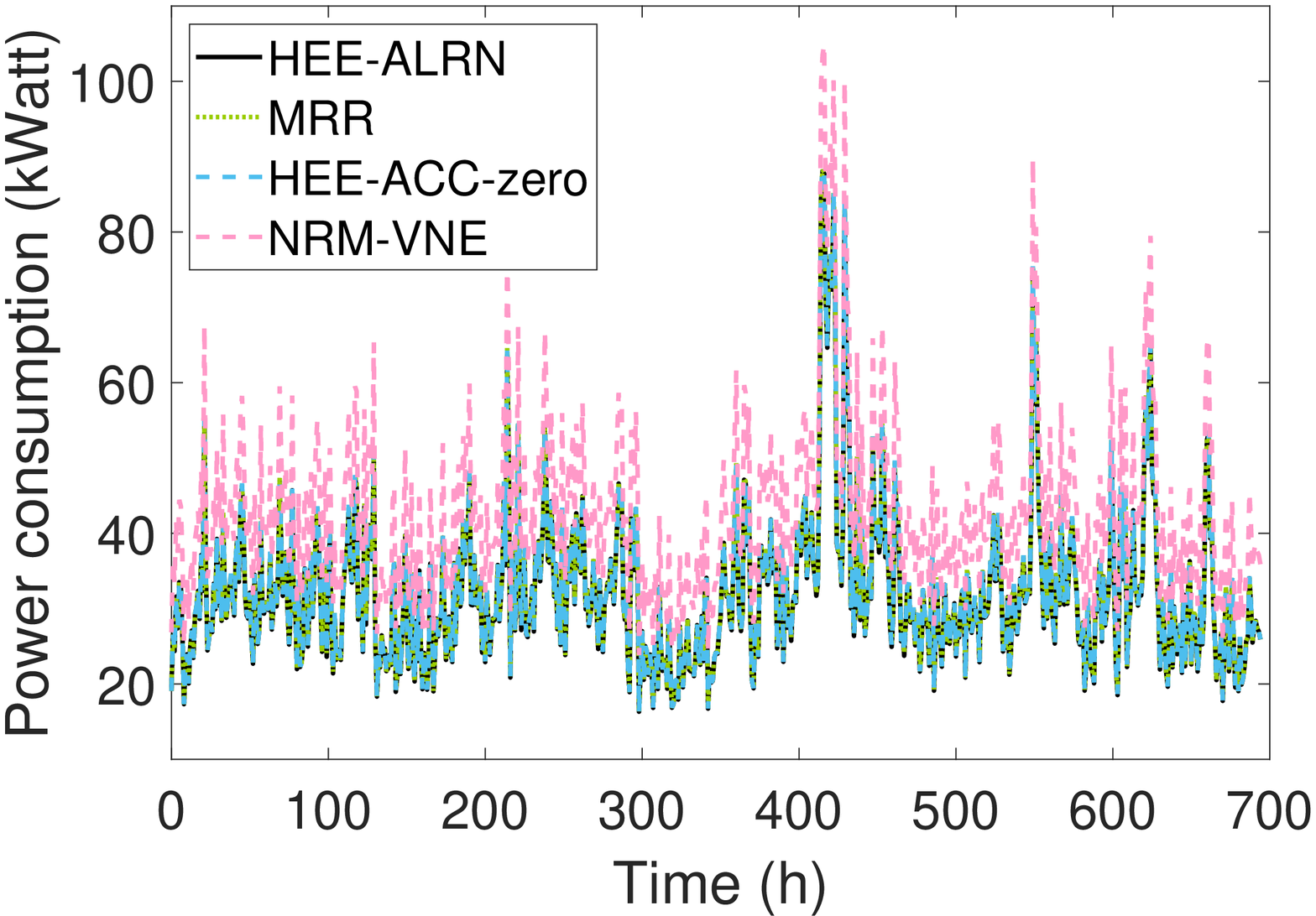}
\label{fig:google-pwer}}
%\end{minipage}
%\begin{minipage}[]{0.66\textwidth}
\subfigure[Relative power conservation]{\includegraphics[width=0.32\linewidth]{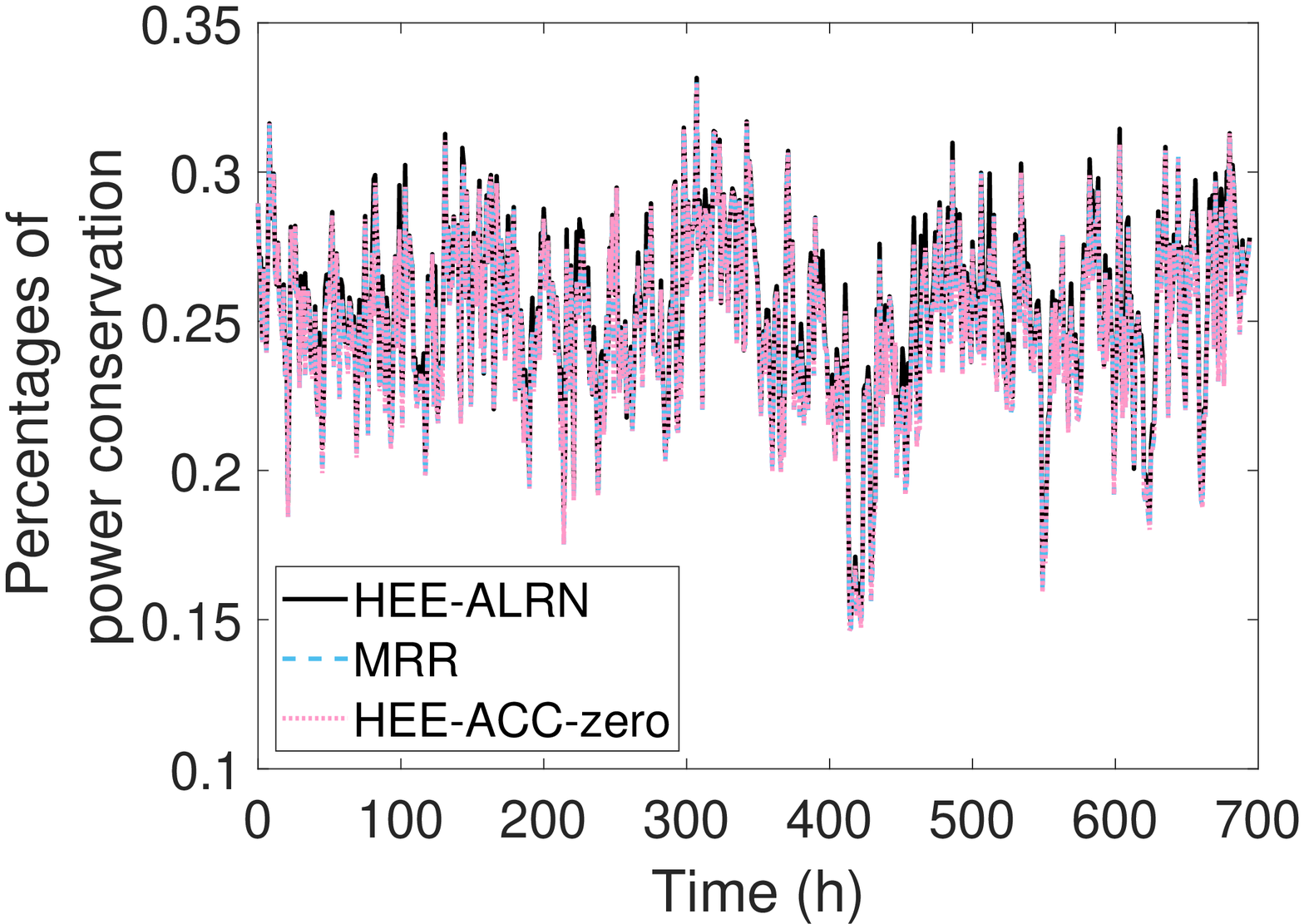}\label{fig:google-pwer-percentage}}
\subfigure[Average delay]{\includegraphics[width=0.32\linewidth]{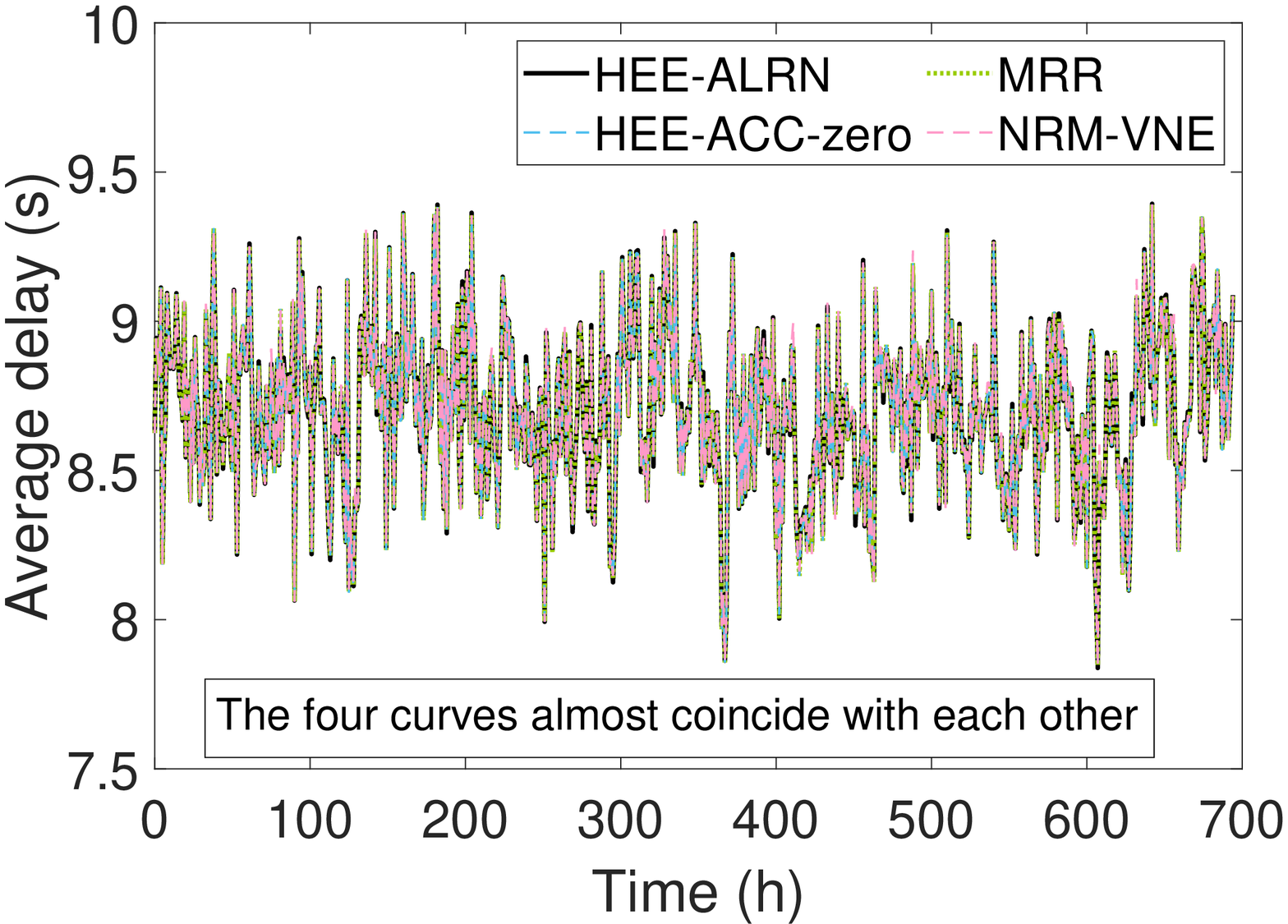}\label{fig:google-delay}}
\caption{Performance evaluation of HEE-ACC-zero and HEE-ALRN with Google traces.}
\label{fig:google}
\end{figure*}

In Figs.~\ref{fig:seed50-rho7.5:timeline} and~\ref{fig:seed50-rho10:timeline}, 
we evaluate the performance of HEE-ACC-zero and HEE-ALRN with respect to power conservation, throughput per unit power, and the average delay of the offloaded tasks.
We examine $\rho=7.5,10$ in the two figures, representing cases with relatively light and heavy traffic, respectively.

Define $\mathcal{E}^{\phi}$ and $\mathcal{T}^{\phi}$ as the long-run average operational power consumption and the long-run average throughput of the edge system, respectively, under policy $\phi$.
In this section, we consider the percentage of power conservation of a policy $\phi$ as the relative difference between $\mathcal{E}^{\text{NRM-VNE}}$ and $\mathcal{E}^{\phi}$; that is,
\begin{equation}
    \frac{\mathcal{E}^{\text{NRM-VNE}}-\mathcal{E}^{\phi}}{\mathcal{E}^{\text{NRM-VNE}}}.
\end{equation}
Similarly, consider the percentage of throughput improvement per unit power of policy $\phi$ as the relative difference between  $\mathcal{T}^{\phi}/\mathcal{E}^{\phi}$ and $\mathcal{T}^{\text{NRM-VNE}}/\mathcal{E}^{\text{NRM-VNE}}$, given by
\begin{equation}
    \frac{\mathcal{T}^{\phi}/\mathcal{E}^{\phi}-\mathcal{T}^{\text{NRM-VNE}}/\mathcal{E}^{\text{NRM-VNE}}}{\mathcal{T}^{\text{NRM-VNE}}/\mathcal{E}^{\text{NRM-VNE}}}.
\end{equation}

In \figurename~\ref{fig:seed50-rho7.5:timeline}, HEE-ALRN, HEE-ACC-zero, and MRR achieve over $15\%$ conservation against NRM-VNE with respect to operational power, while the average delay of the four policies are compatible. 
In this case, with relatively light traffic, HEE-ALRN clearly outperforms HEE-ACC-zero and MRR with respect to power consumption.
When the traffic becomes heavier, in \figurename~\ref{fig:seed50-rho10:timeline}, HEE-ALRN achieves even higher energy conservation compared to HEE-ACC-zero, MRR and NRM-VNE, while keeps compatible average {delay} with the other three policies.

In particular, in Figs.~\ref{fig:rho7.5:throughput-power-seed50:timeline} and~\ref{fig:rho10:throughput-power-seed50:timeline}, we plot the throughput improvement per unit power of HEE-ALRN, MRR and HEE-ACC-zero against NRM-VNE.
Observing these figures, HEE-ALRN, HEE-ACC-zero and MRR all achieve substantially higher throughput per unit power than that of NRM-VNE.
%That is, there is no throughput degradation per unit power but throughput improvement per unit power when HEE-ALRN, HEE-ACC-zero or MRR is employed. 
It is caused by the significant power conservation but negligible throughput degradation for the three policies.
Similarly, in \figurename~\ref{fig:rho10:throughput-power-seed50:timeline} with heavy offered traffic, HEE-ALRN clearly outperforms all the other policies.
It is consistent with our expectation since HEE-ALRN dynamically adjusts the capacity coefficients in a closed-loop manner with feedback from the system, while the other policies consider only offline-determined action variable for each state.

In \figurename~\ref{fig:varying-time:seed50-rho1.5:timeline}, we illustrate the performance of the above-mentioned four policies, where different task groups $j\in[J]$ have different offered traffic intensities, $\rho_j\in \mathbb{R}_+$ and we set $\mu_j(i,i',k)=\lambda_j/\rho_j$. The values of $\rho_j$ for $j\in[J]$ are provided in Appendix~\ref{app:simulation}.
The other system parameters for the simulations presented in \figurename~\ref{fig:varying-time:seed50-rho1.5:timeline} are the same as those for the simulations in \figurename~\ref{fig:seed50-rho7.5:timeline}.
In \figurename~\ref{fig:varying-time:seed50-rho1.5:timeline}, HEE-ALRN remains its clear advantages against all the other policies while achieves the least average delay and compatible throughput per unit power.
The observation is consistent with that in previous figures and our arguments in Section~\ref{sec:policy}.

In Figs.~\ref{fig:seed50-rho7.5:scaler} and~\ref{fig:seed50-rho10:scaler}, we examined the performance of the four policies against the scaling parameter $h$. Recall that the scaling parameter $h$ measures the size of the optimization problem. 
A large $h$ is appropriate for an urban area with highly dense mobile users and many compatible communication and computing capacities, which is the primary concern of this paper.
Similar to Figs.~\ref{fig:seed50-rho7.5:timeline} and~\ref{fig:seed50-rho10:timeline}, HEE-ALRN, HEE-ACC-zero and MRR in general outperforms NRM-VNE in all the tested cases with respect to power consumption and throughput per unit power.
The substantially higher throughput per unit power for HEE-ALRN, HEE-ACC-zero and MRR is guaranteed by their substantial power conservation with negligible degradation in average throughput.
In particular, in Figs.~\ref{fig:delay-seed50-rho7.5:scaler} and~\ref{fig:delay-seed50-rho10:scaler}, we tested the average delay of the offloaded tasks, for which HEE-ALRN achieves the least delay, and the others are slightly higher and similar to each other.
Among these three policies, in Figs.~\ref{fig:seed50-rho7.5:scaler} and~\ref{fig:seed50-rho10:scaler}, HEE-ALRN always outperforms HEE-ACC-zero and MRR with respect to power consumption and average delay, which is consistent with our observations and arguments in the previous paragraphs.

In \figurename~\ref{fig:seed50-varying-rho1.5:scaler}, we consider the performance of the above-mentioned four policies against the scaling parameter $h$ with different $\rho_j$ for each $j\in[J]$. The values of $\rho_j$ for $j\in[J]$ are provided in Appendix~\ref{app:simulation}, and the other system parameters are the same as those for the simulations in \figurename~\ref{fig:seed50-rho7.5:scaler}.
In \figurename~\ref{fig:seed50-varying-rho1.5:scaler}, HEE-ALRN significantly outperforms all the others with respect to power consumption, job throughput and the average delay, which is consistent with our observations and arguments in previous figures and paragraphs.
HEE-ACC-zero achieves some merits in power consumption and has similar job throughput and average delay, compared to NRM-VNE, and it has close performance as MRR.

Apart from the time-invariant arrival rates discussed earlier in this section, in \figurename~\ref{fig:google}, we further consider time-varying arrival rates of tasks.
We incorporate Google cluster traces~\cite{clusterdata:Wilkes2011,clusterdata:Reiss2011} in the same MEC system with $\rho_j=\rho=10$ and $h=1$. 
In Figs.~\ref{fig:google-pwer} and~\ref{fig:google-pwer-percentage}, we present the absolute and relative, respectively, power consumption averaged per hour under the four policies.
Recall that in \figurename~\ref{fig:google-pwer-percentage}, the percentages of power conservation is the relative difference between the dedicated policy and NRM-VNE with respect to power consumption averaged per hour.
As demonstrated in Figs.~\ref{fig:google-pwer} and~\ref{fig:google-pwer-percentage}, HEE-ALRN, HEE-ACC-zero, and MRR still achieves significantly lower power consumption, over $15\%$ lower in all the tested time slots (hours), than that of NRM-VNE (the pink dashes in \figurename~\ref{fig:google-pwer}).
In \figurename~\ref{fig:google-pwer-percentage}, HEE-ALRN saves slightly more power consumption than that of MRR and HEE-ACC-zero, for which the curves almost coincide with each other.
In \figurename~\ref{fig:google-delay}, for the same case, we further test the average delay of the tasks offloading to the edge system, where the four policies have almost the same average delay for each time slot (hour).
\begin{figure}[t]
\centering
\subfigure[HEE-ACC-zero]{\includegraphics[width=0.49\linewidth]{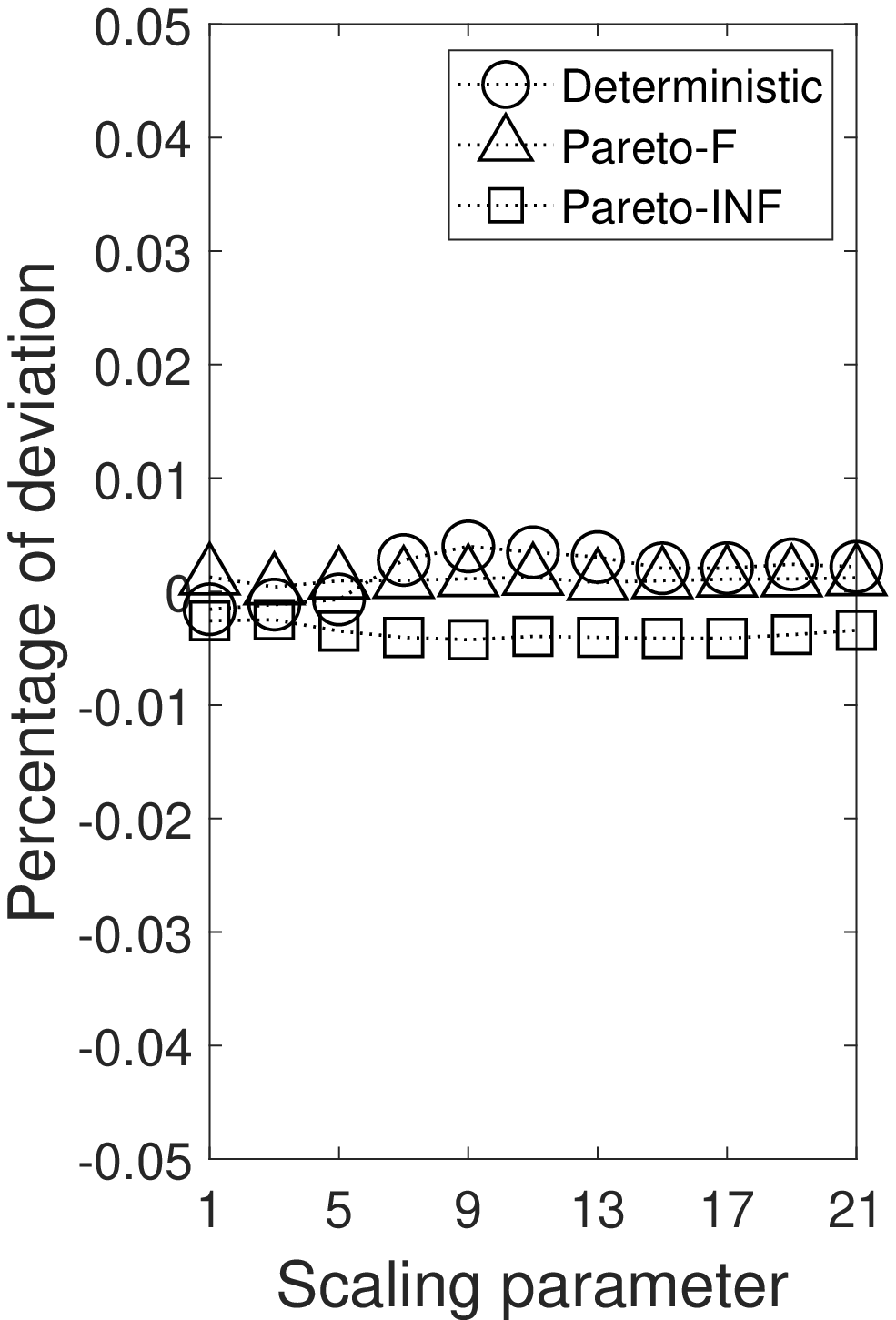}\label{fig:sensitivity-zero}}
\subfigure[HEE-ALRN]{\includegraphics[width=0.49\linewidth]{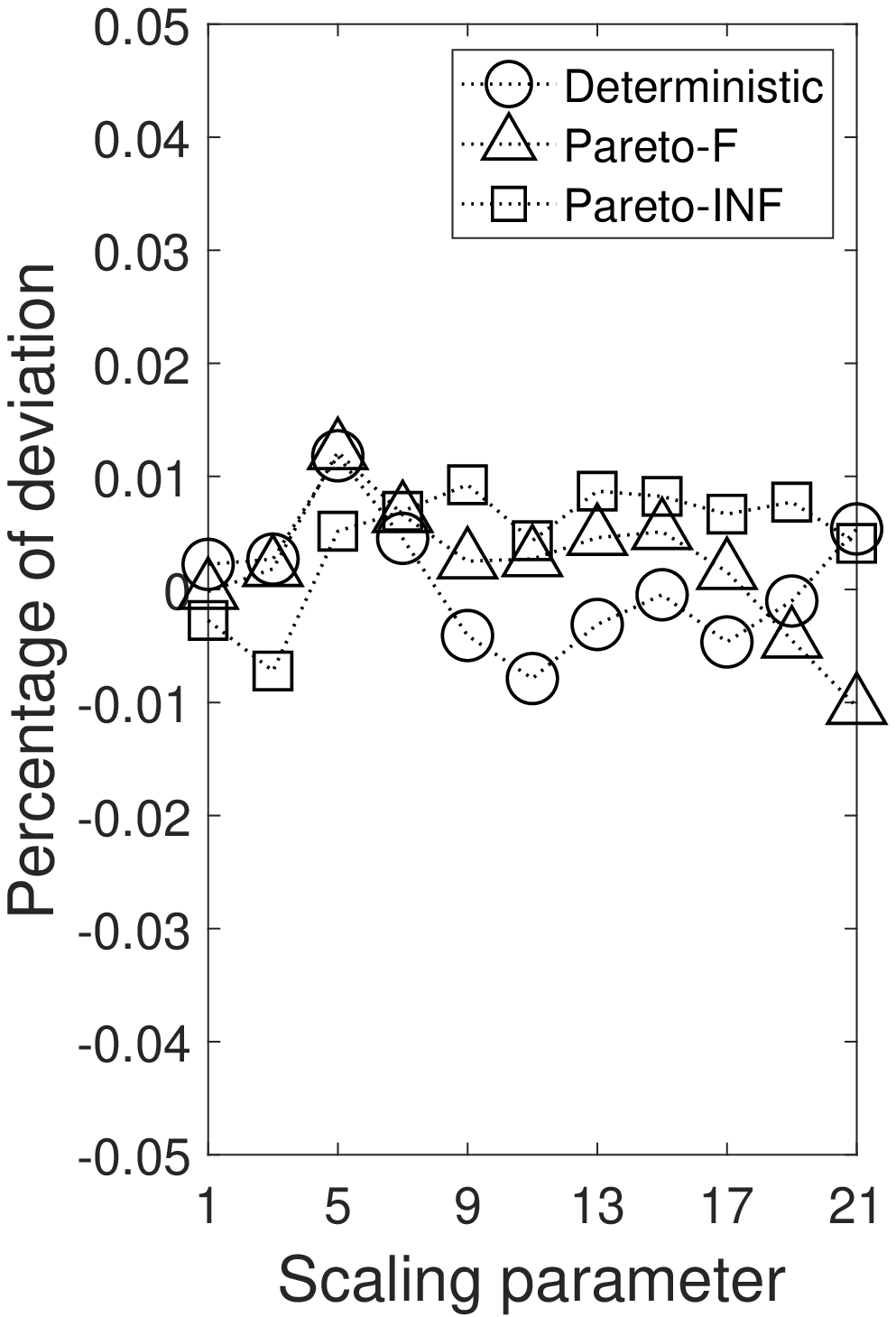}\label{fig:sensitivity-alrn}}
\caption{Deviations of average operational power consumption with non-exponentially distributed task lifespans under HEE-ACC-zero and HEE-ALRN.}
\end{figure}

\subsection{Non-exponential Lifespans}\label{subsec:heavy-tail}

For a large-scale MEC system, the system performance is expected to be robust against different distributions of task lifespans, because, even if some sub-channels and SC units are busy to serve excessively long tasks, the subsequently arrived tasks are still likely to be supported by compatibly many idle sub-channels and SC units. 
The long-run average performance of the entire system is unlikely to be significantly affected by different shapes of the lifespan distributions, such as the heavy-tailed distributions \cite{Crovella1997self,harchol2013performance}.

In Figs.~\ref{fig:sensitivity-zero} and~\ref{fig:sensitivity-alrn}, we evaluate the effects of non-exponentially distributed task lifespans for HEE-ACC-zero and HEE-ALRN, respectively.
Similar to the settings in \cite{fu2020energy}, we simulated four different lifespan distributions for the requests: deterministic, exponential, Pareto distribution with finite variance (Pareto-F), and Pareto distribution with infinite variance (Pareto-INF).
Pareto-F and Pareto-INF are constructed by setting the shape parameter of the Pareto distribution to 2.001 and 1.98, respectively.
Other simulation settings are the same as those for Figs.~\ref{fig:seed50-rho7.5:timeline} and~\ref{fig:seed50-rho7.5:scaler}.

In Figs.~\ref{fig:sensitivity-zero} and~\ref{fig:sensitivity-alrn}, the y-axis stands for the relative deviation between the power consumption for a specified lifespan distribution and that for the exponentially distributed lifespan. 
That is, let $\mathcal{E}^{\phi,D} $ represent the average power consumption under the policy $\phi$ and the lifespan distribution $D$, the y-axis of  Figs.~\ref{fig:sensitivity-zero} and~\ref{fig:sensitivity-alrn} represents
\begin{equation}
\frac{\mathcal{E}^{\phi,D}-\mathcal{E}^{\phi,\text{Exponential}}}{\mathcal{E}^{\phi,\text{Exponential}}},
\end{equation}
for the policies $\phi=$HEE-ACC-zero and HEE-ALRN, respectively, with specified distributions $D=$Deterministic, Pareto-F, and Pareto-INF.
In the figures, the deviations are all the time within $\pm 2\%$, which is negligible given that the confidence intervals of the simulations are considered to be $\pm 3\%$ of the means.
It is consistent with the argument stated at the beginning of this subsection.

\section{Conclusions}\label{sec:conclusion}
We study the energy efficiency of a large-scale MEC offloading problem with task handover. As mentioned earlier in the paper, the complexity and the large size of the problem prevent conventional optimization techniques from being directly applied. We adapt the restless bandit technique to the MEC offloading problem and propose a class of online strategies, the HEE-ACC policies, that are applicable to realistically scaled MEC systems. If appropriate capacity coefficients are provided, the HEE-ACC strategies achieve proven asymptotic optimality – an asymptotically optimal strategy approaches optimality as the scale of the MEC system tends to infinity. This implies that the proposed strategies are likely to be near-optimal in practical scenarios with highly dense mobile users and compatibly many communication channels and computing components. 
Within the HEE-ACC class, we propose two specific strategies: HEE-ACC-zero and HEE-ALRN. The former is neat with proved asymptotic optimality in special cases and the latter is expected to achieve a higher performance by dynamically learning the most appropriate capacity coefficients.
In Section~\ref{sec:simulation}, we validate the effectiveness of HEE-ACC-zero and HEE-ALRN by comparing them with two baseline policies. It is demonstrated that HEE-ALRN outperforms HEE-ACC-zero with respect to power conservation, which is consistent with our discussion in Section~\ref{sec:policy}. We further demonstrate the robustness of the two policies, through numerical simulations, against different lifespan distributions of the proposed policies.

\bibliographystyle{IEEEtran}
\bibliography{IEEEabrv,paperbib}

\begin{thebibliography}{10}
\providecommand{\url}[1]{#1}
\csname url@samestyle\endcsname
\providecommand{\newblock}{\relax}
\providecommand{\bibinfo}[2]{#2}
\providecommand{\BIBentrySTDinterwordspacing}{\spaceskip=0pt\relax}
\providecommand{\BIBentryALTinterwordstretchfactor}{4}
\providecommand{\BIBentryALTinterwordspacing}{\spaceskip=\fontdimen2\font plus
\BIBentryALTinterwordstretchfactor\fontdimen3\font minus
  \fontdimen4\font\relax}
\providecommand{\BIBforeignlanguage}[2]{{%
\expandafter\ifx\csname l@#1\endcsname\relax
\typeout{** WARNING: IEEEtran.bst: No hyphenation pattern has been}%
\typeout{** loaded for the language `#1'. Using the pattern for}%
\typeout{** the default language instead.}%
\else
\language=\csname l@#1\endcsname
\fi
#2}}
\providecommand{\BIBdecl}{\relax}
\BIBdecl

\bibitem{Taleb2017Multi}
T.~Taleb, K.~Samdanis, B.~Mada, H.~Flinck, S.~Dutta, and D.~Sabella, ``On
  {Multi}-access edge computing: A survey of the emerging {5G} network edge
  cloud architecture and orchestration,'' \emph{IEEE Communications Surveys and
  Tutorials}, vol.~19, no.~3, pp. 1657--1681, 2017.

\bibitem{Ho2022joint}
T.~M. Ho and K.-K. Nguyen, ``Joint server selection, cooperative offloading and
  handover in multi-access edge computing wireless network: A deep
  reinforcement learning approach,'' \emph{IEEE Transactions on Mobile
  Computing}, vol.~21, no.~7, pp. 2421--2435, 2022.

\bibitem{Deng2023task}
T.~Deng, Y.~Chen, G.~Chen, M.~Yang, and L.~Du, ``Task offloading based on edge
  collaboration in {MEC}-enabled {IoV} networks,'' \emph{Journal of
  Communications and Networks}, vol.~25, no.~2, pp. 197--207, 2023.

\bibitem{Maleki2023handover}
H.~Maleki, M.~Başaran, and L.~Durak-Ata, ``Handover-enabled dynamic
  computation offloading for vehicular edge computing networks,'' \emph{IEEE
  Transactions on Vehicular Technology}, vol.~72, no.~7, pp. 9394--9405, 2023.

\bibitem{Monir2022seamless}
N.~Monir, M.~M. Toraya, A.~Vladyko, A.~Muthanna, M.~A. Torad, F.~E.~A.
  El-Samie, and A.~A. Ateya, ``Seamless handover scheme for {MEC/SDN}-based
  vehicular networks,'' \emph{Journal of Sensor and Actuator Networks},
  vol.~11, no.~1, 2022.

\bibitem{Shu2023joint}
W.~Shu and Y.~Li, ``Joint offloading strategy based on quantum particle swarm
  optimization for {MEC}-enabled vehicular networks,'' \emph{Digital
  Communications and Networks}, vol.~9, no.~1, pp. 56--66, 2023.

\bibitem{Li020deep}
M.~Li, J.~Gao, L.~Zhao, and X.~Shen, ``Deep reinforcement learning for
  collaborative edge computing in vehicular networks,'' \emph{IEEE Transactions
  on Cognitive Communications and Networking}, vol.~6, no.~4, pp. 1122--1135,
  2020.

\bibitem{Davoli2022flow}
F.~Davoli, M.~Marchese, and F.~Patrone, ``Flow assignment and processing on a
  distributed edge computing platform,'' \emph{IEEE Transactions on Vehicular
  Technology}, vol.~71, no.~8, pp. 8783--8795, 2022.

\bibitem{Wang2020edge}
L.~Wang, J.~Zhang, J.~Chuan, R.~Ma, and A.~Fei, ``Edge intelligence for mission
  cognitive wireless emergency networks,'' \emph{IEEE Wireless Communications},
  vol.~27, no.~4, pp. 103--109, 2020.

\bibitem{Hewa2020multi}
T.~Hewa, A.~Braeken, M.~Ylianttila, and M.~Liyanage, ``Multi-access edge
  computing and blockchain-based secure telehealth system connected with {5G}
  and {IoT},'' in \emph{Proceedings of the IEEE Global Communications
  Conference (GLOBECOM)}, 2020, pp. 1--6.

\bibitem{whittle1988restless}
P.~Whittle, ``Restless bandits: Activity allocation in a changing world,''
  \emph{Journal of Applied Probability}, vol.~25, pp. 287--298, 1988.

\bibitem{fu2021restless}
J.~Fu, B.~Moran, and P.~G. Taylor, ``A restless bandit model for resource
  allocation, competition, and reservation,'' \emph{Operations Research},
  vol.~70, no.~1, pp. 416--431, 2021.

\bibitem{Sun2019online}
Z.~Sun and M.~R. Nakhai, ``An online learning algorithm for distributed task
  offloading in multi-access edge computing,'' \emph{IEEE Transactions on
  Signal Processing}, vol.~68, pp. 3090--3102, 2020.

\bibitem{Zhao2021energy}
M.~Zhao, J.-J. Yu, W.-T. Li, D.~Liu, S.~Yao, W.~Feng, C.~She, and T.~Q.~S.
  Quek, ``Energy-aware task offloading and resource allocation for
  time-sensitive services in mobile edge computing systems,'' \emph{IEEE
  Transactions on Vehicular Technology}, vol.~70, no.~10, pp. 10\,925--10\,940,
  2021.

\bibitem{Liu2023joint}
T.~Liu, D.~Guo, Q.~Xu, H.~Gao, Y.~Zhu, and Y.~Yang, ``Joint task offloading and
  dispatching for {MEC} with rational mobile devices and edge nodes,''
  \emph{IEEE Transactions on Cloud Computing}, vol.~11, no.~3, pp. 3262--3273,
  2023.

\bibitem{Song2022Joint}
H.~Song, B.~Gu, K.~Son, and W.~Choi, ``Joint optimization of edge computing
  server deployment and user offloading associations in wireless edge network
  via a genetic algorithm,'' \emph{IEEE Transactions on Network Science and
  Engineering}, vol.~9, no.~4, pp. 2535--2548, 2022.

\bibitem{Xiang2023joint}
B.~Xiang, J.~Elias, F.~Martignon, and E.~D. Nitto, ``Joint planning of network
  slicing and mobile edge computing: Models and algorithms,'' \emph{IEEE
  Transactions on Cloud Computing}, vol.~11, no.~1, pp. 620--638, 2023.

\bibitem{Xie2023sharingaware}
R.~Xie, J.~Fang, J.~Yao, X.~Jia, and K.~Wu, ``Sharing-aware task offloading of
  remote rendering for interactive applications in mobile edge computing,''
  \emph{IEEE Transactions on Cloud Computing}, vol.~11, no.~1, pp. 997--1010,
  2023.

\bibitem{Zhao2023deep}
P.~Zhao, J.~Tao, K.~Lui, G.~Zhang, and F.~Gao, ``Deep reinforcement
  learning-based joint optimization of delay and privacy in multiple-user {MEC}
  systems,'' \emph{IEEE Transactions on Cloud Computing}, vol.~11, no.~2, pp.
  1487--1499, 2023.

\bibitem{Wang2021Smart}
J.~Wang, L.~Zhao, J.~Liu, and N.~Kato, ``Smart resource allocation for mobile
  edge computing: A deep reinforcement learning approach,'' \emph{IEEE
  Transactions on Emerging Topics in Computing}, vol.~9, no.~3, pp. 1529--1541,
  2021.

\bibitem{hu2021dynamic}
H.~Hu, D.~Wu, F.~Zhou, S.~Jin, and R.~Q. Hu, ``Dynamic task offloading in
  {MEC}-enabled {IoT} networks: A hybrid {DDPG-D3QN} approach,'' in
  \emph{Proceedings of the IEEE Global Communications Conference (GLOBECOM)},
  2021, pp. 1--6.

\bibitem{wakgra2024multi}
F.~G. Wakgra, B.~Kar, S.~B. Tadele, S.-H. Shen, and A.~U. Khan,
  ``Multi-objective offloading optimization in {MEC} and vehicular-fog systems:
  A distributed-{TD3} approach,'' \emph{IEEE Transactions on Intelligent
  Transportation Systems}, pp. 1--13, 2024.

\bibitem{Chen2022dependency}
L.~Chen, J.~Wu, J.~Zhang, H.-N. Dai, X.~Long, and M.~Yao, ``Dependency-aware
  computation offloading for mobile edge computing with edge-cloud
  cooperation,'' \emph{IEEE Transactions on Cloud Computing}, vol.~10, no.~4,
  pp. 2451--2468, 2022.

\bibitem{Chen2021Energy}
Y.~Chen, N.~Zhang, Y.~Zhang, X.~Chen, W.~Wu, and X.~Shen, ``Energy efficient
  dynamic offloading in mobile edge computing for {Internet} of things,''
  \emph{IEEE Transactions on Cloud Computing}, vol.~9, no.~3, pp. 1050--1060,
  2021.

\bibitem{Wu2023computation}
G.~Wu, H.~Wang, H.~Zhang, Y.~Zhao, S.~Yu, and S.~Shen, ``Computation offloading
  method using stochastic games for software-defined-network-based multiagent
  mobile edge computing,'' \emph{IEEE Internet of Things Journal}, vol.~10,
  no.~20, pp. 17\,620--17\,634, 2023.

\bibitem{Su2023Truthful}
Y.~Su, W.~Fan, Y.~Liu, and F.~Wu, ``A truthful combinatorial auction mechanism
  towards mobile edge computing in industrial {Internet} of {Things},''
  \emph{IEEE Transactions on Cloud Computing}, vol.~11, no.~2, pp. 1678--1691,
  2023.

\bibitem{Maleki2023mobility}
E.~F. Maleki, L.~Mashayekhy, and S.~M. Nabavinejad, ``Mobility-aware
  computation offloading in edge computing using machine learning,'' \emph{IEEE
  Transactions on Mobile Computing}, vol.~22, no.~1, pp. 328--340, 2023.

\bibitem{Uniyal2021intelligent}
N.~Uniyal, A.~Bravalheri, X.~Vasilakos, R.~Nejabati, D.~Simeonidou,
  W.~Featherstone, S.~Wu, and D.~Warren, ``Intelligent mobile handover
  prediction for zero downtime edge application mobility,'' in
  \emph{Proceedings of the IEEE Global Communications Conference (GLOBECOM)},
  2021, pp. 1--6.

\bibitem{Yuan2022dynamic}
X.~Yuan, M.~Sun, and W.~Lou, ``A dynamic deep-learning-based virtual edge node
  placement scheme for edge cloud systems in mobile environment,'' \emph{IEEE
  Transactions on Cloud Computing}, vol.~10, no.~2, pp. 1317--1328, 2022.

\bibitem{wang2024adpative}
L.~Wang and J.~Zhang, ``Adaptive multi-armed bandit learning for task
  offloading in mobile edge computing,'' in \emph{Proceedings of the IEEE
  International Conference on Acoustics, Speech and Signal Processing
  (ICASSP)}, 2024, pp. 5285--5289.

\bibitem{li2024a}
H.~Li, L.~Li, S.~Tan, X.~Zhong, and S.~Zhang, ``A multi-user effective
  computation offloading mechanism for {MEC} system: Batched multi-armed
  bandits approach,'' \emph{IEEE Transactions on Network and Service
  Management}, 2024.

\bibitem{Lin2022popularity}
Y.~Lin, Y.~Zhang, J.~Li, F.~Shu, and C.~Li, ``Popularity-aware online task
  offloading for heterogeneous vehicular edge computing using contextual
  clustering of bandits,'' \emph{IEEE Internet of Things Journal}, vol.~9,
  no.~7, pp. 5422--5433, 2022.

\bibitem{jing2011efficient}
N.~Jing, M.~Yang, S.~Cheng, Q.~Dong, and H.~Xiong, ``An efficient {SVM}-based
  method for multi-class network traffic classification,'' in \emph{Proceedings
  of the IEEE International performance computing and communications conference
  (IPCCC)}, 2011, pp. 1--8.

\bibitem{magnano2015novel}
A.~Magnano, X.~Fei, A.~Boukerche, and A.~A. Loureiro, ``A novel predictive
  handover protocol for mobile {IP} in vehicular networks,'' \emph{IEEE
  Transactions on Vehicular Technology}, vol.~65, no.~10, pp. 8476--8495, 2015.

\bibitem{Gillam2018}
L.~{Gillam}, K.~{Katsaros}, M.~{Dianati}, and A.~{Mouzakitis}, ``Exploring
  edges for connected and autonomous driving,'' in \emph{Proceedings of the
  IEEE Conference on Computer Communications Workshops (INFOCOM WKSHPS)}, April
  2018, pp. 148--153.

\bibitem{Mukherjee2018}
M.~{Mukherjee}, L.~{Shu}, and D.~{Wang}, ``Survey of fog computing:
  Fundamental, network applications, and research challenges,'' \emph{IEEE
  Communications Surveys Tutorials}, vol.~20, no.~3, pp. 1826--1857,
  thirdquarter 2018.

\bibitem{Lee2017}
G.~{Lee}, W.~{Saad}, and M.~{Bennis}, ``An online secretary framework for fog
  network formation with minimal latency,'' in \emph{Proceedings of the IEEE
  International Conference on Communications (ICC)}, May 2017, pp. 1--6.

\bibitem{You2017}
C.~{You}, K.~{Huang}, H.~{Chae}, and B.~{Kim}, ``Energy-efficient resource
  allocation for mobile-edge computation offloading,'' \emph{IEEE Transactions
  on Wireless Communications}, vol.~16, no.~3, pp. 1397--1411, March 2017.

\bibitem{Deng2015}
R.~{Deng}, R.~{Lu}, C.~{Lai}, and T.~H. {Luan}, ``Towards power
  consumption-delay tradeoff by workload allocation in cloud-fog computing,''
  in \emph{Proceedings of the IEEE International Conference on Communications
  (ICC)}, June 2015, pp. 3909--3914.

\bibitem{Wu2020}
J.~Wu, E.~W.~M. Wong, Y.-C. Chan, and M.~Zukerman, ``Power consumption and
  {GoS} tradeoff in cellular mobile networks with base station sleeping and
  related performance studies,'' \emph{IEEE Transactions on Green
  Communications and Networking}, vol.~4, no.~4, pp. 1024--1036, 2020.

\bibitem{gittins2011multiarmed}
J.~C.~Gittins, K.~Glazebrook, and R.~R.~Weber, \emph{Multi-armed bandit
  allocation indices: 2nd edition}.\hskip 1em plus 0.5em minus 0.4em\relax
  Wiley, Mar. 2011.

\bibitem{wang2018energy}
Q.~Wang, J.~Fu, J.~Wu, B.~Moran, and M.~Zukerman, ``Energy-efficient
  priority-based scheduling for wireless network slicing,'' in
  \emph{Proceedings of the IEEE Global Communications Conference (GLOBECOM)},
  Abu Dhabi, UAE, Dec. 2018.

\bibitem{fu2020energy}
J.~Fu and B.~Moran, ``Energy-efficient job-assignment policy with
  asymptotically guaranteed performance deviation,'' \emph{IEEE/ACM
  Transactions on Networking}, vol.~28, no.~3, pp. 1325--1338, 2020.

\bibitem{boyd2004convex}
S.~P. Boyd and L.~Vandenberghe, \emph{Convex optimization}.\hskip 1em plus
  0.5em minus 0.4em\relax Cambridge University Press, 2004.

\bibitem{fu2020resource}
J.~Fu, B.~Moran, P.~G. Taylor, and C.~Xing, ``Resource competition in virtual
  network embedding,'' \emph{Stochastic Models}, vol.~37, no.~1, pp. 231--263,
  2020.

\bibitem{zhang2017virtual}
P.~Zhang, H.~Yao, and Y.~Liu, ``Virtual network embedding based on computing,
  network, and storage resource constraints,'' \emph{IEEE Internet of Things
  Journal}, vol.~5, no.~5, pp. 3298--3304, 2017.

\bibitem{clusterdata:Wilkes2011}
J.~Wilkes, ``More {Google} cluster data,'' Google research blog, Nov. 2011,
  posted at
  \url{http://googleresearch.blogspot.com/2011/11/more-google-cluster-data.html},
  accessed at Jul. 8, 2019.

\bibitem{clusterdata:Reiss2011}
C.~Reiss, J.~Wilkes, and J.~L. Hellerstein, ``{Google} cluster-usage traces:
  format + schema,'' Google Inc., Mountain View, CA, USA, Technical Report,
  Nov. 2011, revised 2014-11-17 for version 2.1. Posted at
  \url{https://github.com/google/cluster-data}, accessed at Jul. 8, 2019.

\bibitem{Crovella1997self}
M.~E. Crovella and A.~Bestavros, ``Self-similarity in {World Wide Web} traffic:
  evidence and possible causes,'' \emph{IEEE/ACM Transactions on Networking},
  vol.~5, no.~6, pp. 835--846, Dec. 1997.

\bibitem{harchol2013performance}
M.~Harchol-Balter, \emph{Performance Modeling and Design of Computer Systems:
  Queueing Theory in Action}.\hskip 1em plus 0.5em minus 0.4em\relax Cambridge
  University Press, 2013.

\end{thebibliography}

%\newpage
\begin{IEEEbiography}
[{\includegraphics[width=1in,height=1.25in,clip,keepaspectratio]{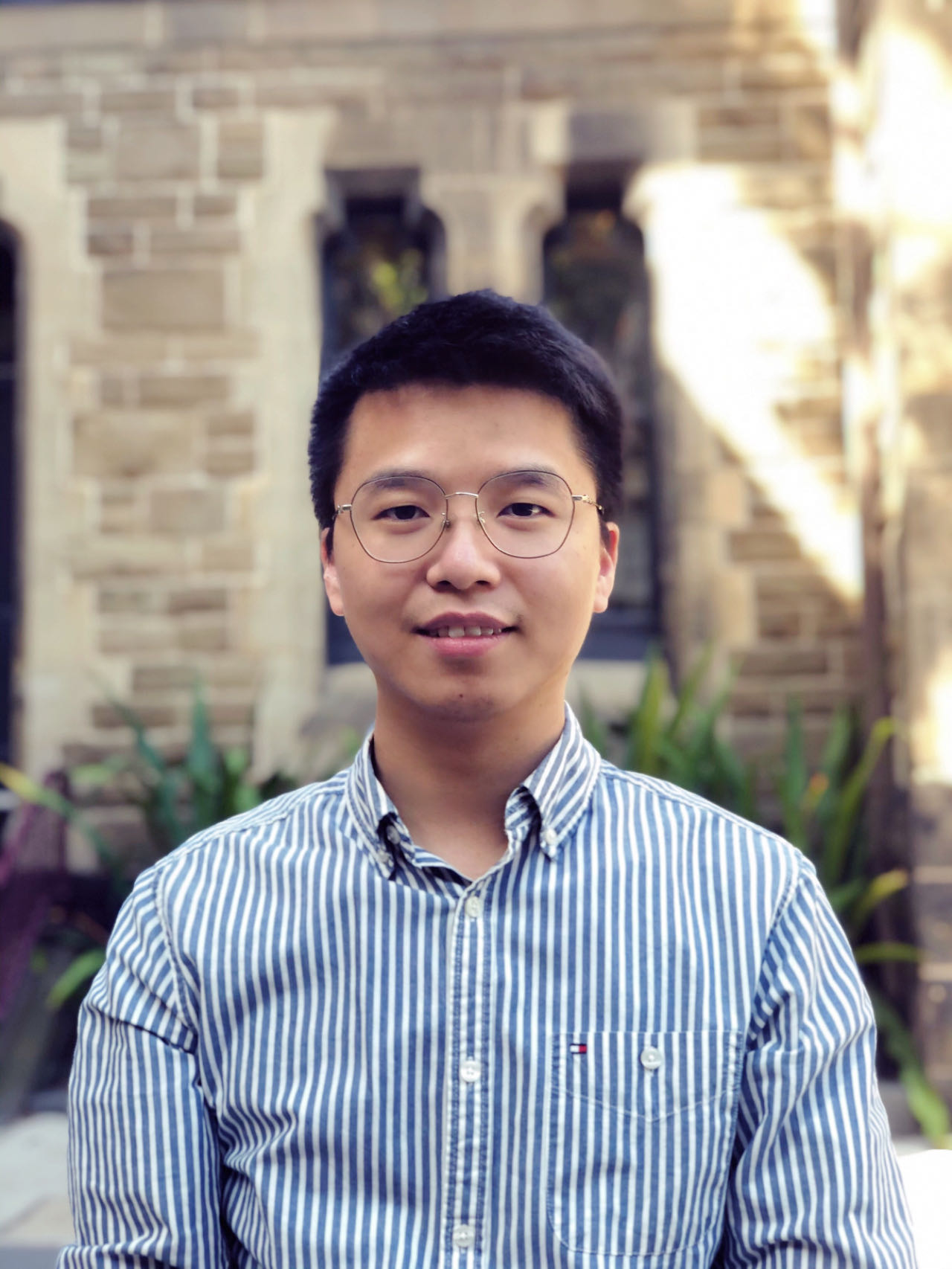}}]
{Ling Hou} received Bachelor of Electrical and Electronic Engineering degree from Southwest Petroleum University, China, in 2013, and Master of Electrical and Electronic Engineering degree from RMIT University, Melbourne, Australia, in 2018. He received Ph.D, School of Engineering, STEM College, RMIT University, Melbourne, Australia in 2024. His research interests include 5G/6G communications networks, multi-access edge computing, vehicular networks, and machine learning.
\end{IEEEbiography}

\begin{IEEEbiography}[%\vspace{-1mm}
{\includegraphics[width=1in,height=1.25in,clip,keepaspectratio]{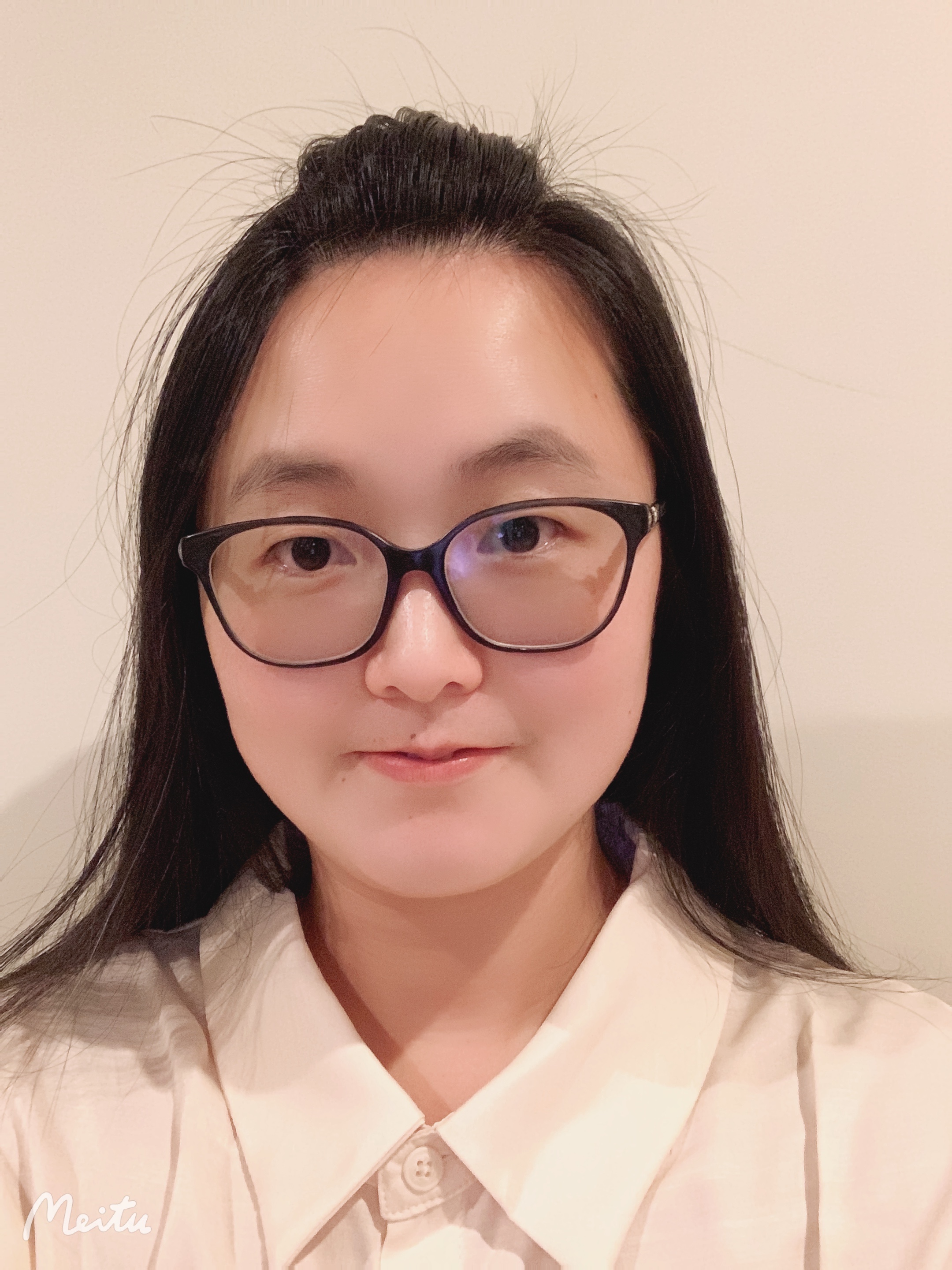}}]{Shi Li} (Student Member, IEEE) received the B.E. degree with Honours in Electrical and Electronic Engineering (2014) and M.E degree in Telecommunications Engineering (2015) from the University of Melbourne, VIC, Australia. She is currently pursuing the Ph.D. degree in Information and Communication Technology with the School of Science, Computing and Engineering Technologies, Swinburne University of Technology, Melbourne, VIC, Australia. Her research interests include multi-agent cloud robotics and deep reinforcement learning.
\end{IEEEbiography}

%\vskip -2\baselineskip plus -1fil

\begin{IEEEbiography}[{\includegraphics[width=1in,height=1.25in,clip,keepaspectratio]{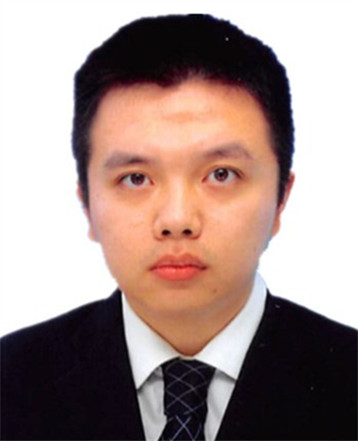}}]{Zhishu Shen}
received the B.E. degree from the School of Information Engineering at the Wuhan University of Technology, Wuhan, China, in 2009, and the M.E. and Ph.D. degrees in Electrical and Electronic Engineering and Computer Science from Nagoya University, Japan, in 2012 and 2015, respectively. He is currently an Associate Professor in the School of Computer Science and Artificial Intelligence, Wuhan University of Technology. From 2016 to 2021, he was a research engineer of KDDI Research, Inc., Japan. His major interests include network design and optimization, data learning, edge computing and the Internet of Things.
\end{IEEEbiography}

\begin{IEEEbiography}
[{\includegraphics[width=1in,height=1.25in,clip,keepaspectratio]{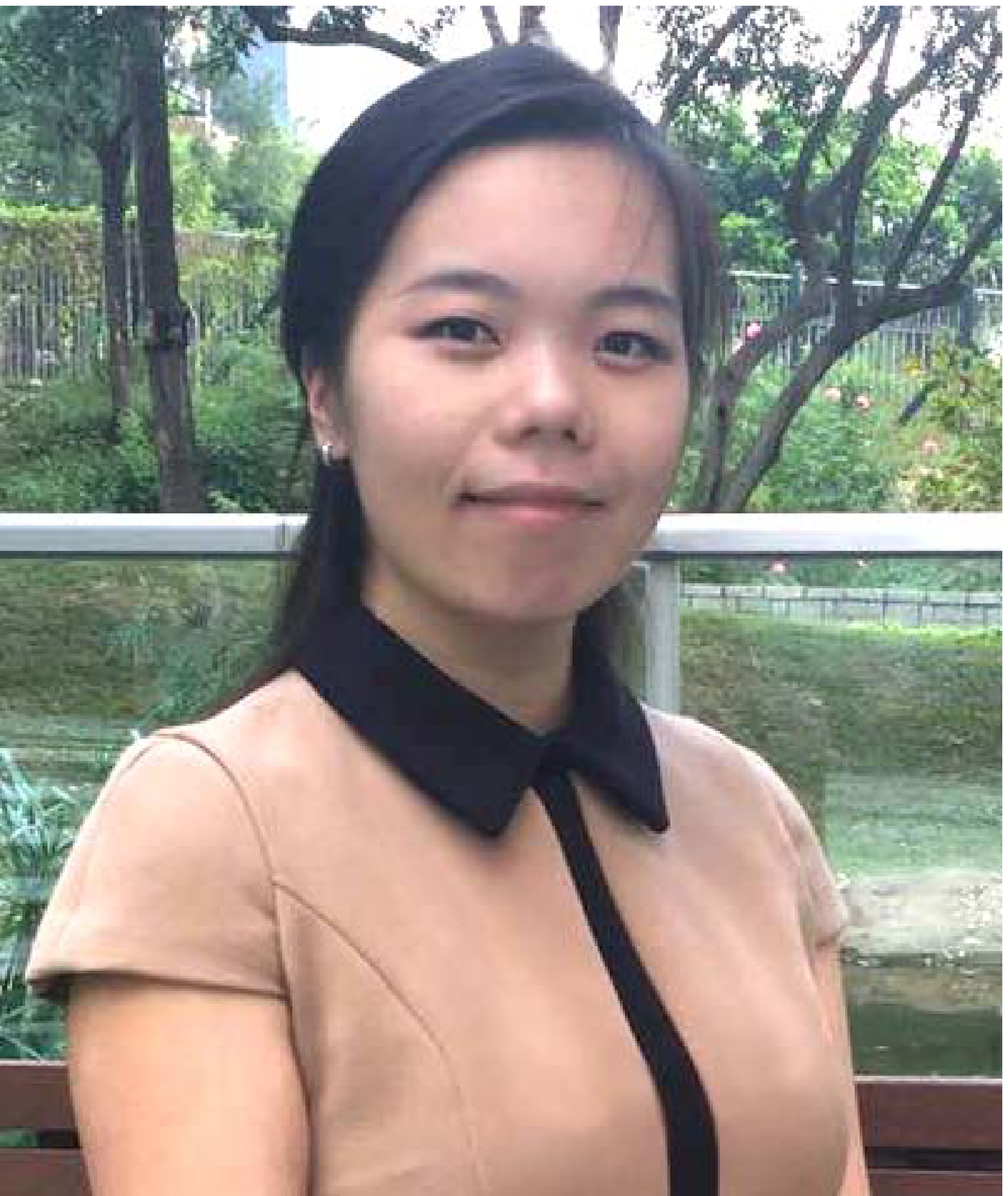}}]
{Jing Fu} (S'15-M'16) received the B.Eng. degree in computer science from Shanghai Jiao Tong University, Shanghai, China, in 2011, and the Ph.D. degree in electronic engineering from the City University of Hong Kong in 2016. She has been with the School of Mathematics and Statistics, the University of Melbourne as a Post-Doctoral Research Associate from 2016 to 2019. She has been a lecturer in the discipline of Electronic \& Telecommunications Engineering, RMIT University, since 2020. Her research interests now include energy-efficient networking/scheduling, resource allocation in large-scale networks, semi-Markov/Markov decision processes, restless multi-armed bandit problems, stochastic optimization.
\end{IEEEbiography}
\begin{IEEEbiography}
[{\includegraphics[width=1in,height=1.25in,clip,keepaspectratio]{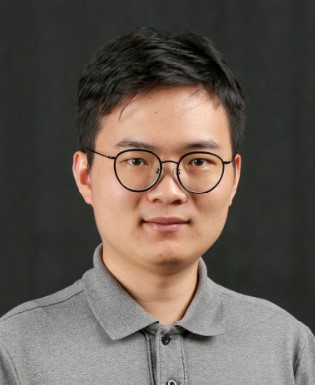}}]
{Jingjin Wu} (S'15-M'16) received the B.Eng. degree (with first class honors) in information engineering in 2011, and the Ph.D. degree in electronic engineering in 2016, both from City University of Hong Kong, Hong Kong SAR. Since 2016,  he has been with the Department of Statistics  and  Data Science, BNU-HKBU United International College, Zhuhai, Guangdong, China, where he is currently an Associate Professor. His current research focuses on design, performance analysis, and optimization of wireless communication networks.
\end{IEEEbiography}
\begin{IEEEbiography}[%\vspace{-1mm}
{\includegraphics[width=1in,height=1.25in,clip,keepaspectratio]{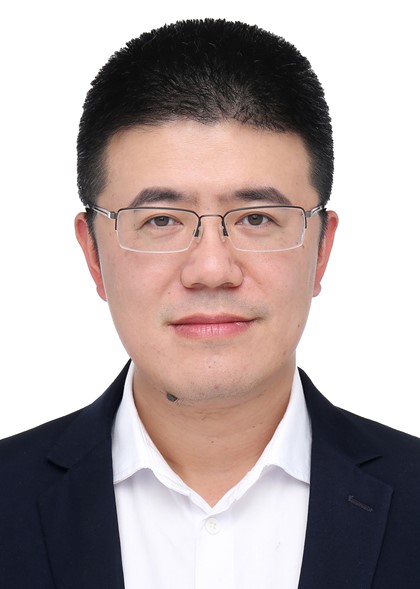}}]{Jiong Jin} (Member, IEEE) received the B.E. degree with First Class Honours in Computer Engineering from Nanyang Technological University, Singapore, in 2006, and the Ph.D. degree in Electrical and Electronic Engineering from the University of Melbourne, VIC, Australia, in 2011. He is currently a full Professor in the School of Science, Computing and Engineering Technologies, Swinburne University of Technology, Melbourne, VIC, Australia. His research interests include network design and optimization, edge computing and networking, robotics and automation, and cyber-physical systems and Internet of Things as well as their applications in smart manufacturing, smart transportation and smart cities. He was recognized as an Honourable Mention in the AI 2000 Most Influential Scholars List in IoT (2021 and 2022). He is currently an Associate Editor of IEEE Transactions on Industrial Informatics.
\end{IEEEbiography}
\clearpage
\appendices
\setcounter{page}{1}
\IncMargin{1em}

%\IncMargin{1em}

\begin{algorithm*}[b]\small
\linespread{1}\selectfont

%\SetKwFunction{IndexPolicy}{$(\bm{\eta}_e)\gets$ IndexPolicy}
\SetKwProg{Fn}{Function}{}{End}
\SetKwInOut{Input}{Input}\SetKwInOut{Output}{Output}
\SetAlgoLined
\DontPrintSemicolon

\Input{$\pmb{\gamma}\in\mathbb{R}_0^K$ and $\bm{\eta}\in\mathbb{R}_0^{IJL}$.}
\Output{$\pmb{\nu}^*(\pmb{\gamma},\bm{\eta})$ defined in Eq.~\eqref{eqn:nu_star}, and action variables $\pmb{\alpha}^{\varphi^*(\pmb{\gamma},\bm{\eta})}\coloneqq \Bigl(\alpha^{\varphi^*(\pmb{\gamma},\bm{\eta})}_{i,i',j,k}(x):~i,i'\in[I],j\in[J],k\in[K],x\in\mathscr{X}_{i,i',j,k}\Bigr)$.}

\Fn{Multipliers}{
	Initialize $\pmb{\nu}^*(\pmb{\gamma},\bm{\eta})\gets \bm{0}$\;
    $\alpha^{\varphi^*(\pmb{\gamma},\bm{\eta})}_{i,i',j,k}(x)\gets 0$ for all $i,i'\in[I]$, $j\in[J]$, $k\in\mathscr{K}$, and $x\in\mathscr{X}_{i,i',j,k}$.\;
    Rank all the tuples $(i,i',j,k)$ according to the ascending order of their $\psi_j(i,i',j)$, where tie cases are broken by selecting the shortest lifespans.\;
    $N\gets I^2J(K+1)$\;    
    Let $(i_{\iota},i'_{\iota},j_{\iota},k_{\iota})$ represent the $\iota$th tuple in the above mentioned ranking.\;
    $f_j\gets -1$ for all $j\in[J]$\;
    \For{$\iota\in[N]$}{
        \If{$f_{j_{\iota}}=0$}{
            {\bf Break}\;
        }
        \For{$x\in\mathscr{X}_{i_{\iota},i'_{\iota},j_{\iota},k_{\iota}}$}{
            $a_1\gets \min\Bigl\{a\in[0,1]~\Bigl| \text{ \eqref{eqn:action_constraint:relax} achieves equality for $j=j_{\iota}$ by setting } \alpha_{i_{\iota},i'_{\iota},j_{\iota},k_{\iota}}^{\varphi^*(\pmb{\gamma},\bm{\eta})}(x) =a\Bigr\}\cup \Bigl\{1\Bigr\}$\;
            $a_2\gets \min\Bigl\{a\in[0,1]~\Bigl| \text{ \eqref{eqn:capacity_constraint1:relax} achieves equality for $k=k_{\iota}$ by setting } \alpha_{i_{\iota},i'_{\iota},j_{\iota},k_{\iota}}^{\varphi^*(\pmb{\gamma},\bm{\eta})}(x) =a\Bigr\}\cup \Bigl\{1\Bigr\}$\;
            $a_3\gets \min\Bigl\{a\in[0,1]~\Bigl| \text{ \eqref{eqn:capacity_constraint2:relax} achieves equality for $(i,j,\ell)=\bigl(i_{\iota},i'_{\iota},\ell(k_{\iota})\bigr)$ by setting } \alpha_{i_{\iota},i'_{\iota},j_{\iota},k_{\iota}}^{\varphi^*(\pmb{\gamma},\bm{\eta})}(x) =a\Bigr\}\cup \Bigl\{1\Bigr\}$\;
            
            $\alpha_{i_{\iota},i'_{\iota},j_{\iota},k_{\iota}}^{\varphi^*(\pmb{\gamma},\bm{\eta})}(x) \gets \min\{a_1,a_2,a_3\}$\;
            \If{$\min\{a_1,a_2,a_3\}=0$}{
                $f_{j_{\iota}} \gets 0$\;
                $\nu^*_{j_{\iota}}(\pmb{\gamma},\bm{\eta})\gets\psi_{j_{\iota}}(i_{\iota},i'_{\iota},k_{\iota})$\;
                {\bf Break}\;
            }
        }
    }
    \Return $\pmb{\nu}^*(\pmb{\gamma},\bm{\eta})$ and $\pmb{\alpha}^{\varphi^*(\pmb{\gamma},\bm{\eta})}$\;
}
\caption{Computing $\pmb{\nu}^*(\pmb{\gamma},\bm{\eta})$.}\label{algo:nu_star}
\end{algorithm*}
We provide summary of important variables in Tables~\ref{table:symbol1}.
\begin{table*}[b]
	\caption{Important Symbols}\label{table:symbol1}
	\begin{tabular}{p{1.4cm}p{7cm}p{1.4cm}p{6.9cm}}
		\hline
		\multicolumn{4}{l}{Real Numbers and Vectors}\\
		\hline
		$K$&The number of SC groups in the edge network&
		$L$&The number of destination areas\\
		$I$&The number of channels&
        $J$&The number of task classes\\
		$C_k$&The capacity of SC group $k$&
        $N_i$& The number of sub-channels of channel $i$\\
		$w_{j,k}$&The number of SC units occupied by a $j$-task if it is served by SC group $k$ &
        $\mu_{i,j}$&The transmission rate for transmitting a $j$-task through channel $i$\\
		$\lambda_j$&The arrival rate of $j$-tasks&
		$\mu_j(i,i',k)$&The reciprocal of the expected lifespan of a $j$-task served by channels $i,i'$ and SC group $k$\\
		$\varepsilon_k$& The operational power of an SC unit of group $k$&
		$\varepsilon_k^0$&The  static power of an SC unit of group $k$\\
		$\bar{\varepsilon}_j$&The expected energy consumption per $j$-task processed by the cloud&
		$\psi_j(i,i',k)$&The index for assigning a $j$-task to resource tuple $i,i',k)$, defined in \eqref{eqn:indexability:2}, used for constructing the HEE-ACC policy\\	$D_0$ & The transmission time between the edge network and the cloud&  $\mathcal{E}^{\phi}_c$ &The long-run average power consumption for computing tasks offloaded to the cloud\\	$\pmb{\nu}\in\mathbb{R}^J_0$&Lagrange multipliers for the relaxed constraints in \eqref{eqn:action_constraint:relax}&		
        $\pmb{\gamma}\in\mathbb{R}_0^K$&Lagrange multipliers for the relaxed constraints in \eqref{eqn:capacity_constraint1:relax}\\
		$\bm{\eta}\in\mathbb{R}_0^I$&Lagrange multipliers for the relaxed constraints in \eqref{eqn:capacity_constraint2:relax}&
        $a^{\phi}_{i,i',j,k}(\bm{x})$
        &The action variable of the task offloading problem, taking values in $\{0,1\}$.
        Given the state vector $\bm{X}^{\phi}(t)=\bm{x}\in\mathscr{X}$, it indicates whether or not a newly arrived $j$-task is assigned to the resource tuple $(i,i',k)$ at time $t$ under policy $\phi$\\
        $\bm{a}^{\phi}(\bm{x})$&
        $\bm{a}^{\phi}(\bm{x})\coloneqq (a^{\phi}_{i,i',j,k}(\bm{x}): i,i'\in[I], j\in[J], k\in\mathscr{K})$, the action vector of the task offloading problem &
        $\alpha^{\phi}_{i,i',j,k}(x)$
        &The action variable of the sub-problem associated with $(i,i',j,k)$, taking values in $[0,1]$.
        Given the state of the sub-problem $X^{\phi}_{i,i',j,k}(t)=x\in\mathscr{X}_{i,i',j,k}$, it indicates the probability of assigning a newly arrived $j$-task to the resource tuple $(i,i',k)$ at time $t$ under policy $\phi$\\
        \hline
		\multicolumn{4}{l}{Important Labels}\\
		\hline
         $i\in[I]$& Label of a channel&
        $j\in[J]$& Label of a task class \\
        $k\in\mathscr{K}$ & Label of an SC group &
        
        $\ell\in[L]$ & Label of a destination area which locates a set of SC groups\\
		\hline
		\multicolumn{4}{l}{Sets}\\
		\hline
		$\mathscr{K}$&The set of all SC groups in the edge and cloud&
		$\mathscr{K}_{\ell}$&The set of SC groups located in the area $\ell\in[L]$\\
		$\mathscr{X}$&The state space of the entire task offloading problem&
		$\mathscr{X}_{i,i',j,k}$&The state space of the sub-problem associated with $(i,i',j,k)$\\
		$\Phi$&The set of all policies determined by action variables $a^{\phi}_{i,i',j,k}(\bm{x})$ for all $i,i'\in[I]$, $j\in[J]$, $k\in\mathscr{K}$, and $\bm{x}\in\mathscr{X}$&
		$\tilde{\Phi}$&The set of all policies determined by action variables $\alpha^{\phi}_{i,i',j,k}(x)$ for all $x\in\mathscr{X}_{i,i',j,k}$, $i,i'\in[I]$, $j\in[J]$, and $k\in\mathscr{K}$.\\
		\hline
		\multicolumn{4}{l}{Random Variables}\\
		\hline
		$X^{\phi}_{i,i',j,k}(t)$&The number of $j$-tasks that are being served by resource tuple $(i,i',k)$ at time $t$ under policy $\phi$&
        $\bm{X}^{\phi}(t)$& $\bm{X}^{\phi}(t)\coloneqq(X^{\phi}_{i,i',j,k}(t): i,i'\in[I], j\in[J], k\in\mathscr{K})$, the state vector of the entire task offloading problem at time $t$ under policy $\phi$\\
		\hline
	\end{tabular}
\end{table*}
\section{Diagram for Decision Making}\label{app:diagram}
\begin{figure}[t]
\centering
\includegraphics[width=\linewidth]{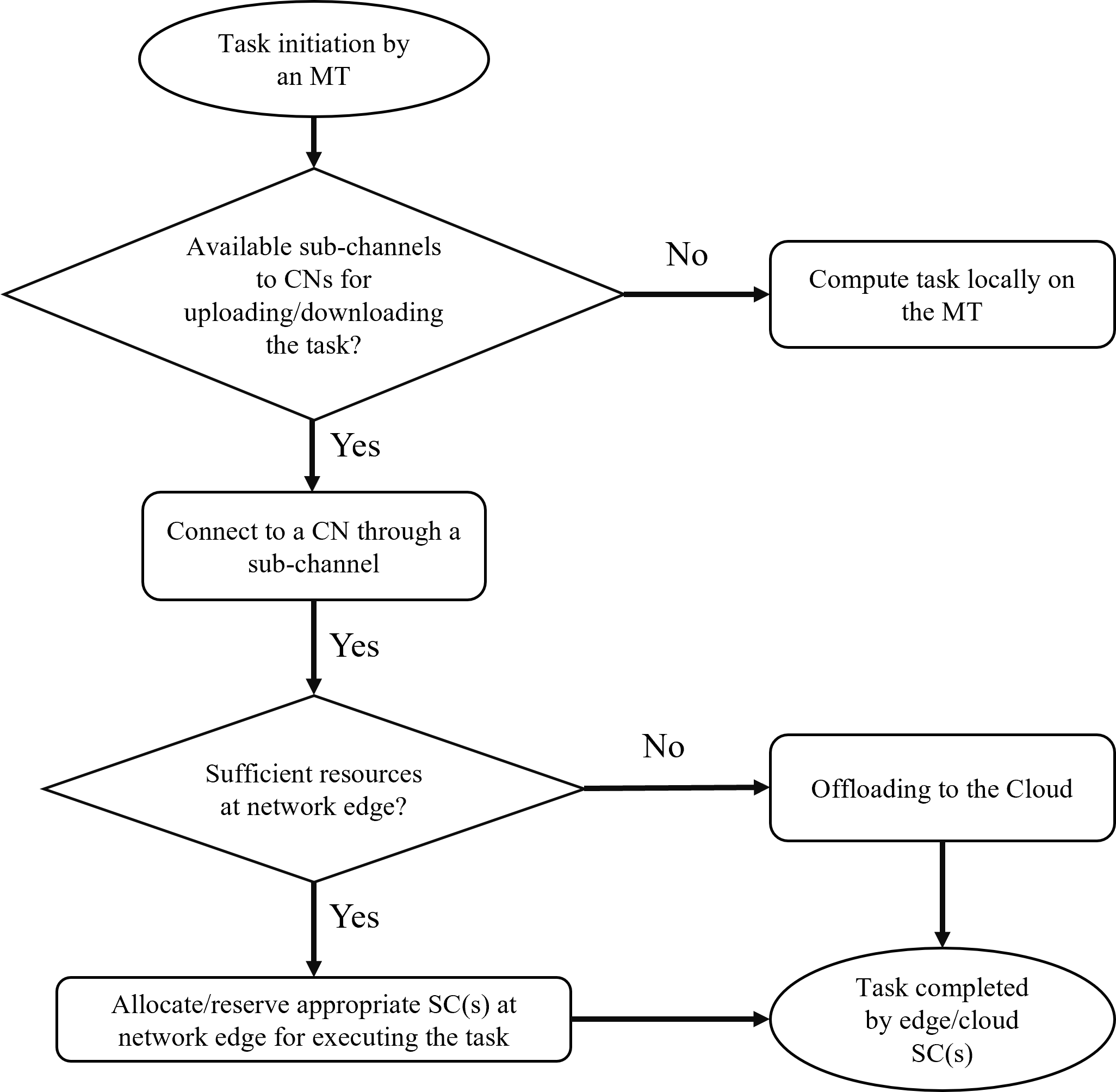}
\caption{Diagram for decision making process.}
\label{diagram}
\end{figure}

We provide a diagram in \figurename~\ref{diagram} to demonstrate the process of deciding channel-SC tuples to serve newly arrived tasks.

\section{Proof of Proposition~\ref{prop:asym_opt}}\label{app:prop:asym_opt}

\begin{proof}{Proposition~\ref{prop:asym_opt}}
Let $\Gamma^{\phi}$ represent the long-run average power consumption (energy consumption rate), as described in \eqref{eqn:obj}, under policy $\phi\in\tilde{\Phi}$, where recall $\tilde{\Phi}$ is the set of the policies with randomized action variables and includes those policies applicable to the original and/or the relaxed problem.
Let $\Gamma^*$ and $\Gamma^{\text{R},*}$ represent the minimized objective function of the original and the relaxed problem, respectively.

If, for $\bm{\nu}=\bm{\nu}^*$, $\pmb{\gamma}=\pmb{\gamma}^*$ and $\bm{\eta}=\bm{\eta}^*$, a policy $\phi\in\tilde{\Phi}$ satisfying \eqref{eqn:indexbility:1}  is optimal to the relaxed problem described in \eqref{eqn:obj:relax}, \eqref{eqn:action_constraint:relax}, \eqref{eqn:capacity_constraint1:relax} and \eqref{eqn:capacity_constraint2:relax}, then, together with Corollary~\ref{coro:indexability},
we obtain that, for such a policy $\phi$, $\Gamma^{\phi} = \Gamma^{\text{R},*} = L(\bm{\nu}^*,\pmb{\gamma}^*,\bm{\eta}^*)$.

Without loss of generality, for $k\in[K]$ and $i\in[I]$, we rewrite $C_k = hC_k^0$, $N_i = hN_i^0$, and $\lambda_j = h \lambda_j^0$ for $h\in\mathbb{N}_+$, $C_k^0,N_i^0 \in \mathbb{N}_+$, and $\lambda_j^0 \in\mathbb{R}_+$.
From \cite[Theorem EC.1]{fu2021restless}, for a policy $\phi$ satisfying \eqref{eqn:indexbility:1}, there exists a policy $\varphi$ derived based on $\psi_j(i,i',k)$ such that
\begin{equation}\label{eqn:asym_opt:1}
    \lim\nolimits_{
    h \rightarrow +\infty} 
    \lvert \Gamma^{\varphi}-\Gamma^{\phi}\rvert = 0.
\end{equation}
Note that $\Gamma^{\varphi}$ and $\Gamma^{\phi}$ in \eqref{eqn:asym_opt:1} are dependent on $h$ through $C_k$, $N_i$, and $\lambda_j$. 
It follows that
\begin{equation}\label{eqn:asym_opt:2}
    \lim\nolimits_{
    h \rightarrow +\infty} 
    \lvert \Gamma^{\varphi}-\Gamma^{\text{R},*}\rvert = 0,
\end{equation}
which proves the proposition.\qed
\end{proof}

%\section{Pseudo-code for implementing HEE-ACC}\label{app:algorithm 1}
%The pseudo-code for implementing HEE-ACC is given in Algorithm~\ref{algo:HEE-ACC}.

\section{Proof of Proposition~\ref{prop:asym_opt_zero}}\label{app:prop:asym_opt_zero}

\begin{proof}{Proposition~\ref{prop:asym_opt_zero}}
If, for each $j\in[J]$ and resource tuple $(i,i',k)\in[I]^2\times[K]$ with $\mu_j(i,i',k)>0$, the SC $k$, channel $i$ or channel $i'$ is dominant, then the system is \emph{weakly coupled}~\cite[Section 3.3.1]{fu2021restless}. 
If the blocking probabilities of all task classes $j\in[J]$ are positive in the asymptotic regime, then the system is in \emph{heavy traffic}~\cite[Section 3.3.2]{fu2021restless}. 
By invoking \cite[Corollary EC.1]{fu2021restless}, a policy $\phi\in\Phi$ given by
\begin{multline*}
    a^{\phi}_{i,i',j,k}\bigl(\bm{X}^{\phi}(t)\bigr) \\= \begin{cases}
        1,& \begin{array}{l}\text{if } (i,i',k) = \\~~~\arg\min\nolimits_{(i,i',k)\in\mathscr{T}_j(\bm{X}^{\text{HEE-ACC}}(t))}\frac{\psi_j(i,i',k)}{w_{j,k}(1+\frac{\lambda_j}{\mu_j(i,i',k)})},\end{array}\\
        0, & \text{otherwise},
    \end{cases}
\end{multline*}
is asymptotically optimal - it approaches optimality as $h\rightarrow +\infty$. 
When $w_{j,k}(1+\frac{\lambda_j}{\mu_j(i,i',k)})$ is a constant for all $j\in[J]$ and $(i,i',k)\in[I]^2\times[K]$ with $\mu_j(i,i',k)$, the above described policy $\phi$ coincides with HEE-ACC-zero.
It proves the proposition. \qed
\end{proof}

\section{Pseudo-code for Computing $\bm{\nu}^*$ and the associated action variables for $\varphi^*$}\label{app:algo:nu_star}
The pseudo-code for computing $\bm{\nu}^*(\pmb{\gamma},\bm{\eta})$ and the associated action variables for $\varphi^*(\pmb{\gamma},\bm{\eta})$ with given $\pmb{\gamma}\in\mathbb{R}^K_0$ and $\bm{\eta}\in\mathbb{R}^I_0$ is given in Algorithm~\ref{algo:nu_star}.

\section{Simulation Settings}\label{app:simulation}
In the system discussed in Section~VI, 
\begin{itemize}
\item for $i\in[I]$, the channel capacities are $N_i = \bar{N}_{\lfloor (i-1)/h\rfloor+1}$ where $(\bar{N}_1,\bar{N}_2,\ldots,\bar{N}_{10}) = (8,5,5,7,6,5,5,6,5,7)$ and $(\bar{N}_{11},\bar{N}_{12},\ldots,\bar{N}_{20})=(9,6,6,5,5,9,9,9,5,7)$;
\item the SC capacities are $(C_1^0,C_2^0,C_{3}^0) = (5,5,8)$;
\item for $k\in[K]$, the requested AC units $w_{1,k}=3$, $w_{2,k} = 4$, $w_{3,k}=2$ and $w_{4,k} =1$ except that $w_{1,2}=w_{3,1} = w_{3,2}=w_{4,1}=w_{4,2}=+\infty$;
\item the operational power $(\varepsilon_1,\varepsilon_2,\varepsilon_3) = (3.362,3.996 ,8.979)$, and the power consumption per task processed in the cloud $\bar{\varepsilon}_j$ ($j\in[J])$ are set to be $20.1\rho/\lambda_j$ (or $20.1\rho_j/\lambda_j$ for the case with heterogeneous traffic intensities presented in \figurename~\ref{fig:varying-time:seed50-rho1.5:timeline} and \ref{fig:seed50-varying-rho1.5:scaler}), of which the unit is Watt;
\item and the arrival rates of requests  are $(\lambda_1,\lambda_2,\ldots,\lambda_4) = (1.097, 1.026, 1.456,1.383)$, which represent the numbers of arrived requests per second; 
\end{itemize}
All the above-listed numbers are instances generated by a pseudo-random generator in C++.
Let $\mathscr{I}_j^{\text{start}}$ and $\mathscr{I}_j^{\text{end}}$ represent the sets of eligible candidates for the starting and ending channels of requests in class $j\in[J]$, respectively.
We set 
$(\mathscr{I}_1^{\text{start}},\mathscr{I}_1^{\text{end}}) = (\{3,	5,	6,	7,	10,	12,	13,	14,	20\}, \{3,	5,	7,	8,	9,\\	10,	11,	12,	13,	14,	15,	16\})$,
$(\mathscr{I}_2^{\text{start}},$ 
$\mathscr{I}_2^{\text{end}})=(\{2,	9,	17,	19\},\\\{2, 7,	8,	9,	10,	11,	12,	16,	19\})$,
$(\mathscr{I}_3^{\text{start}},\mathscr{I}_3^{\text{end}}) = (\{2,	9,	17,\\	19\},\{2,	8,	9,	11,	19\})$, and
$(\mathscr{I}_4^{\text{start}},\mathscr{I}_4^{\text{end}}) = (\{1,	4,	6,	17,	18\},\\\{1,	4,	17,	18,	19\})$. 
In the simulated system, each CN in the system is considered to connect with all the ACs through wired links with sufficiently large capacities.

For the HEE-ALRN policy, we set the hyper-parameters $\Delta_{\lambda}=\Delta_{\eta}=\Delta^+_{\lambda} = \Delta^+_{\eta}=2$ and $\bar{M}=100$.

For the simulations in \figurename~\ref{fig:varying-time:seed50-rho1.5:timeline} and \ref{fig:seed50-varying-rho1.5:scaler}, the offered traffic intensities are $\rho_1 = 3.876$, $\rho_2=9.115$, $\rho_3 = 7.042$, and $\rho_4 = 8.150$.

\end{document}